\documentclass{article}

\usepackage{arxiv}

\usepackage[T1]{fontenc}    % use 8-bit T1 fonts
\usepackage{hyperref}       % hyperlinks
\usepackage{url}            % simple URL typesetting
\usepackage{booktabs}       % professional-quality tables
\usepackage{amsfonts}       % blackboard math symbols
\usepackage{nicefrac}       % compact symbols for 1/2, etc.
\usepackage{microtype}      % microtypography
\usepackage{lipsum}
\usepackage{graphicx}
\graphicspath{ {./images/} }
\usepackage{amsmath}
\usepackage{multirow}
\usepackage[square,numbers,super]{natbib}
\usepackage{float}
\usepackage{placeins}

\providecommand{\normaldistn}{\mathrm{Normal}}
\providecommand{\lognormaldistn}{\mathrm{logNormal}}

\providecommand{\expect}{\mathrm{E}}
\providecommand{\uisd}{\sigma_{\mathrm{u}}}
\providecommand{\Var}{\mathrm{Var}}
\providecommand{\Cov}{\mathrm{Cov}}
\providecommand{\young}{\mathrm{Y}}
\providecommand{\old}{\mathrm{O}}
\providecommand{\male}{\mathrm{M}}
\providecommand{\female}{\mathrm{F}}
\usepackage{threeparttable}

\title{Utilizing subgroup information in random-effects meta-analysis of few studies}

\author{
Ao Huang \\
Graduate School of Information Sciences\\
Tohoku University\\
Sendai, Miyagi 980-8579, Japan\\
Department of Medical Statistics\\
University Medical Center G\"{o}ttingen\\
Humboldtallee 32, 37073 G\"{o}ttingen, Germany\\
\texttt{ao.huang.b4@tohoku.ac.jp}\\
\And
Christian R\"{o}ver \\
Department of Medical Statistics\\
University Medical Center G\"{o}ttingen\\
Humboldtallee 32, 37073 G\"{o}ttingen, Germany\\
\texttt{christian.roever@med.uni-goettingen.de}\\
\And
Tim Friede \\
Department of Medical Statistics\\
University Medical Center G\"{o}ttingen\\
Humboldtallee 32, 37073 G\"{o}ttingen, Germany\\
DZHK (German Center for Cardiovascular Research), partner site Lower Saxony\\
G\"{o}ttingen, Germany\\
DZKJ (German Center for Child and Adolescent Health)\\
G\"{o}ttingen, Germany\\
\texttt{tim.friede@med.uni-goettingen.de}\\
}

\begin{document}
\maketitle
\begin{abstract}
Random-effects meta-analyses are widely used for evidence synthesis in medical research. However, conventional methods based on large-sample approximations often exhibit poor performance in case of very few studies (e.g., 2~to~4), which is very common in practice. Existing methods aiming to improve small-sample performance either still suffer from poor estimates of heterogeneity or result in very wide confidence intervals. Motivated by meta-analyses evaluating surrogate outcomes, where units nested within a trial are often exploited  when the number of trials is small, we propose an inference approach based on a common-effect estimator synthesizing data from the subgroup level instead of the study level. Two DerSimonian-Laird type heterogeneity estimators are derived using the subgroup-level data, and are incorporated into the Henmi-Copas type variance to adequately reflect variance components. We considered $t$-quantile based intervals to account for small-sample properties and used flexible degrees of freedom to reduce interval lengths. A comprehensive simulation is conducted to study the performance of our methods depending on a subgroup selection procedure and various magnitudes of subgroup interaction effects as well as subgroup prevalences. Some general recommendations are provided on how to select the subgroups, and methods are illustrated using two example applications.
\end{abstract}

% keywords can be removed
\keywords{Meta-analysis, heterogeneity, subgroups, small samples }

\section{Introduction}
In a meta-analysis all studies on a particular topic are combined to yield a more reliable and valid answer for a specific research question than from a single trial. There are two commonly used types of models to achieve this purpose, the \emph{common-effect} model (also known as the \emph{fixed-effect} model), and the \emph{random-effects} model. The former assumes the true treatment effects are identical across all the included studies, which is often questionable or unrealistic in practice. 
In a random-effects model, on the other hand, treatment effects are assumed to vary from study to study around a grand mean~$\mu$ with a normally distributed term of variance~$\tau^2$. 
Considering the different backgrounds of individual trials (e.g., types of patients, different investigators), and given that heterogeneity is often found to be present in practical applications,\cite{KontopantelisSpringateReeves2013} it is reasonable to take these differences into account and hence the random-effects model is commonly applied in practice; for a detailed discussion see e.g. Borenstein \emph{et~al}.\cite{BorensteinEtAl2010}

To perform a random-effects meta-analysis, the commonly-used two-stage approach first requires estimation of the between-study heterogeneity~$\tau$, and only then the overall treatment effect~($\mu$) may be estimated using the plug-in estimate of the between-study heterogeneity. However, a major challenge lies in the estimation of heterogeneity, in particular, in meta-analyses of few (e.g., 2 to 4) studies.\cite{HigginsThompsonSpiegelhalter2009,TurnerEtAl2012} In this context, the heterogeneity frequently fails to be identified by classical methods (e.g., the DerSimonian-Laird method),\cite{FriedeRoeverWandelNeuenschwander2017a} which in turn results in an underestimation of the variance of the treatment effect. Some methods have been developed to account for this imprecision, such as the Knapp-Hartung method\cite{HartungKnapp2001a,SidikJonkman2003} or its modification.\cite{rover2015} Bender \emph{et~al.}\cite{BenderEtAl2018} performed a comprehensive comparison of available methods, and concluded that no universally satisfactory (frequentist) method is currently available to
perform meta-analyses in the case of (very) few studies.

On the other hand, meta-analyses of individual participant data (IPD) are considered a more reliable option to meta-analyses using aggregated study-level data.\cite{CooperPatall2009,riley2010,BenderEtAl2018} However, a number of obstacles to sharing or obtaining detailed data still pose a major challenge for performing IPD meta-analyses.\cite{ventresca2020} Alternatively, aggregated data at a subpopulation level (e.g., study population subgroups based on gender or age), also known as subgroup-level data, may be easier to obtain and are routinely reported in clinical trials.\cite{brankovic2019} Therefore, it can be regarded as a compromise for evidence synthesis in between the aggregated study-level data and IPD\@. Actually, synthesizing such kind of subpopulation data is not new in the meta-analysis field. In meta-analyses evaluating surrogate outcomes in clinical trials, models have been developed with units nested within a trial (e.g., center, country, investigator) when the number of trials itself is insufficient to apply meta-analytic methods.\cite{burzykowski2005,buyse2016} In this paper, motivated by a similar idea, we propose synthesizing the evidence from subgroup level instead of the study level when the number of studies is very small. For the estimation of the overall treatment effect~$\mu$, we rely on the common-effect estimator, which guarantees unbiasedness and consistency as the conventional estimator based on study-level data. We show that the resulting between-study heterogeneity estimators based on subgroup-level data show a smaller risk of returning zero estimates than the classic heterogeneity estimators based on the study-level data (e.g., the DerSimonian-Laird estimator). These may be incorporated into variance calculation within the Henmi-Copas type variance framework to correctly reflect the variance components.\cite{henmi2010} Confidence intervals (CIs) are constructed based on the Student-$t$ distribution for the consideration of small samples as proposed by Follmann and Proschan,\cite{FollmannProschan1999} while the degrees of freedom are inherently based on the number of subgroups rather than the number of studies, hence the resulting width of the CI may be substantially shorter.

The remainder of the paper is organized as follows. In Section~2, we introduce two motivating examples representing typical applications of meta-analysis with few studies. In Section~3, we briefly review existing methodological developments for meta-analysis with very few studies. In Section~4, we introduce our proposed method under a data-generating model for subgroup-level data, including the derivation of two proposed heterogeneity estimators and the construction of CIs. In Section~5, we discuss the practical issue of subgroup selection when implementing our method. In Section~6, we present the results of a simulation study to evaluate the performance of our proposed method and other competing approaches. In Section~7, we revisit the two motivating examples and discuss the implications of our findings. A brief discussion and concluding remarks are provided in Section~8.

\section{Motivating examples}
\label{sec:examples}
\subsection{The RESPIRE trials}
\label{sec:respire}
\begin{figure}[h]
\centering\includegraphics[width=0.6\textwidth]{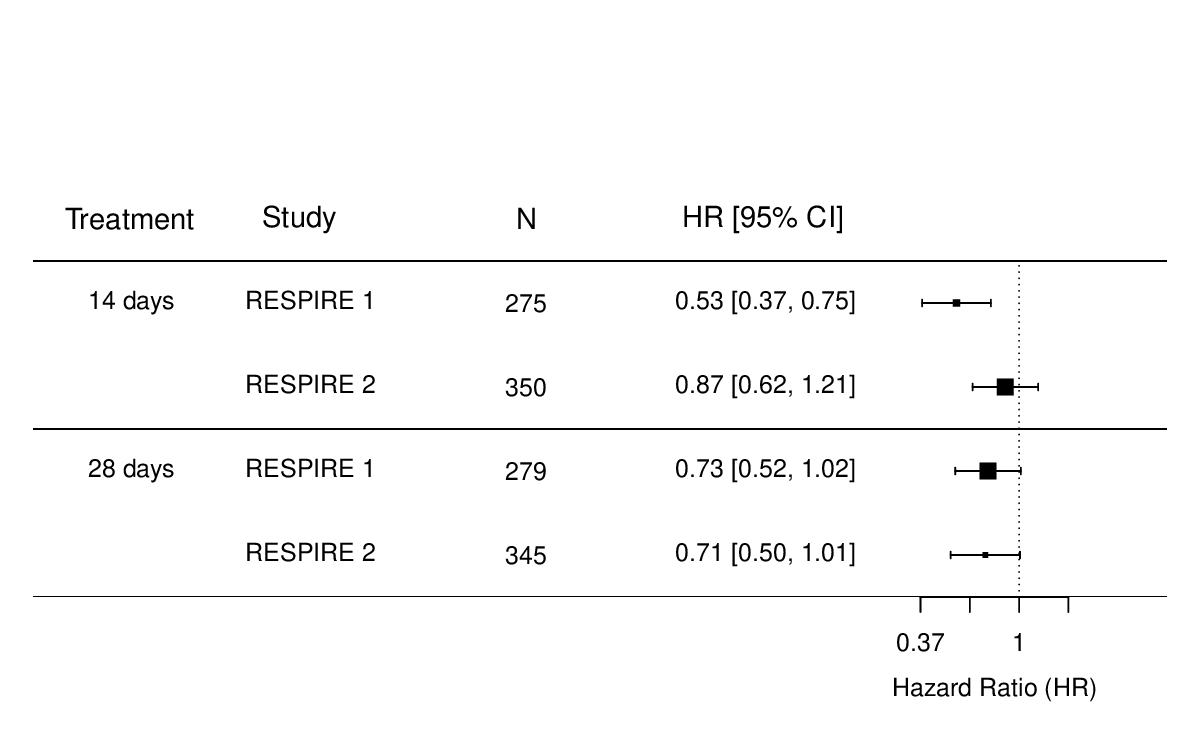}
\caption{Data on the two (14-day and 28-day) regimens reported in the two \textsc{respire} trials.\cite{de2018respire,aksamit2018}}
\label{fig:RESPIRE}
\end{figure}

Non-cystic fibrosis bronchiectasis (NCFB) is a chronic respiratory disease and is known for its lack of effective licensed therapies for long-term disease management. The \textsc{respire} trials were conducted to investigate the potential of ciprofloxacin dry powder for inhalation (DPI) to fill this gap, which involved two international phase~III prospective, parallel-group, randomized, double-blinded, multicentre, placebo-controlled trials of the same design.\cite{de2018respire,aksamit2018} In both trials, there are two active treatment groups (14~days on/off regimen as well as 28~days on/off regimen), both with their own matching placebo groups. The primary efficacy endpoints include the \emph{time to first exacerbation within 48~weeks} after the start of treatment for ciprofloxacin DPI versus pooled placebo (which was quantified using a hazard ratio), and the \emph{frequency of exacerbations during the 48-week study} for ciprofloxacin DPI versus matching placebo (quantified using an incidence rate ratio). A controversial issue in this experiment was that, despite the similarity in study designs of these two trials, the findings from the two independent trials appeared inconsistent. Meta-analytic techniques could be applied to synthesize the results from both trials for a more precise conclusion. Chotirmall \emph{et~al.}\cite{ChotirmallChalmers2018} and R\"{o}ver and Friede\cite{RoeverFriede2024} both investigated the endpoint of frequency of exacerbations and discussed the role of between-study heterogeneity for the (apparent) inconsistency. Thus, this is a typical meta-analysis in practice facing the problems of very few studies and substantial uncertainty in the heterogeneity. In the following, we focus on the primary endpoint of time to first exacerbation. A similar phenomenon was observed in the treatment group of 14~days on/off regimen, where a significant result was only reported for the \textsc{respire~I} trial. For the treatment group of 28~days on/off therapy, both trials reported non-significant results (see Figure~\ref{fig:RESPIRE}).
We applied two separate meta-analyses for the two active treatments to show the specific problems in a typical meta-analysis of few studies and how we may resolve these problems using our new proposal.

\subsection{SGLT2 inhibitor studies}
\label{sec:SGLT2}
\begin{figure}[h]
\centering\includegraphics[width=0.6\textwidth]{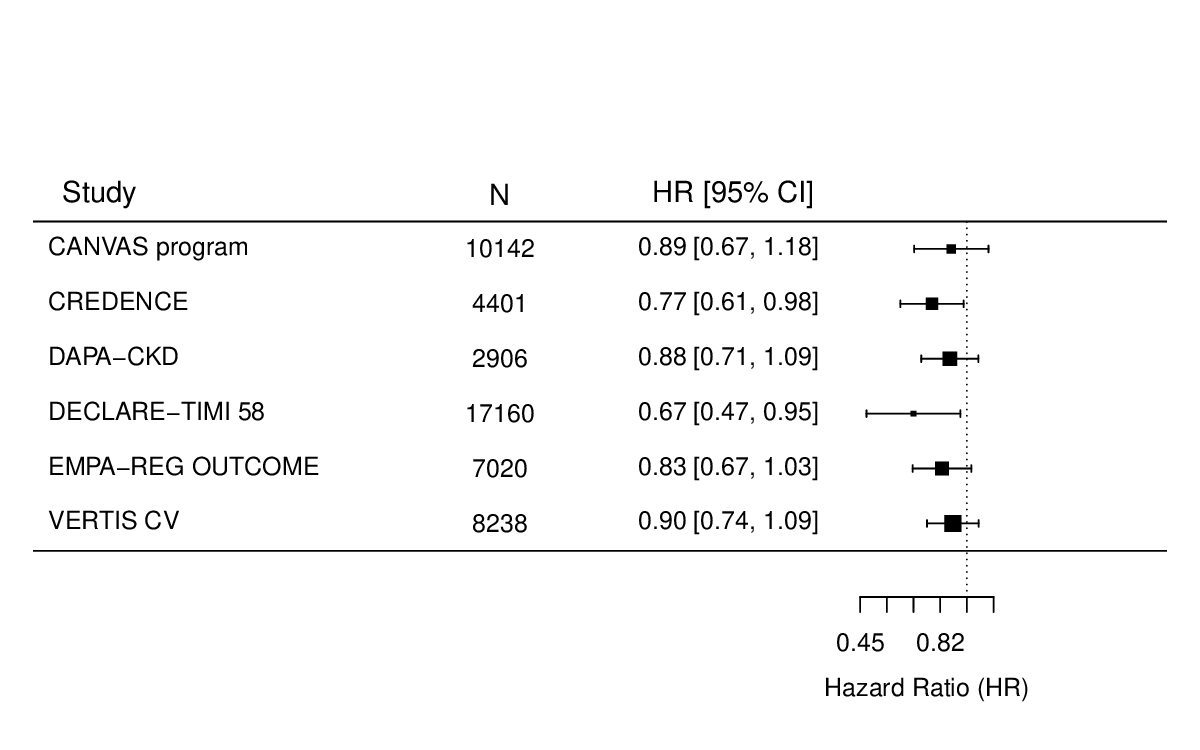}
\caption{Data from the meta-analysis of sodium-glucose cotransporter~2 (SGLT2) inhibitor studies due to Neuen \emph{et~al.}\cite{neuen2022}}
\label{fig:SGCExample}
\end{figure}

For patients with type~2 diabetes, especially those at high cardiovascular risk or with chronic kidney
disease (CKD), the risk of serious hyperkalemia might limit the utilization of renin-angiotensin-aldosterone system inhibitors
and mineralocorticoid receptor antagonists (MRAs). Evidence from some small studies of
relatively short duration showed that sodium-glucose cotransporter~2 (SGLT2) inhibitors have the potential to reduce the risk of hyperkalemia.\cite{lo2018} To investigate the long-term effect of SGLT2 inhibitors on the risk of serious hyperkalemia, Neuen \emph{et~al.}\cite{neuen2022} conducted a meta-analysis of six randomized, double-blind, placebo-controlled clinical trials (see Figure~\ref{fig:SGCExample}). While the original meta-analysis was conducted with individual participant data (IPD), we will make use of the aggregated subgroup data from its supplementary material to illustrate our proposed methods.
The primary outcome was the incidence of serious hyperkalemia, defined as time to the first central laboratory-determined serum potassium $\ge$ 6.0 mmol/L and quantified as a log-hazard ratio. Although the original motivation in this review is to evaluate the long-term effect of SGLT2 inhibitors in large trials in patients with varying background therapies that may potentially be heterogeneous, the standard random-effects model failed to identify any heterogeneity and hence effectively resulted in a common-effect analysis. Neuen \emph{et~al.}\cite{neuen2022} concluded that SGLT2 inhibitors reduced the risk of serious hyperkalemia by 16\% with an estimated hazard ratio of 0.84 (95\% CI: [0.76, 0.93]). However, this result seems to be sensitive to potential heterogeneity (see the upper boundary of the CI). Considering the small number of studies included in this meta-analysis, a zero estimate of heterogeneity might not be uncommon even in the presence of substantial heterogeneity. The challenging question then concerns the robustness of this conclusion given the uncertainty around the presence of heterogeneity, and how to adequately estimate this between-study heterogeneity as well as the associated overall effect.

\section{Standard random-effects model and approaches for inference}\label{sec:MAModels}
 
The main focus of a meta-analysis usually is to estimate the grand population mean effect~$\mu$ using the observed estimates~$y_i$ from $k$~existing studies on the same topic, and their associated standard errors~$s_i$, which are commonly taken to be known without uncertainty. Under a random-effects model, we assume the variation of each~$y_i$ consists of two parts: the within-study sampling variance~$s_i^2$ and the between-study heterogeneity~$\tau^2$ which is unknown and needs to be estimated. Then the observed data can be explained within the popular normal-normal hierarchical model (NNHM) framework. For each study, the estimates~$y_i$ are assumed to be approximately normally distributed with the mean of a study-specific effect $\theta_i$ as 
\begin{eqnarray}
        y_i|\theta_i,s_i  & \sim & \normaldistn(\theta_i,\, s_i^2),\qquad \mbox{($i=1,\ldots,k$).}
\end{eqnarray}
Furthermore, these study-specific effects are assumed to vary around the grand population mean~$\mu$ with an additive (random-effects) variance component of~$\tau^2$,
  \begin{eqnarray}
       \theta_i|\mu,\tau & \sim & \normaldistn(\mu,\, \tau^2) \mbox{.}
       \label{norm_stu}
\end{eqnarray}
Integration over the distribution of the study-specific effects $\theta_i$ yields a simplified marginal model as
\begin{eqnarray}
y_i|\mu,\tau&\sim& \normaldistn(\mu,\tau^2+s_i^2).
\label{marg_stu}
\end{eqnarray}
When~$\tau=0$, it reduces to the special case of a common-effect model. In a frequentist framework, parameter estimation within the NNHM is commonly approached in two steps. First, the nuisance parameter~$\tau$ is estimated; although many options are available,\cite{petropoulou2017} here we focus on the widely-used moment estimator (also known as the \emph{DerSimonian-Laird (DL)} estimator), which is defined by
\begin{eqnarray}
\hat{\tau}_{\mathrm{DL}}^2=\max\left\{0,\,
\frac{Q-(k-1)}
{\sum_{i=1}^{k}s_i^{-2}-\sum_{i=1}^{k}s_i^{-4}/\sum_{i=1}^{k}s_i^{-2}
}
\right\},
\label{dl}
    \end{eqnarray}
where $Q =\sum_{i=1}^{k}(y_i-\hat{\mu}_{\mathrm{CE}})^2/s_i^2$ is the test statistic for homogeneity (``Cochran's~$Q$''), which follows a $\chi^2$~distribution with $k-1$~degrees of freedom if~$\tau=0$.\cite{HigginsThompson2002} $\hat{\mu}_{CE}$ is the common-effect estimator assuming $\tau=0$. Next, conditioning on~$\tau$ (effectively assuming $\tau$~was known and fixing its value at $\hat{\tau}_{\mathrm{DL}}$), the grand mean is estimated using the inverse-variance weighted average \citep{cochran1954Bio} as
\begin{eqnarray}
\hat{\mu}_{\mathrm{RE}}=\frac{\sum_{i=1}^{k}\omega_{i,\mathrm{RE}} \, y_i}{\sum_{i=1}^{k}\omega_{i,\mathrm{RE}}},\nonumber
\label{mu}
\end{eqnarray}
where $\omega_{i,\mathrm{RE}}=(s_i^2+\hat{\tau}_{\mathrm{DL}}^2)^{-1}$ is the weight of study~$i$ under the random-effects model, which reduces to $\omega_{i,\mathrm{CE}}=s_i^{-2}$ when the between-study heterogeneity estimator $\hat{\tau}_{\mathrm{DL}}$ turns out as zero, which is often the case in meta-analysis of few studies even in the presence of heterogeneity,\cite{FriedeRoeverWandelNeuenschwander2017a,RoeverFriede2024} motivating a number of methodological developments to account for this imprecision.

Several approaches have been well-established to proceed with inferences and construct confidence intervals in a frequentist fashion. These approaches are either based on a normal approximation, consider an adjusted CI using Student-$t$ quantiles,\cite{HartungKnapp2001a,SidikJonkman2003,zejnullahi2024} or rely on non-parametric resampling procedures.\cite{michael2019} Here, we briefly review these methods.

{\bf Normal approximation}: The marginal model~(\ref{marg_stu}) may be utilized by plugging in the heterogeneity estimate~$\hat{\tau}_{\mathrm{DL}}$ (and ignoring potential estimation uncertainty). The resulting CI is given by
$\hat{\mu}_{\mathrm{RE}}\pm z_{1-\alpha/2}\sqrt{\hat{V}_{\mathrm{RE}}}$, where $z_{\gamma}$ denotes the $\gamma$-quantile of the standard normal distribution and $\hat{V}_{\mathrm{RE}}=\frac{1}{\sum_{i=1}^k\omega_{i,\mathrm{RE}}}$ is the variance estimator of~$\hat{\mu}_{\mathrm{RE}}$. In case $\hat{\tau}_{\mathrm{DL}}=0$, this leads to the variance of a common-effect model, which is given by $\hat{V}_{\mathrm{CE}}=\frac{1}{\sum_{i=1}^k\omega_{i,\mathrm{CE}}}$. This CI may work well as long as the number of studies is large, or when heterogeneity is small.\cite{FriedeRoeverWandelNeuenschwander2017a}

\textbf{Hartung-Knapp-Sidik-Jonkman (HKSJ) method}: CIs based on the Student-$t$ distribution have independently been proposed by Hartung and Knapp\cite{HartungKnapp2001a} and Sidik and Jonkman\cite{SidikJonkman2003} with the aim of yielding valid intervals also in case of small numbers~($k$) of studies. Compared to the previous expression, the variance estimator of $\hat{\mu}_{\mathrm{RE}}$ is adjusted via a multiplicative factor~$q=\frac{1}{k-1}\sum_{i=1}^k\omega_{i,RE}(y_i-\hat\mu_{RE})^2$, while a Student-$t$ quantile is used in place of the normal quantile. The resulting CI is given by $\hat{\mu}_{RE}\pm t_{k-1;1-\alpha/2}\,\sqrt{\hat{V}_{\mathrm{HKSJ}}}$, where $t_{k-1,\gamma}$ denotes the $\gamma$-quantile of a Student-$t$ distribution with $k-1$ degrees of freedom and $\hat{V}_{\mathrm{HKSJ}} =q\hat{V}_{\mathrm{RE}}$. 
These CIs are generally wider and exhibit better coverage; however, in some cases, these may also turn out \emph{shorter} than the intervals based on a normal approximation.\cite{wiksten2016} Knapp and Hartung\cite{KnappHartung2003} hence already suggested an \emph{ad~hoc} modification, 
replacing $q$~by the truncated quantity $q^\star=\max\left\{1,q\right\}$, 
%therefore the variance $\hat{V}_{\mathrm{mHK}}=q^\star\hat{V}_{\mathrm{RE}}$,
therefore the variance estimator $\hat{V}_{\mathrm{mKH}} =q^\star\hat{V}_{\mathrm{RE}} \geq \hat{V}_{\mathrm{RE}}$,
which will ensure a more conservative procedure.
When $q^\star=1$, the resulting CIs are the same as the \emph{ad~hoc} procedure proposed by Follmann and Proschan,\cite{FollmannProschan1999}. They are also analogous to CIs obtained from the HKSJ method combined with the \emph{Paule-Mandel (PM)}  estimator when the PM heterogeneity estimate is positive ($\hat{\tau}^2_{\mathrm{PM}}>0$).\cite{JacksonEtAl2017}

\textbf{Zejnullahi and Hedges' method}: To account for the small-sample properties in variance estimation of small meta-analysis with the standardized mean difference, Zejnullahi and Hedges \cite{zejnullahi2024} borrowed ideas from the econometrics area to develop a robust variance estimator for $\hat{\mu}_{\mathrm{RE}}$ under the random-effects model as 
$\hat{V}_{\mathrm{ROB}}=\sum_{i=1}^k\frac{\omega_{i,\mathrm{RE}}^2(y_i-\hat{\mu}_{\mathrm{RE}})^2}{(\sum_{\ell=1}^k\omega_{i,\mathrm{RE}})^2}\Bigg(1-\frac{\omega_{i,\mathrm{RE}}}{\sum_{\ell=1}^k\omega_{i,\mathrm{RE}}}\Bigg)^{-C}$, where $C$~constitutes a scale factor for the penalty term. Sidik and Jonkman\cite{sidik2006} considered $C=1$, while 
Zejnullahi and Hedges proposed the use of $C=2$ to further penalize the squared sample residuals when the sample sizes across studies are substantially different and $C=1$ may be insufficient. 
The proposed CIs based on the Student-$t$ distribution are given by  $\hat{\mu}_{\mathrm{RE}}\pm t_{k-1;1-\alpha/2}\sqrt{\hat{V}_{\mathrm{ZH}}}$, where
$\hat{V}_{\mathrm{ZH}}$ is the 
robust variance estimator based on~$C=2$ . 

\textbf{Michael et al method}: Alternatively, Michael \emph{et~al}.\cite{michael2019} proposed a Monte Carlo sampling based approach to construct exact confidence intervals for random-effects meta-analyses in case of few studies. This method involves finding the confidence limits with approximated test statistics from Monte Carlo samples, hence is computationally intensive, and it is known to provide overcoverage as well as wide CIs.\cite{hanada2023}

In case of meta-analyses with only few studies, the above-mentioned extensions of the simple intervals based on a normal approximation have been reported to improve performance in terms of coverage probability. 
However, these may yield disturbingly wide CIs, due to their reliance on the $t$-quantiles; in the extreme (yet common) case of combining only $k=2$~studies, a quantile of $t_{1,0.975}=12.7$ is used instead of the corresponding normal quantile of $z_{0.975}=1.96$, often rendering the resulting quantile effectively useless. Also, these methods do not solve the common problems with poor heterogeneity estimates. In the following, we will try to improve the poor behavior of heterogeneity estimates and the derived CIs.

\section{Proposed method using subgroup information}\label{sec:subgroups}
\subsection{Motivation}
In meta-analyses evaluating surrogate outcomes in clinical trials,
splitting each study into multiple units (e.g., by center) is suggested as a useful approach when the number of studies is insufficient to apply a hierarchical analysis model at the aggregated study-level data.\cite{burzykowski2005} 
Based on this motivation, we consider extending the study-level framework (NNHM) from Section~\ref{sec:MAModels} to aggregated subgroup-level data settings for random-effects meta-analysis of few studies. 
Estimation of the overall treatment effect~$\mu$ and in particular the heterogeneity~$\tau$ then is based on subgroup-level rather than study-level data.

Since subgroup analysis plays an important role in investigating interactions of main (treatment) effects and subgroups, it is routinely reported as a secondary objective in clinical trials (quite commonly, in fact, without appropriate clinical trial design and prespecification,  such analyses are generally not regarded as providing strong evidence for effect modification).\cite{brankovic2019,IQWiG2025} Previous studies have shown that subgroup analyses are commonly published, especially in high-impact general medical journals.\cite{sun2011,gabler2009} Data on endpoints at the level of study subgroups (e.g., by age or gender) are hence likely to be readily available in many cases.
By considering subgroup-level data, we may expect a positive bias in the heterogeneity estimator, which may be considered acceptable if this leads to improved overall inferential performance,
such as improved coverage and reduced lengths of the resulting CI. 

In the following section, we introduce a data-generating model for the subgroup-level data. While the NNHM from Section~\ref{sec:MAModels} is essentially still used for inference, the model extension will enable the investigation of some consequences of data scenarios on the performance of the resulting estimators, analytically as well as in simulations.

\subsection{Data-generating model for subgroup-level data}
\label{sec:subgroupmodel}
We first describe the data structure (a three-level hierarchy) in a meta-analysis using subgroup information. In practice, two subgroups 
(e.g., gender, age) may be common, and more than two may also occur, but for simplicity, we focus on the simple case of two subgroups within each study. To index the new hierarchy at the subgroup level, we introduce a subscript~$j \in\{1, 2\}$. Note that $j$~here is just used to distinguish the different groups within one study; the same subscript~$j$ may not necessarily represent the same type of group across different studies (e.g., study $i=1$ may use subgroups by gender, while study $i=2$ may use subgroups by ethnicity). This will become more concrete when we revisit the two examples, and in the Discussion section. 
The proposed data-generating model is a generalization of the NNHM introduced in Section~\ref{sec:MAModels} and is designed to serve several purposes: 
(a)~besides a study's overall sample size ($n_i$), the subgroup prevalences~($p_i=\frac{n_{i,1}}{n_i})$ need to be accommodated, 
(b)~the average across the two subgroup effects should again yield the overall effect (collapsibility: $\theta_{i}=p_i\,\theta_{i,1} + (1-p_i)\,\theta_{i,2}$), 
(c)~subgroup differences (treatment-by-subgroup interaction effects~$\Delta=\expect[\theta_{i,2}-\theta_{i,1}]$) need to be specified, and 
(d)~interactions may be considered fixed~($\sigma_\Delta=0$) or random~($\sigma_\Delta>0$).

Let $y_{i,j}$ denote the observed outcome from the $j$th subgroup of $i$th study, and $s_{i,j}$ the corresponding standard error.
For the study-level standard errors, we assume that
\begin{eqnarray}
   s_i & = & \frac{\uisd}{\sqrt{n_i}} \mbox{,}
   \label{uisd_stu}
\end{eqnarray}
where $\uisd$ is the \emph{unit information standard deviation (UISD)} and $n_i$~is the $i$th study's sample size.

Analogously, the standard error of the effect in a subgroup then is given by
\begin{eqnarray}
   s_{i,j} & = & \frac{\uisd}{\sqrt{n_{i,j}}} \mbox{,}
   \label{uisd_sub}
\end{eqnarray}
where $n_{i,j}$~is the sample size of the $j$th subgroup in the $i$th study. 
For some more concrete examples of endpoints where (approximate) inverse proportionality of sample size and (squared) standard error may be motivated, see e.g. Spiegelhalter \emph{et~al.},\cite{SpiegelhalterEtAl} R\"{o}ver,\cite{Roever2020} or R\"{o}ver \emph{et~al.}\cite{rover2021}
In certain cases, the ``sample size''~$n_{i}$ does not necessarily correspond to the number of subjects; for example, for time-to-event endpoints (hazard ratios), the standard errors scale with the number of \emph{events} instead.
The assumption of a constant UISD~$\uisd$ here is introduced to formulate a reasonable and coherent connection between sample sizes and standard errors in the subgroup-level model. UISDs might also vary between studies. In the following analytical and numerical considerations, however, we keep these constant across studies as well, since varying the UISD would have the same effect as varying the total sample size~$n_i$ (which is kept variable already).

Suppose that $p_i$~is the proportion (prevalence) of patients in subgroup~$1$ of $i$th study; then $n_{i,1}=p_in_i$ and $n_{i,2}=(1-p_i)n_i$. The case of $p_i=0.5$ corresponds to equally-sized subgroups, for different prevalences, the subgroups are somewhat imbalanced. 
Allowing for the new hierarchy at subgroup level, we extend the NNHM from Section~\ref{sec:MAModels} as
\begin{eqnarray}
  y_{i,j}      & \sim & \normaldistn(\theta_{i,j},\, s_{i,j}^2) \mbox{, where} \label{obs_sub}\\
  \theta_{i,j} & =    & \left\{\begin{array}{ll} \theta_i - (1-p_i)\,\delta_i & \mbox{if $j=1$} \\ 
\theta_i + p_i\,\delta_i     & \mbox{if $j=2$, and}\end{array}\right.\label{eqn:sub_specific2}\\
  \delta_i     & \sim & \normaldistn(\Delta,\,\sigma_\Delta^2)\mbox{.} \label{eqn:sub_specific3}
\end{eqnarray}
In this model, the observed estimates~$y_{i,j}$ from each subgroup are assumed to be normally distributed with the mean of a subgroup-specific effect~$\theta_{i,j}$ and its sampling variance~$s_{i,j}^2$.
The~$\theta_{i,j}$ are associated with the corresponding study-specific effect~$\theta_i$ from Equation~(\ref{norm_stu}) by considering the overall study-specific effect to result as a weighted mean of the subgroup-specific treatment effects (with weights corresponding to subgroup prevalences~$p_i$ and~$(1 - p_i)$ for subgroups~1 and~2, respectively). 
Here, $\delta_i$ in Equation~(\ref{eqn:sub_specific3}) denotes the (treatment-by-subgroup) \emph{interaction effect}, the difference between subgroups ($\theta_{i,2}-\theta_{i,1}$) within the $i$th study.
This particular model formulation allows us to vary~$\delta_i$ around a mean (interaction) effect of~$\Delta$ with a non-zero variance of~$\sigma_\Delta^2$, thereby allowing for differences between subgroups. Setting $\delta_i=\Delta$ to a constant (i.e., assuming $\sigma_\Delta=0$) results in a homogeneous (``common-effect'') model with identical interaction effects across studies. This interaction effect is decomposed with respect to the study-specific prevalence~$p_i$ so that
\begin{eqnarray}
p_i\,\theta_{i,1}+(1-p_i)\,\theta_{i,2} &=& \theta_i\mbox{,}
\label{eqn:Effectsrelationship}
\end{eqnarray}
which means the average effect at the study level remains the same as in the ``classical'' NNHM from Section~\ref{sec:MAModels} (collapsibility). 

\subsection{Heterogeneity estimate using subgroup information} \label{sec:subgroupHeterogeneity}
Next, we would like to introduce a new heterogeneity estimator based on the above-mentioned model and some desirable properties of it.
To begin with, we first summarize the marginal model as
\begin{equation}
\label{subModel}
\scalebox{0.8}{$\displaystyle
\left(
\begin{array}{c}
  y_{i,1} \\ y_{i,2}
\end{array}
\right) \Big| \, \mu,\tau,\Delta,\sigma_\Delta
\;\sim\;
\normaldistn \left(
\left(
\begin{array}{c}
  \mu \\
  \mu
\end{array}
\right)
+
\left(
\begin{array}{c}
 -(1-p_i)\Delta \\
 p_i\Delta
\end{array}
\right),
\left(
\begin{array}{cc}
\tau^2+s_{i,1}^2 & \tau^2 \\
\tau^2 & \tau^2+s_{i,2}^2
\end{array}
\right)
+
\left(
\begin{array}{cc}
(1-p_i)^2\sigma_\Delta^2 &
-p_i(1-p_i)\sigma_\Delta^2 \\
-p_i(1-p_i)\sigma_\Delta^2 &
p_i^2\sigma_\Delta^2
\end{array}
\right)
\right).
$}
\end{equation}
For the vector of subgroup-specific treatment effect estimates~$(y_{i,1},y_{i,2})^\prime$, both mean and variance-covariance matrix can be clearly divided into two contributions, from the study level as well as from the subgroup level.
To utilize the (``DerSimonian-Laird'') moment estimator, we define the $Q$-statistic based on subgroup information as
\begin{eqnarray} \label{eqn:QS}
        Q_{\mathrm{S}}&=&\sum_{i=1}^k\sum_{j=1}^2\omega_{i,j}\,(y_{i,j}-\hat{\mu}_{\mathrm{CE}})
\end{eqnarray}
where 
\begin{equation}
\hat{\mu}_{\mathrm{CE}}\;=\;\frac{\sum_{i=1}^k\sum_{j=1}^2\omega_{i,j}\,y_{i,j}}{\sum_{i=1}^k\sum_{j=1}^2\omega_{i,j}}\;=\;\frac{\sum_{i=1}^k\omega_{i,\mathrm{CE}}\,y_i}{\sum_{i=1}^k\omega_{i,\mathrm{CE}}}, \quad \mbox{and} \quad \omega_{i,j}\;=\;\frac{1}{s_{i,j}^2}.
   \label{eqn:CE.estimators}
\end{equation}
The consistency of these two common-effect estimates in Equation (\ref{eqn:CE.estimators}) is due to the relationship in~(\ref{uisd_stu}) and~(\ref{uisd_sub}), that
\begin{equation}
    \omega_{i,1}\;=\;p_i\frac{1}{s_i^2}\;=\;p_i\,\omega_{i,\mathrm{CE}}
    \quad \mbox{and} \quad
    \omega_{i,2}\;=\;(1-p_i)\,\frac{1}{s_i^2}\;=\;(1-p_i)\,\omega_{i,\mathrm{CE}}.
    \nonumber
 \end{equation}     
Consequently, we have
\begin{equation}
\omega_{i,\mathrm{CE}}\;=\;\sum_{j=1}^2\omega_{i,j} 
\quad \mbox{and}\quad 
y_i\;=\;\frac{\sum_{j=1}^2\omega_{i,j}\,y_{i,j}}{\sum_{j=1}^2\omega_{i,j}}\mbox{.}
\label{cal_stu}
\end{equation}     
In Web Appendix~A, we show that
\begin{equation}
\scalebox{0.9}{$\displaystyle
\expect(Q_{\mathrm{S}})=(2k-1)+\tau^2\left\{\sum_{i=1}^k\sum_{j=1}^2\omega_{i,j}-\frac{\sum_{i=1}^k\sum_{j=1}^2\omega_{i,j}^2}{\sum_{i=1}^k\sum_{j=1}^2\omega_{i,j}}\right\}-\tau^2\left\{\frac{2\sum_{i=1}^k\omega_{i,1}\,\omega_{i,2}}{\sum_{i=1}^k\sum_{j=1}^2\omega_{i,j}}\right\}+(\Delta^2+\sigma_\Delta^2)\sum_{i=1}^k\sum_{j=1}^2\omega_{i,j}\,p_i\,(1-p_i)
$}
\label{EQ_sub}
\end{equation}   
The resulting DerSimonian-Laird estimator using subgroup-level data (DLS) is given by
\begin{eqnarray}
\hat{\tau}_{\mathrm{DLS}}^2&=&\frac{Q_{\mathrm{S}}-(2k-1)}{\sum_{i=1}^k\sum_{j=1}^2\omega_{i,j}-\frac{\sum_{i=1}^k\sum_{j=1}^2\omega_{i,j}^2}{\sum_{i=1}^k\sum_{j=1}^2\omega_{i,j}}} \mbox{,}%\nonumber
\end{eqnarray}
where, as usual, negative estimates are truncated to zero (as in Equation~(\ref{dl})). According to expectation of~(\ref{EQ_sub}), we can show that
\begin{equation}
\begin{aligned}
\expect(\hat{\tau}_{\mathrm{DLS}}^2)
={}&
\tau^2\left\{
1-
\frac{
2\sum_{i=1}^k\omega_{i,1}\omega_{i,2}
}{
\left(\sum_{i=1}^k\sum_{j=1}^2\omega_{i,j}\right)^2
-\sum_{i=1}^k\sum_{j=1}^2\omega_{i,j}^2
}
\right\}\\
&+
(\Delta^2+\sigma_\Delta^2)
\frac{
\left(\sum_{i=1}^k\sum_{j=1}^2
\omega_{i,j}p_i(1-p_i)\right)
\left(\sum_{i=1}^k\sum_{j=1}^2\omega_{i,j}\right)
}{
\left(\sum_{i=1}^k\sum_{j=1}^2\omega_{i,j}\right)^2
-\sum_{i=1}^k\sum_{j=1}^2\omega_{i,j}^2
}\\
={}&A\tau^2+B.
\end{aligned}
\label{exp:tau.dls}
\end{equation}
which is a function of the true value of~$\tau^2$ as well as the unknown subgroup parameters $\Delta$, $\sigma_\Delta$ and~$p_i$.
In Web Appendix~A, we show that the term~$A$ approaches unity as the number of studies~$k$ increases. However, when $k$~is small, $A$~might down-scale~$\tau^2$ in a certain range, depending on the study sizes~$n_i$ (see Figure S28 in Web Appendix~C). On the other hand, the term~$B$ is always positive.
Since very little information is available about $\tau^2$ in meta-analyses of few studies, it is common to get zero estimates from the DL approach.\cite{FriedeRoeverWandelNeuenschwander2017a} Alternative methods that may include a positive bias for~$\tau$ have shown good performance in simulation studies.\cite{KontopantelisSpringateReeves2013,ChungEtAl2013a} Using a similar motivation, we consider to adjust the potential impacts of term \emph{A} on the \emph{DLS} estimator as
\begin{eqnarray}
\hat{\tau}_{\mathrm{DLS.adj}}^2=\frac{\hat{\tau}_{\mathrm{DLS}}^2}{A}\mbox{,}
\end{eqnarray}
which can ensure a positive bias (via~$\frac{B}{A}$) asymptotically. 
To further reduce the number of zero estimates and avoid the underestimation of the between-study heterogeneity, thereby preserving the variance of $\hat\mu_{\mathrm{RE}}$, we propose two naive but empirically useful alternatives based on the above-mentioned two DerSimonian-Laird type estimators as
\begin{eqnarray}
\hat{\tau}_{\mathrm{max}1}^2 &=& \max\left\{\hat{\tau}_{\mathrm{DL}}^2,\,\hat{\tau}_{\mathrm{DLS}}^2\right\}\label{adhoc1},\\
\mbox{and} \quad \hat{\tau}_{\mathrm{max}2}^2 &=& \max\left\{\hat{\tau}_{\mathrm{DL}}^2,\hat{\tau}_{\mathrm{DLS.adj}}^2\right\}\label{adhoc2},
\end{eqnarray}
which are hybrid versions of DL-type estimators using study-level data~(\ref{dl}) and subgroup-level data~(\ref{adhoc1}) and~(\ref{adhoc2}) that can always achieve the maximum. For these two estimators, we have the relationship that $\hat{\tau}_{\mathrm{max}1}^2 \le \hat{\tau}_{\mathrm{max}2}^2$, since the adjustment term is a proportion always less than~1.

\subsection{Confidence interval with Henmi-Copas type variance}
As shown in (\ref{eqn:CE.estimators}), the common-effect model leads to a consistent estimator of the grand population mean~$\mu$ that remains the same for \emph{both} weighted averages, across subgroup-level~($y_{i,j}$) as well as study-level~($y_i$) data.
%The subgroup effects do not appear in this estimate, since they cancel in (\ref{eqn:Effectsrelationship}).
Therefore, in our method, we rely on the (unbiased) effect estimate~$\hat{\mu}_{\mathrm{CE}}$ based on common-effect weights (see also Section~\ref{sec:subgroupHeterogeneity}).
For inference, to adequately account for the total variance components, we use the Henmi-Copas (HC) type variance which is derived for the common-effect point estimator while also accounting for heterogeneity.\cite{henmi2010} Under the marginal model for subgroup data described in Section~\ref{sec:subgroupHeterogeneity}, the HC-type variance is given by
\begin{eqnarray}         
   V_{\mathrm{HCS}}&=&\frac{\sum_{i=1}^k\omega_{i,1}^2\,\Var(y_{i,1})+\omega_{i,2}^2\,\Var(y_{i,2})+2\,\omega_{i,1}\,\omega_{i,2}\,\Cov(y_{i,1},y_{i,2})}{(\sum_{i=1}^k\sum_{j=1}^2\omega_{i,j})^2}\nonumber\\
      &=&\frac{\tau^2\,\sum_{i=1}^k\omega_{i,\mathrm{CE}}^2+\sum_{i=1}^k\omega_{i,\mathrm{CE}}}{(\sum_{i=1}^k\omega_{i,\mathrm{CE}})^2}\nonumber.
\end{eqnarray}
As one can see, its form is akin to the original estimator developed for meta-analysis using study-level data ($y_i, s_i$), the only difference being the choice of heterogeneity estimator plugged in here. In the original paper, $\tau^2$~is replaced by the empirical point estimates $\hat{\tau}_{\mathrm{DL}}^2$ using study-level data.\cite{henmi2010} We use the aforementioned two hybrid versions of DL stimators instead. When these estimators give zero estimates, the variance reduces to the variance of a common-effect model; otherwise, it can help us inflate the variance reasonably to account for the imprecision from between-study variance. 

To derive CIs, we utilize  Student-$t$ quantiles, defining the limits
\begin{equation}
\hat{\mu}_{\mathrm{CE}} \,\pm\, t_{\nu;1-\alpha/2}\,\sqrt{\hat{V}_{\mathrm{HCS}}},\quad \mbox{where} \quad
\nu = \left\{\begin{array}{ll} k-1 & \mbox{if $\hat{\tau}_{\mathrm{max}}^2\leq\hat{\tau}_{\mathrm{DL}}^2$} \\ 2k-1 & \mbox{if $\hat{\tau}_{\mathrm{max}}^2>\hat{\tau}_{\mathrm{DL}}^2$}\end{array}\right.\mbox{.}
\end{equation}
The degrees of freedom~($\nu$) used in deriving the CIs are flexible to reflect the choice of heterogeneity estimates in the two hybrid estimators. When a heterogeneity estimator based on subgroups is adopted (e.g.,$\hat{\tau}_{DLS}^2>\hat{\tau}_{DL}^2$), one may benefit from the extended data structure with subgroup information and use the larger number ($2k-1$) of degrees of freedom. Otherwise, we resort to the meta-analysis using study-level data, with degrees of freedom as in the approaches relying on Student-$t$-distribution (e.g., HKSJ, mKH). Recall that, in the meta-analysis using study-level data, when the between-study heterogeneity $\tau^2$ is estimated as zero ($Q<k-1$), the original HKSJ-adjusted CIs might be unfavorable (for more discussion see Wiksten \emph{et~al.}\cite{wiksten2016} or Jackson \emph{et~al.}\cite{JacksonEtAl2017}). Here, we show the potential benefits of our proposed method under this situation of a zero estimate of~$\tau$ from the perspective of variance estimation as follows:
\begin{equation}
\hat{V}_{\mathrm{HKSJ}} \;\leq\; \hat{V}_{\mathrm{CE}}\nonumber\;=\;\hat{V}_{\mathrm{mKH}}
             \le\hat{V}_{\mathrm{HCS}(\mathrm{max}1)}
             \;\leq\;\hat{V}_{\mathrm{HCS}(\mathrm{max}2)}.
\end{equation}
From this inequality, one can see that our proposed method will lead to more conservative variance estimates in the problematic situation commonly encountered in practice. However, increasing the degrees of freedom from $(k\!-\!1)$ to $(2k\!-\!1)$ also has a counteracting effect. We will investigate how this eventually works out in the simulation section.

\section{Subgroup selection and implementation issues}
\label{sec:subgroup selection}
Next, we discuss some issues in the practical application of our proposed methods. As already suggested in Section~\ref{sec:subgroups}, the study subgroup definitions may not necessarily need to be the same (i.e., based on the same subject characteristics) across all studies. This is because 1)~in reality, different studies may report different types of aggregated subgroup data, and 2)~our interest is not in estimating the actual subgroup interaction effects, but rather in enhancing heterogeneity variance estimation. Therefore, when several subgroups are available for each study, as is often the case, one needs to make a choice of subgroups.

Subgroup meta-analysis is a closely related application area aiming at investigating potential sources of between-study heterogeneity,\cite{CochraneHandbook} and it is included for consideration in the 27-item checklist of the \emph{Preferred Reporting Items for Systematic Reviews and Meta-Analyses (PRISMA) 2020 statement}.\cite{page2021} In contrast to the selection of study characteristics in the context of a subgroup meta-analysis, where explanatory variables, potential effect modifiers or covariates need to be pre-specified and carefully interpreted, subgroup selection here rather constitutes a \emph{post hoc} selection based on observed differences. In the spirit of a conservative specification as we will see in the simulation study, we suggest aiming for study subgroups (empirically) exhibiting increased heterogeneity, or suggesting greater interaction effects. However, known effect modifiers or dramatic effect differences should also be avoided due to the potential of over-coverage:
the idea here is that subgroups are considered in order to enhance variance estimation by tapping into a closely related source of residual noise.
Obvious and strong effect modifiers on the other hand would only serve to inflate variance by some positive amount (such factors would, however, most likely have been adjusted for in the original analysis already).
When $p$-values of each subgroup comparison within the studies are available (these commonly appear in forest plots for an individual study), one idea is to simply select the subgroups associated with the smallest $p$-values. Note that, as commented by Rasch \emph{et~al.},\cite{rasch2015} such subgroup tests usually are underpowered due to the \emph{post hoc} nature and multiple testing problem, and are therefore not appropriate for rigorously demonstrating treatment effect modification. For a clear demonstration of how to establish evidence of a treatment effect modifier based on an appropriate subgroup analysis, one may follow the guidance provided by the German Institute for Quality and Efficiency in Health Care (IQWiG) \cite{IQWiG2025}. If such information is not provided in the original publication, one might alternatively make a selection among candidate subgroups within each study using the empirical $Q$-statistics ($Q_{\mathrm{S}}$, (\ref{eqn:QS}), evaluated within each study by fixing~$i$), since this is equivalent to the square of the $t$-test, and hence the $F$-test comparing two subgroups; for a more detailed discussion, see Friede \emph{et~al.} (2009).\cite{friede2009} 

Specifically, suppose there are $m_i$~subgroup results reported in the $i$th study. The subgroup combinations can then be determined via $\sum_i m_i$ evaluations of~$Q_{\mathrm{S}}$. 
Since our focus is on small meta-analyses, such computations are typically straightforward. We provide \textsf{R}~functions in the supplementary material to implement our method, including subgroup selection; the performance of this strategy is evaluated in the subsequent simulation study. For readers familiar with widely used \textsf{R}~packages for meta-analysis (e.g., \texttt{metafor}), the required data format is easy to organize, as one only needs to provide all subgroup-level and study-level data within a single data table (see example data in Web Appendix~D). We emphasize that our function is intended only for prespecified subgroup candidates, meaning that meta-analysts should carefully assess the quality and appropriateness of choices of subgroups in advance. We illustrate this based on our two example applications in Section~\ref{exp:reanalysis}.

\section{Simulation}
\subsection{Data generation}\label{sec:DataGeneration}
We conducted a comprehensive simulation study to evaluate the performance of the proposed methods in comparison with common standard approaches. 
To reflect the subgroup selection process when employing our methods in practice, we first expand the subgroup structure within a study as outlined in Table \ref{tab:simulSubgroup}. 
Suppose study~$i$ has two subgroup options, say age (young or old, Y/O) and gender (male or female, M/F).
Note again that the subgroups considered may differ across studies in practice. 
Age and gender together define four mutually exclusive subgroups that together constitute the entire study population. 
For the simulations we assume independence of the two features, so that the prevalences of subgroup combinations result as products; for example, for given prevalences of young patients~($p_{\young,i}$) and males~($p_{\male,i}$), the corresponding prevelance of young males is $p_{\young\male,i}=p_{\young,i} \times p_{\male,i}$.
Given the two subgroup interaction effects 
%(e.g., $\delta_{i,Age}$ and $\delta_{i,Gender}$) 
($\delta_{\mathrm{age},i}$ and $\delta_{\mathrm{gender},i}$) 
generated from Equation (\ref{eqn:sub_specific3}), 
the four subgroup-specific effects then result as 
\begin{eqnarray}
    \theta_{\young\male,i}&=&\theta_i-(1-p_{\young,i})\,\delta_{\mathrm{age},i}-(1-p_{\male,i})\,\delta_{\mathrm{gender},i}\nonumber\\
    \theta_{\young\female,i}&=&\theta_i-(1-p_{\young,i})\,\delta_{\mathrm{age},i}+p_{\male,i}\,\delta_{\mathrm{gender},i}\nonumber\\
    \theta_{\old\male,i}&=&\theta_i+p_{\young,i}\,\delta_{\mathrm{age},i}-(1-p_{\male,i})\,\delta_{\mathrm{gender},i}\nonumber\\
    \theta_{\old\female,i}&=&\theta_i+p_{\young,i}\,\delta_{\mathrm{age},i}+p_{\male,i}\,\delta_{\mathrm{gender},i}\label{eqn:simulSubCal}
\end{eqnarray}
Based on these conditional subgroup-specific effects, the marginal subgroup effects are obtained as weighted averages, for example, 
$\theta_{\young,i}=\frac{p_{\young\male,i}\,\theta_{\young\male,i}+p_{\young\female,i}\,\theta_{\young\female,i}}{p_{\young,i}}$. 
It can be shown that these derived marginal subgroup effects satisfy Equation~(\ref{eqn:sub_specific2}); a detailed proof is provided in the Web Appendix~A.

\begin{table}[t] 
\centering
\caption{\label{tab:simulSubgroup}A table illustrating the (conditional and marginal) treatment effects within subgroups of a study in the simulation setup. The effects result from the overall treatment effect~($\theta_i$), the \emph{age} and \emph{gender} interaction effects ($\delta_{\mathrm{age},i}$ and $\delta_{\mathrm{gender},i}$) as well as the subgroup prevalences; see Section~\ref{sec:DataGeneration} for details.}

  \begin{tabular}{llcccc}
    \toprule
    & & \multicolumn{2}{c}{gender} & \\
    \cmidrule(rl){3-4}
    & & male  & female  & marginal \\
    \midrule
    age & young  & $\theta_{\young\male,i}$ & $\theta_{\young\female,i}$ & $\theta_{\young,i}$\\
        & old  & $\theta_{\old\male,i}$ & $\theta_{\old\female,i}$ & $\theta_{\old,i}$\\[1ex]
        & marginal & $\theta_{\male,i}$ & $\theta_{\female,i}$ & $\theta_i$\\
    \bottomrule
  \end{tabular}
\end{table}

Based on the above definitions, we then simulated studies' continuous outcome data motivated by a setting involving log-odds ratios as outcomes.
We first generated the true study-specific treatment effects according to Equation~(\ref{norm_stu}), and the magnitudes of heterogeneity were considered to reflect the situations from homogeneous to substantially heterogeneous ($0.0 \leq \tau \leq 1.0$), on the log-odds ratio scale.\cite{rover2021}  
We then generated subgroup interaction effects for the two subgroups~1 and~2 according to Equation~(\ref{eqn:sub_specific3}). 
For a given combination of the nuisance parameters ($\Delta_1$, $\sigma_{\Delta_1}$, $\Delta_2$, $\sigma_{\Delta_2}$, $p_1$, $p_2$), where the subscripts~1 and~2 denote the corresponding 
dichotomizing patient features (such as age and gender), we generated interaction effects ($\delta_{1,i}$, $\delta_{2,i}$) and subsequently subgroup-specific and marginal effects as defined above~(\ref{eqn:simulSubCal}). We restricted our settings for the two subgroup interaction effects such that $\Delta_1 \geq \Delta_2$ and $\sigma_{\Delta_1}\geq\sigma_{\Delta_2}$, 
thereby making subgroup~1 more likely to exhibit larger interaction effects and to be selected by $Q_{\mathrm{S}}$. The prevalence of subgroup~1 ($p_1$) was set to one of three levels ($\frac{1}{4}$, $\frac{1}{3}$ or $\frac{1}{2}$), whereas the prevalence of subgroup~2 ($p_2$) was kept fixed at~$\frac{1}{2}$. 
This allowed us to investigate how subgroup prevalence influenced the performance of the proposed methods. Although subgroup prevalences may vary across studies in practice, we focused on the biases that might potentially be caused by consistent effects in a common direction. Following R\"{o}ver \emph{et~al.},\cite{rover2021} we generated the sampling variance of each subgroup using $\uisd=4$, which is motivated by the common example setting of a log-odds ratio endpoint. The observed subgroup effects were then generated using Equation~(\ref{obs_sub}). With the generated subgroup-level data, we calculated the aggregated study-level data as in Equation~(\ref{cal_stu}) to perform the study-level meta-analysis. The number of studies was set to $k\in\{2,3,5\}$. Study sizes~($n_i$) were rounded to multiples of~24 in order to ensure integer subgroup sizes. Table~\ref{tab:simulScenarios} summarizes the parameter values considered in the simulation study.
Including the restrictions of $\Delta_1\geq \Delta_2$ and $\sigma_{\Delta_1}\ge\sigma_{\Delta_2}$
resulted in a total of 10125~scenarios. For each scenario, 1000~meta-analyses were simulated.

\begin{table}[h] 
\centering
\caption{\label{tab:simulScenarios}Parameter settings considered in the simulation study.}
\begin{tabular}{lc}
 \toprule
 parameter  & values \\
 \midrule 
 number of studies ($k$) & 2, 3, 5\\
 population mean ($\mu$) & 0.0\\
 between-study heterogeneity ($\tau$) & 0.0, 0.1, 0.2, 0.5, 1.0\\
 (mean) subgroup interaction effect ($\Delta_1$ and $\Delta_2$) & 0.0, 0.1, 0.2, 0.5, 1.0\\
 (SD) subgroup interaction effect ($\sigma_{\Delta_1}$ and $\sigma_{\Delta_2}$) & 0.0, 0.1, 0.2, 0.5, 1.0\\
 subgroup prevalence ($p_1$ and $p_2$) & $p_1\in \{1/2, 1/3, 1/4\},\; p_2=1/2$ \\
 study size ($n_i$)\textsuperscript{$\ast$} & $\lognormaldistn(5,1)$\\
\bottomrule 
  \multicolumn{2}{r}{\footnotesize\textsuperscript{$\ast$}study sizes~$n_i$ were rounded to multiples of 24 (minimum 24) to ensure integer subgroup sizes.}
\end{tabular}
\end{table}

\subsection{Methods to be compared}
To assess the empirical performance of our proposed methods for heterogeneity variance estimation, we compared the two proposed estimators (max1 and max2) based on the subgroup-level data to several standard estimators using the study-level data, including the widely used DerSimonian–Laird estimator (DL), the Paule–Mandel (PM) estimator,\cite{paule1982} and the restricted maximum likelihood (REML) estimator.\cite{Viechtbauer2005} In addition, we implemented the Bayes Modal estimator (BM) proposed by Chung \emph{et al.} \citep{chung2013} 
%as a Bayesian benchmark.
as an alternative approach aimed at ensuring non-zero heterogeneity estimates.
To further evaluate the impact of these heterogeneity estimators on inference on~$\mu$, we compared their statistical properties using three confidence intervals: the standard random-effects model based on a normal approximation (Normal), the Hartung–Knapp–Sidik–Jonkman method (HKSJ), and its modified version (mKH).\cite{rover2015} Results are reported according to the combination of heterogeneity estimator and confidence interval method employed. As noted by Weber \emph{et al.},\cite{weber2021} computational difficulties may arise when using the Monte Carlo sampling approach proposed by Michael \emph{et al.}\cite{michael2019} We encountered the same issue and therefore excluded this method from our comparisons. The standard study-level meta-analyses, together with the Normal and HKSJ CIs, were implemented using the \texttt{metafor} R~package. The mKH, ZH, and BM methods were implemented using our own code; the corresponding \textsf{R}~scripts are provided in the supplementary material.

\subsection{Results}
In total, we considered 10125~scenarios in the simulation study. In the following, we summarize the results by reporting marginal performance measures for each combination of the first subgroup's parameters ($\Delta_1$, $\sigma_{\Delta_1}$). 
Since our methods incorporate a subgroup selection step based on the $Q_{\mathrm{S}}$-guided procedure, we expect it to be dominated by the first subgroup's features (due to the constraints of $\Delta_1\geq\Delta_2$ and $\sigma_{\Delta_1}\geq\sigma_{\Delta_2}$).
Since all methods performed very similarly across different prevalence values, 
here we focus on the (1125)~settings with~$p_1=\frac{1}{2}$ and only~$k=2$ studies. In the case of only $k=2$ studies, the DL, REML and PM heterogeneity estimators coincide.\cite{Rukhin2012} Therefore, we present only the results based on the DL~estimator below. The complete simulation results showing the performance of all heterogeneity estimators and remaining settings are provided in Web Appendix~C\@.

\begin{figure}[h]
\centering\includegraphics[width=0.95\textwidth]{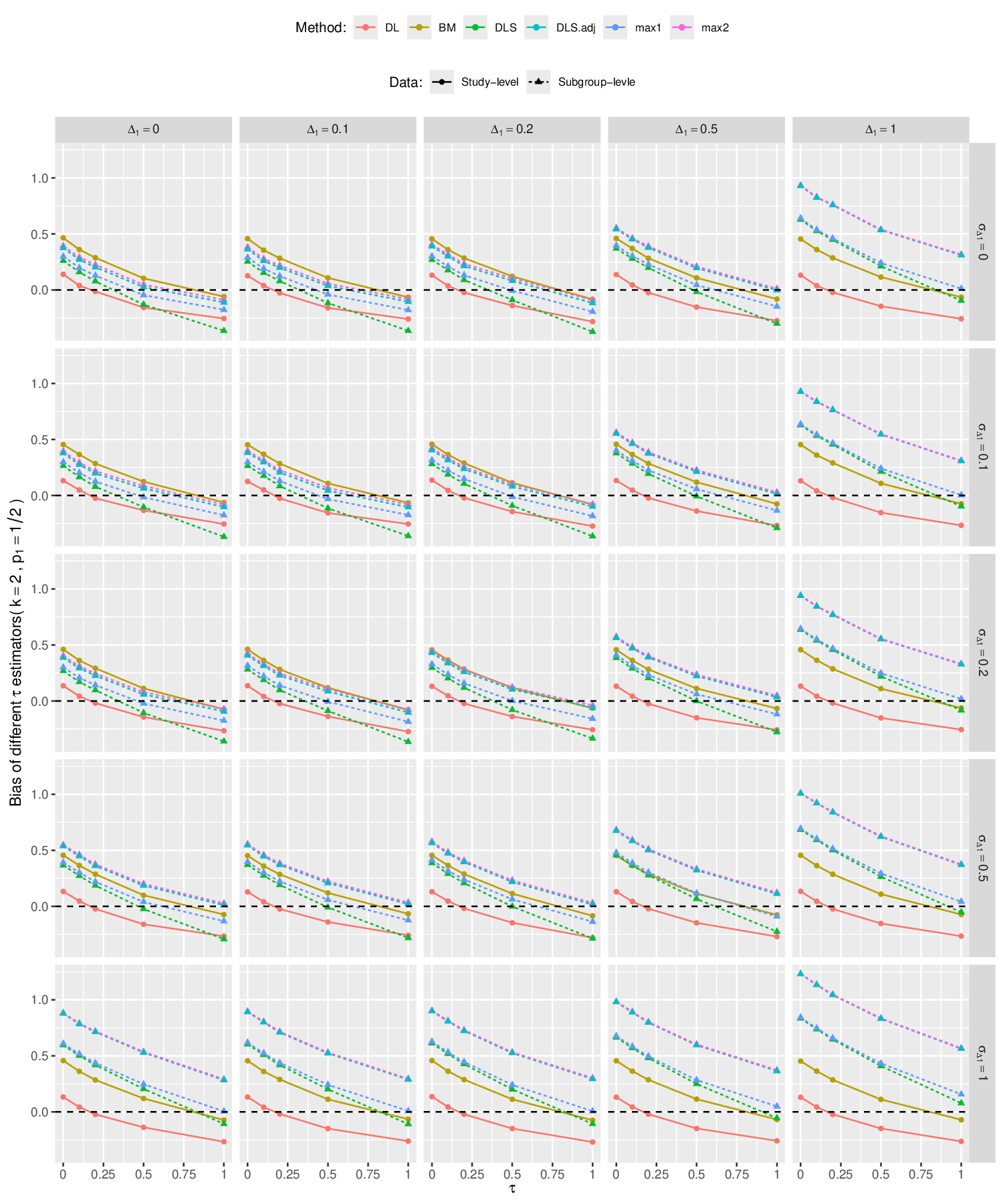}
\caption{Bias in estimating the between-study heterogeneity~$\tau$ for several study-level estimators as well as newly proposed estimators based on subgroup-level data ($k=2$~studies, subgroup prevalence~$p_1=\frac{1}{2}$).}
\label{fig:tauBias}
\end{figure}

We first focus on estimation of the heterogeneity~$\tau$, which is only an intermediate step, since primary interest is usually in the subsequent inference on the overall effect~$\mu$. It should however be noted that over- and underestimation of~$\tau$ have different consequences; underestimation is usually a greater concern, as it leads to overconfidence in the effect estimate (too small standard errors), while overestimation may be considered a ``conservative'' form of bias.
Figure~\ref{fig:tauBias} presents the bias of several $\tau$~estimators across a range of true $\tau$~values. All estimators based on study-level data exhibit increasing underestimation as the true~$\tau$ increases, although the BM~estimator is more robust to a moderate magnitude of heterogeneity values ($\tau\le 0.5$) than the alternatives. The DLS~estimator, which is derived solely from the subgroup-level data, may be substantially influenced by the shrinkage term~A (see equation~(\ref{exp:tau.dls}), and Figure~S28 in Web Appendix~C for a detailed illustration of its behavior). Although adjusting for this shrinkage substantially reduces the negative bias when the true heterogeneity is moderate to large ($\tau>0.5$), the DLS.adj estimator may still exhibit negative bias when the subgroup interaction effects are relatively small ($\Delta_1\le 0.5$ and $\sigma_{\Delta_1}\le 0.5$), corresponding to the nine panels in the upper-left corner. On the other hand, we also observed that the hybrid estimator max2 overlaps with the DLS.adj estimator across most scenarios, whereas max1 and the naive DLS estimators are more likely to diverge from each other. This suggests that the DLS.adj estimator is selected more frequently within max2, whereas max1 tends to adapt the classical DL estimator based on study-level data more often when the true heterogeneity is moderate to large. Consequently, max1 is less likely to benefit from the increased degrees of freedom arising from the extended hierarchical structure. Besides, even in the absence of subgroup interaction effects ($\Delta_1 = 0$ and $\sigma_{\Delta_1} = 0$), both proposed estimators outperform DL\@. A clear trend of the bias against the subgroup interaction effects for both max1 and max2 can be captured from left to right and from top to the bottom in Figure~\ref{fig:tauBias}. 

\begin{figure}[h]
\centering\includegraphics[width=0.95\textwidth]{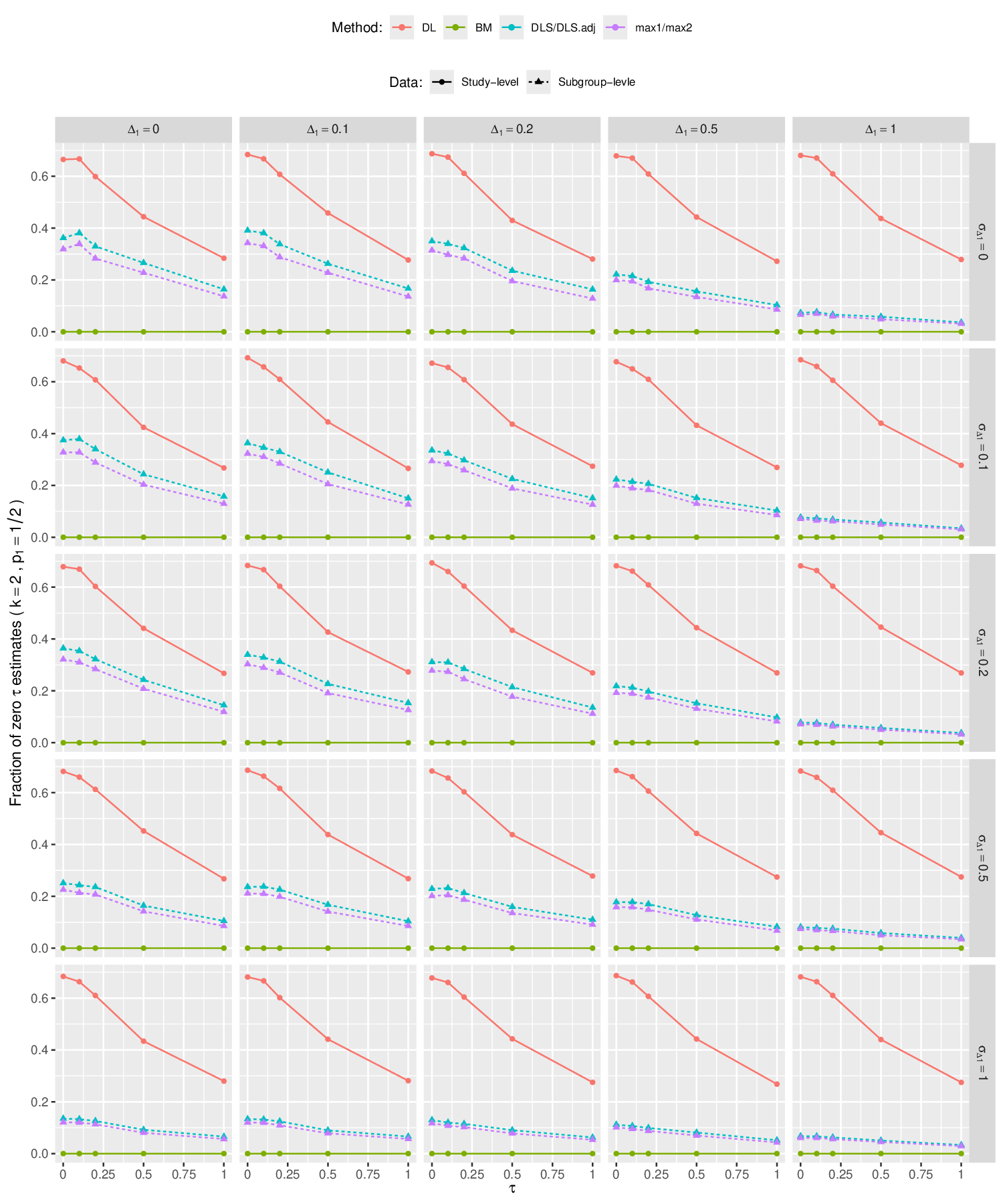}
\caption{Proportion of zero estimates for the between-study heterogeneity~$\tau$ for classic estimators using study-level data and proposed estimators using subgroup-level data ($k=2$~studies, subgroup prevalence~$p_i=\frac{1}{2}$).}
\label{fig:tauZeroes}
\end{figure}

Another important property of different $\tau$~estimators is how frequently they will yield zero estimates, since zero estimates of~$\tau$ will push the inference procedure to a usually overoptimistic common-effect model, hence may heavily impact the performance of CIs for the overall effect~$\mu$. From Figure~\ref{fig:tauZeroes} we can see that, except for the BM~estimator, which produces non-zero estimates in all scenarios, the DL estimator yields the highest proportions of zero estimates in all scenarios. On the other hand, the subgroup-level DLS estimator and its adjusted version (DLS.adj) produce substantially fewer zero estimates than those classical study-level estimators (roughly about half as many or less). The proposed estimators, max1 and max2, can further reduce the occurrence of zero estimates, this again highlights the necessity of introducing these hybrid version estimators. This benefit largely depends on the magnitude of subgroup interaction effects, represented by the values of~$\Delta$ and~$\sigma_\Delta$ in our model. When these parameters are sufficiently large (e.g., $\Delta_1=1$ or $\sigma_{\Delta_1}=1$), max1 and max2 only yield very few zeroes. A similar pattern in the impacts of $\Delta$ and $\sigma_\Delta$ on the number of zero estimates can be observed from the left to the right and from the top to the bottom in Figure~\ref{fig:tauZeroes}. Even when the subgroups are homogeneous ($\Delta_1=\sigma_{\Delta_1}=0$, top left panel of Figure~\ref{fig:tauZeroes}), these yield substantially fewer zero estimates than the corresponding study-level estimators.

Next, we investigate the performance of overall effect~($\mu$) estimates based on the different heterogeneity estimators. As noted by Henmi and Copas (2010),\cite{henmi2010} assuming normality of the treatment effects' distribution as in~(\ref{marg_stu}), both $\hat{\mu}_{\mathrm{CE}}$ and $\hat{\mu}_{\mathrm{RE}}$ are unbiased estimators to the population mean regardless of their different weighting scales, hence the focus here is on the coverage probabilities and lengths of CIs of various methods. 
In realistic applications, complications might however arise if effect estimates and (estimated) standard errors are correlated.\cite{KulinskayaHoaglinBakbergenuly2021} 

\begin{figure}[h]
\centering\includegraphics[width=0.95\textwidth]{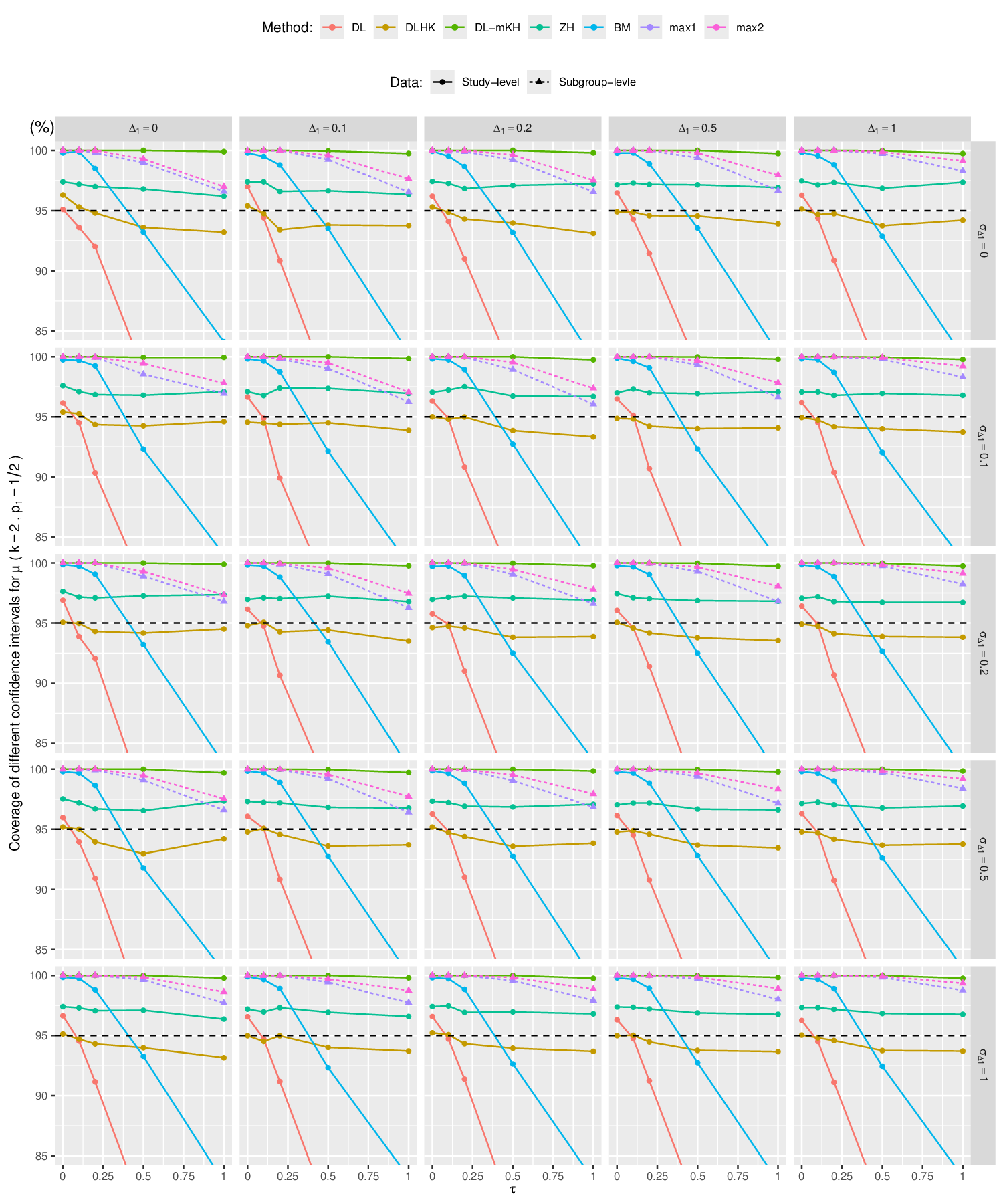}
 \caption{Coverage percentages for the overall effect~$\mu$ with 95\%~confidence intervals based on study-level data or based on subgroup-level data ($k=2$~studies, subgroup prevalence~$p_1=\frac{1}{2}$).}
\label{fig:muCoverage}
\end{figure}
Figure~\ref{fig:muCoverage} shows the coverage percentages of different CIs for~$\mu$ depending on the different values of between-study heterogeneity. To distinguish, we use solid lines to represent the methods using study-level data, and dashed lines for our proposed methods using subgroup-level data. Without small-sample adjustment, the CIs based on a normal approximation (DL and BM) exhibit substantial undercoverage. The CIs based on the HKSJ adjustment (DL-HKSJ) generally yield coverage probabilities below the nominal level of 95\% when the true heterogeneity is non-negligible ($\tau\geq 0.1$). In contrast, the mKH adjustment (DL-mKH) tends to produce substantial overcoverage, with coverage probabilities often close to 1. The ZH method also tends to moderately inflate the coverage, with coverage probabilities generally around 97.5\%. Our proposed methods (max1 and max2) generally lie between DL-mKH and ZH, with coverage closer to DL-mKH when $\tau\leq 0.5$ and closer to ZH otherwise.

\begin{figure}[ht]
\centering\includegraphics[width=0.95\textwidth]{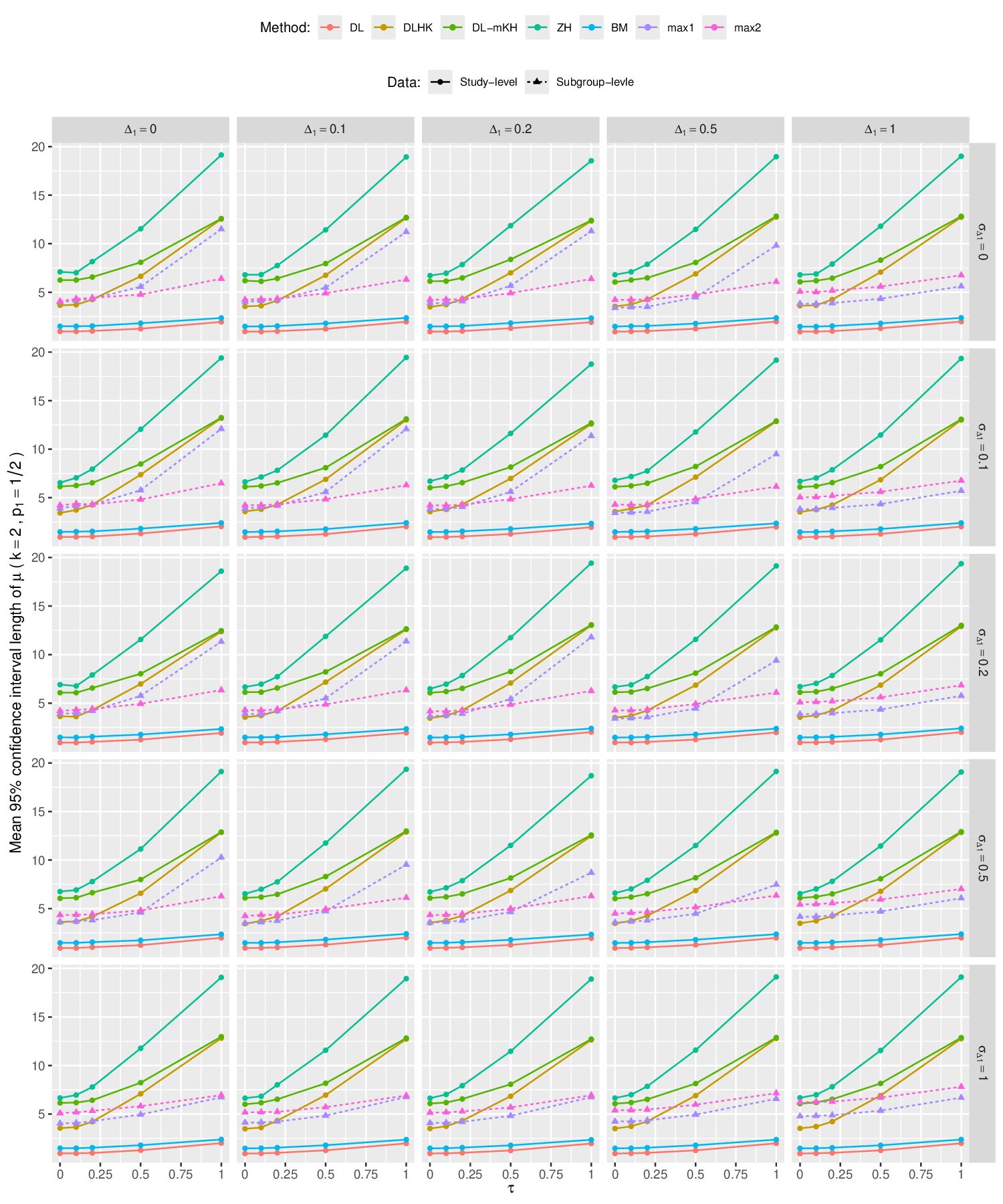}
\caption{Mean lengths of confidence intervals for the overall mean effect~$\mu$ ($k=2$~studies, subgroup prevalence~$p_1=\frac{1}{2}$).}
\label{fig:CiLengths}
\end{figure}

Next, we would like to investigate whether we may gain precision (in terms of the length of CIs) compared to the methods based on study-level data. 
Figure~\ref{fig:CiLengths} addresses this question, indicating a clear divergence between the study-level and subgroup-level approaches. Across all scenarios, the ZH method produces the widest CIs, whereas the CIs based on HKSJ and mKH adjustments yield similarly wide CIs and generally result in the second-largest interval lengths among the methods using study-level data. We see that the proposed methods (max1 and max2) have the ability to substantially reduce the length of CIs when moderate to large between-study heterogeneity is present (e.g., $\tau\geq 0.5$). In particular, CIs based on the max2 estimator provide a further reduction in CI length relative to max1 when the subgroup interaction effects are smaller than the true heterogeneity (e.g., $\tau \geq 0.5$, $\Delta_1 \leq 0.5$, and $\sigma_{\Delta_1} \leq 0.5$). We also observed that this advantage of our methods in reducing the length of CIs diminishes with increasing~$k$ (see the simulation results for $k>2$ in Web Appendix~C). 

\begin{figure}[ht]
\centering\includegraphics[width=0.95\textwidth]{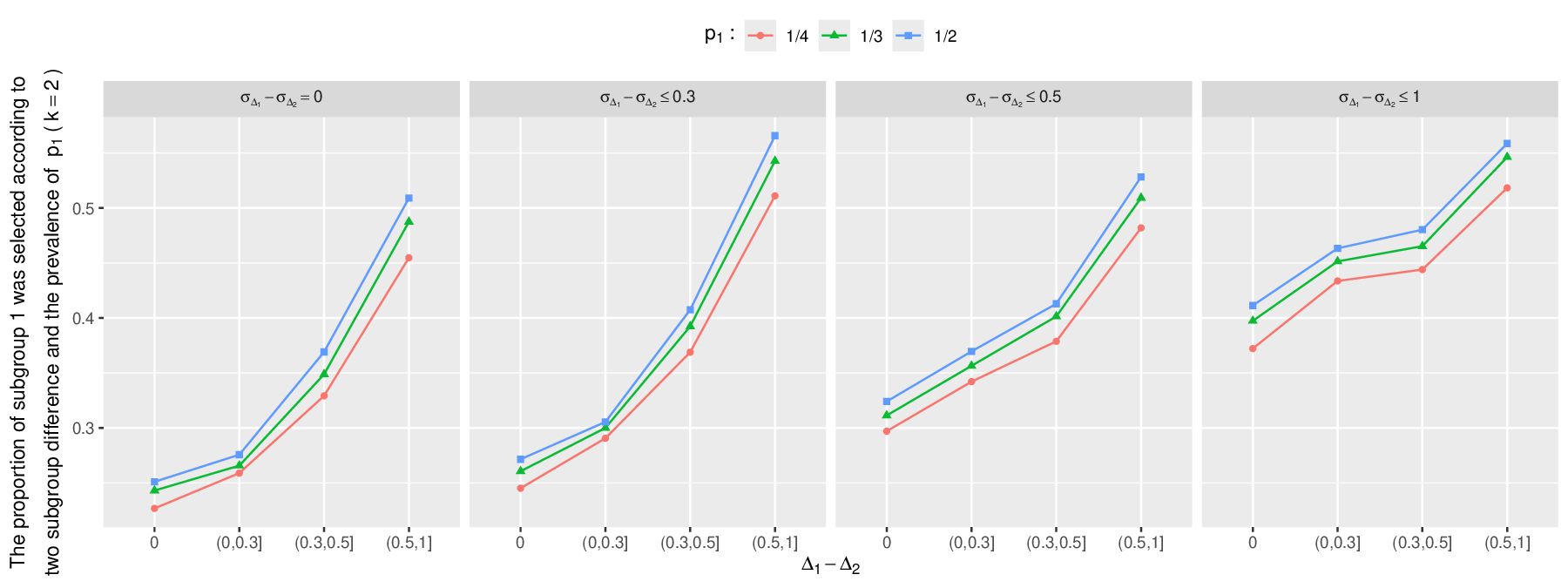}
\caption{Selection rates of Subgroup~1 under varying (treatment-by-subgroup) interaction effect differences and prevalences~$p_1$ of subgroup~1 ($k=2$~studies).}
\label{fig:subsel}
\end{figure}

Finally, we investigate the potential impacts of the subgroup selection procedure and subgroup prevalence on the performance of our method. Figure~\ref{fig:subsel} shows the frequency with which subgroup~1 is selected under varying interaction effects differences between the two subgroups and prevalences (of subgroup~1). Since the selection patterns are generally similar across different levels of heterogeneity, we present only the results as a function of the interaction effects differences ($\delta_1$ and $\delta_2$) and the prevalence of Subgroup~1 here. The corresponding heterogeneity-specific results are provided in Web Appendix~C\@. Note that we count only those cases in which all included studies select subgroup~1, since our simulation design assumes that subgroup~1 tends to exhibit larger subgroup interaction effects than subgroup~2. Overall, the selection rates of subgroup~1 are relatively low and tend to decrease as the number of studies increases. Unsurprisingly, subgroup~1 is more likely to be selected when the treatment difference between the two subgroups is substantial. In addition, a more balanced subgroup prevalence appears to facilitate the selection of subgroup~1. The level of heterogeneity appears to have little impact on the subgroup selection rates, as illustrated in Web Appendix~C.\@ 
Next, we investigate the potential influence of subgroup prevalence on coverage performance. We observe that prevalence of subgroup~1 ($p_1$) has little impact on the coverage probabilities of both proposed method (max1 or max2), although the scenarios with $p_1=\frac{1}{4}$ generally exhibit the lowest coverage among the three levels of~$p_1$. This may be explained by the fact that subgroup~1 is not selected particularly frequently across most studies. Since the observations are generally similar across different combinations of~$\Delta_1$ and~$\sigma_{\Delta_1}$, we only present the result under a moderate subgroup difference in Figure~\ref{fig:subsel}. The corresponding results stratified by~$\Delta_1$ and~$\sigma_{\Delta_1}$ are provided in Web Appendix~C\@.

\begin{figure}[ht]
\centering\includegraphics[width=0.45\textwidth]{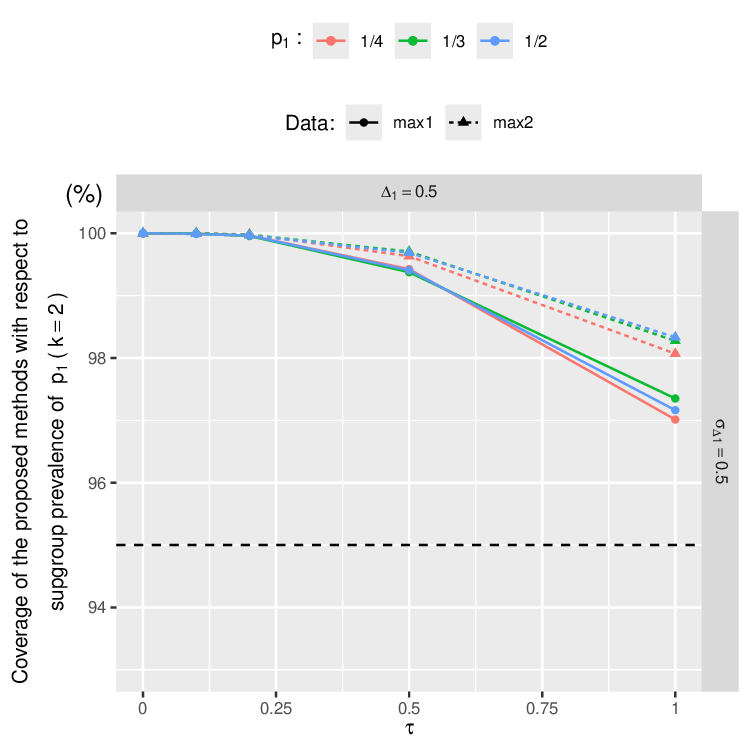}
\caption{Coverage percentages for the overall mean effect~$\mu$ with 95\%~confidence intervals based on subgroup-level data depending on the subgroup prevalence~$p_1$ ($k=2$~studies).}
\label{fig:muCoverPrev}
\end{figure}

\section{Re-analysis of motivating examples using subgroup information}
\label{exp:reanalysis}
\subsection{The \textsc{Respire} trials}
The two \textsc{Respire} trials introduced in Section~\ref{sec:respire} investigated the effects of DPI in long-term disease management of NCFB; the original data for the two (14-day and 28-day) regimens are shown in Figure~\ref{fig:RESPIRE}. For the primary analysis, a Cox proportional hazards model was used to compare the time to first exacerbation between the ciprofloxacin and placebo groups.
Since the treatment effects were estimated from covariate-adjusted Cox proportional hazards models, we assume that the UISD assumption applies to the adjusted treatment effect estimates. Specifically, the standard error of the adjusted treatment effect estimate is assumed to remain approximately inversely proportional to the corresponding square root of the number of events. To apply our method, we collected subgroup information from the FDA briefing document of the \textsc{Respire} trials,\citep{FDA2017b} as such information is generally available for trials submitted for regulatory approval. For the subgroup options, we considered three dichotomous subgroup variables based on patient \emph{age}, \emph{sex}, and \emph{race}. Only subgroup analyses in which subgroup sample sizes summed to the total sample size in the primary analysis were included. Consequently, the \emph{race} subgroup from RESPIRE~1 was excluded because its subgroup sample sizes were inconsistent with those reported in the primary analysis. Further details of the data are provided in Web Appendix~D. 

We synthesized the estimates from the two (14-day and 28-day) regimens from this pair of studies using classical as well as the newly proposed approaches; the results are presented in Table~\ref{tab:respire}. 
Since for $k=2$ studies the confidence intervals obtained using different heterogeneity estimators are identical, we report only those based on the DL~estimator. 
For the 14-days on/off regimen, random-effects meta-analyses using study-level data all produced inconclusive results, even when the standard normal approximation was used. 
CIs with small-sample adjustments were considerably wider, with the ZH method producing an extremely wide interval that may be of limited practical value. Given the small number of studies and the estimated magnitude of the between-study heterogeneity, this finding is not surprising. 
The BM~estimator yielded a slightly larger heterogeneity estimate of~0.360 in contrast to the frequentist estimates of~0.304, resulting in a slightly wider CI than the standard normal-approximation method. 
Using the same selection strategy we employed in the simulation, we selected the subgroup combination that maximized the empirical~$Q_{\mathrm{S}}$. 
Consequently, the \emph{sex} subgroup was selected for both studies to synthesize the treatment effects for the 14-day on/off regimen. 
This yielded a slightly larger heterogeneity estimate of~$\hat{\tau}_{\mathrm{DLS}}=0.387$ than both the ``classical'' DL and BM estimates using study-level data. 
The adjusted version $\hat{\tau}_{\mathrm{DLS.adj}}$ was even slightly larger at~$0.468$. 
However, with the increased number of degrees of freedom (3~instead of~1), the resulting CIs using our methods are much shorter and still hold the same conclusion. 
For the $\mu$~estimate, the common-effect estimate based on the subgroup-level data ($\hat{\mu}_\mathrm{CE}=0.689$) also differed slightly from the corresponding random-effects estimate based on the study-level data ($\hat{\mu}_\mathrm{DL}=0.680$).

We did the same analysis for the second treatment regimen (28-day on/off), using the combination of \emph{sex} for \textsc{Respire~1} and \emph{age} for \textsc{Respire~2} based on~$Q_{\mathrm{S}}$. 
In this case, all frequentist heterogeneity estimators, including our proposed estimators, yielded zero estimates, leading to common-effect analyses. 
In particular, for our proposed method, we would analyze the data in a study-level fashion, using the same degrees of freedom as mKH, and ZH methods. 
All methods except BM and mKH yielded statistically significant results. 
The CIs obtained using the mKH and our proposed methods were of similar length, whereas the BM method produced a relatively shorter CI despite its non-zero heterogeneity estimate ($\hat\tau_{\mathrm{BM}}=0.175$). The treatment effects of both regimens may warrant further consideration.

\begin{center}
\begin{table*}[h] 
\centering
\caption{Summary of meta-analyses of the two \textsc{Respire} trials using a range of approaches. The endpoint is the hazard ratio (HR) of the time to first exacerbation (see also Section~\ref{sec:respire}). Both trials included two treatment regimens (14-day and 28-day on/off).}
\label{tab:respire}
\begin{tabular}{lccccc}
 \toprule
 treatment regimen & method  & data & HR (95\%CI) & DF & $\hat{\tau}$  \\
 \midrule 
 14 days on/off & DL      & study-level    & 0.681 [0.419, 1.107]& &0.304\\
                & DL-HKSJ & study-level    & 0.681 [0.029, 15.868]&1&0.304\\
                & DL-mKH  & study-level    & 0.681 [0.029, 15.868]&1&0.304\\
                & BM      & study-level    & 0.681 [0.391, 1.185]& &0.360\\
                & ZH      & study-level    & 0.681 [0.008, 58.503]&1 &0.304\\[1ex]
                & max1    & subgroup-level & 0.689 [0.263, 1.803] &3&0.387\\
                & max2    & subgroup-level & 0.689 [0.223, 2.130]&3&0.468\\
  \cmidrule(rl){1-6}
 28 days on/off & DL      & study-level    & 0.720 [0.566, 0.917]&&0.000\\
                & DL-HKSJ & study-level    & 0.720 [0.604, 0.859]&1&0.000\\
                & DL-mKH  & study-level    & 0.720 [0.150, 3.451]&1&0.000\\
                & BM      & study-level    & 0.720 [0.511, 1.014]&&0.175\\
                & ZH      & study-level    & 0.720 [0.561, 0.925]&1&0.000\\[1ex]
                & max1    & subgroup-level & 0.705 [0.143, 3.469]&1&0.000\\
                & max2    & subgroup-level & 0.705 [0.143, 3.469]&1&0.000\\
\bottomrule 
\end{tabular}
\end{table*}
\end{center}

\subsection{The SGLT2 inhibitor meta-analysis}
\label{sec:SGLT2reanalysis}
The original meta-analysis of sodium-glucose cotransporter~2 (SGLT2) inhibitor studies (see Section~\ref{sec:SGLT2}) was implemented in two stages; first, Cox proportional hazard regression models were applied within each individual study using the intention-to-treat population, then the estimated log-hazard ratios were pooled using a random-effects model.
The significance of the SLGT2 inhibitor effect was eventually based on a zero heterogeneity estimate, effectively yielding a common-effect analysis.
Since subgroup summary effects are provided in the supplementary material, as is often the case in practice, we utilized the aggregated subgroup information for implementing our method, based on four pairs of subgroups that provided complete information across all studies (\emph{HbA1c}, \emph{eGFR}, \emph{Heart failure}, and \emph{Diuretic use}). For details of the data, see Web Appendix~D.
We again only considered those subgroups with numbers of patients consistent with those reported in the primary analysis.
Except for DECLARE-TIMI~58 and EMPA-REG OUTCOME, which each had two available subgroup options, all other studies had three subgroup options. 
The subgroup combination identified by~$Q_{\mathrm{S}}$ selected \emph{diuretic~use} for \textsc{CANVAS}~program, DAPA-CKD and DECLARE-TIMI~58, \emph{heart~failure} for CREDENCE, VERTIS~CV and EMPA-REG OUTCOME\@.
Except for the BM~estimator, which yielded a non-zero estimate of  
$\hat\tau_{BM}=0.069$, all other methods produced zero heterogeneity estimates based on the study-level data 
(therefore, only the results based on the DL estimator are shown here).
In contrast, based on the selected subgroups, the DLS method indicated non-zero heterogeneity, with $\hat{\tau}_{\mathrm{DLS}}=0.099$, which increased slightly after adjustment ($\hat{\tau}_{\mathrm{DLS.adj}}=0.104$).
Both estimates exceed the BM~estimate, resulting in wider 95\% confidence intervals whose upper bounds are close to the null value. Consequently, the evidence for a beneficial effect of sodium-glucose cotransporter~2 inhibitors appears more marginal, suggesting that the treatment effect should be interpreted with caution.

\begin{table}[h]
\centering
\caption{Summary of meta-analysis of sodium-glucose cotransporter~2 inhibitor studies (see also Section~\ref{sec:SGLT2}) using different analysis methods. The endpoint is the hazard ratio (HR) of time to serious hyperkalemia. }
\label{tab:sglt2}
\begin{threeparttable}
\begin{tabular}{ccccc}
 \toprule
  method  & data & HR (95\%CI) & DF & $\hat{\tau}$  \\
 \midrule 
  DL      & study-level    & 0.840 [0.763, 0.925]&&0.000\\
  DL-HKSJ & study-level    & 0.840 [0.762, 0.925]&5&0.000\\
  DL-mKH  & study-level    & 0.840 [0.740, 0.953]&5&0.000\\
  BM      & study-level    & 0.836 [0.749, 0.933]&&0.069\\
  ZH      & study-level    & 0.840 [0.764, 0.923]&5&0.000\\[1ex]
  max1    & subgroup-level & 0.843 [0.730, 0.973]&11&0.099\\
  max2    & subgroup-level & 0.843 [0.728, 0.976]&11&0.104\\
\bottomrule 
\end{tabular}
\end{threeparttable}
\end{table}

\section{Discussion}
The accuracy of meta-analysis of few studies often suffers from underestimation of the total variance, due to the fact that between-study heterogeneity cannot be reliably estimated in such a case. Some existing methods (e.g., HKSJ, mKH, ZH) aim to account for this issue, while considering the total variance as a whole and implementing heavier-tailed specifications (e.g., the Student-$t$ distribution) generally lead to wider intervals. Consequently, these efforts usually come at the cost of some unfavorable statistical properties, such as over-conservative coverage and wide CIs that may not be meaningful in practice. In the present investigation, we approached this problem by carefully nudging the empirical variance estimates by reasonable amounts, based on so far commonly ignored study information, while using the additional data granularity to yield less conservative inferences. This has been achieved in three aspects: 1) by utilizing the potential subgroup interaction effects within each study, we derived a heterogeneity estimator based on subgroup-level data, which can dramatically reduce the number of zero estimates compared to classic estimators; 2) the new data structure based on the subgroup-level (three-level) hierarchy allowed to increase the degrees-of-freedom in the $t$-quantile based CIs, which in turn may shorten the length of CIs as shown in the numerical study and empirical examples;
3) We utilized the HC-type variance under the common-effect framework because, when subgroups are collapsible, the resulting $\mu$ estimator is invariant to whether study-level or subgroup-level data are used. Changing the heterogeneity estimators affects only the confidence interval width, thus improving the coverage. Obtaining and checking of subgroup data requires some amount of ``detective work'', while our simulations demonstrated the greatest benefit in meta-analyses of~2 or~3 studies, yielding fewer zero estimates of heterogeneity, shorter confidence intervals, and less conservative coverage than standard methods based on study-level data, which may fail in such settings. Given this trade-off, we regard our methods as an important alternative for \emph{small} meta-analyses. In particular, our approach can be used as a sensitivity analysis when standard methods fail to detect heterogeneity but substantial heterogeneity remains a concern regarding the validity of conclusions under a common-effect model. Alternatively, when researchers seek greater precision in drawing conclusions, our methods may provide a useful alternative to the HKSJ- and mKH-adjusted methods.
 
Although the data structure is very similar, there are some distinct differences between our approach and subgroup meta-analysis.\cite{PanaroRoeverFriede2025} Firstly, the estimands are different. In our methods, we still focus on inference for the overall treatment effect~$\mu$, as in a ``classical'' meta-analysis using study-level data (while in a subgroup meta-analysis, the interest commonly is in the estimation of the subgroup interaction effect, denoted as~$\Delta$ above). The subgroup interaction effect(s) considered in our methods only contribute to the estimation of between-study heterogeneity, since potential overestimation may be helpful in preventing the notorious zero estimates as we showed in the simulation, and hence may altogether improve statistical properties. This introduces the second crucial difference, that is, we utilize the subgroups with the most pronounced differences for analysis in practice (as identified via $p$-values or empirical $Q_{\mathrm{S}}$-statistics). Therefore, the subgroups are not necessarily based on the same patient characteristics for all studies 
(recall the motivating examples discussed in Sections~\ref{sec:examples} and~\ref{exp:reanalysis}).

One might also wonder whether the search through combinations of subgroups across studies to maximize heterogeneity might be improved. We have described a ``local'' search strategy here, where $Q_{\mathrm{S}}$~is maximized for each study one-by-one. We have also investigated a ``global'' search across all possible combinations; if $m_i$~subgroups are reported for the $i$th study, this leads to a total of~$\prod_{i=1}^k m_i$ combinations to optimize over, compared to $\sum_i m_i$ evaluations for the local search. We did not notice any advantage of a global search, so that a local search that is much easier to perform and report appears to be preferrable.

We would also like to note several limitations of the proposed methods. The data-generating model considered here focused on the generic case of \emph{collapsible} subgroups (where the study's overall effect results as the average subgroup effect). In general, this may not necessarily hold,\cite{Greenland2010} which might add an additional source of variability to the subgroup-level data. This might in fact be accommodated in the data-generating model by adding an extra (fixed or random) term in~(\ref{eqn:sub_specific2}), however, we would expect the effect to be similar to that of increasing~$\Delta$ or~$\sigma_\Delta$, and this should again only affect heterogeneity estimation.
Furthermore, the current work is developed under the widely used normal-normal hierarchical model (NNHM), which assumes that the treatment effects follow a normal distribution. While normal approximations simplify computation and are appropriate in many cases, in particular when studies themselves are large, normality assumptions also have their limitations.\cite{JacksonWhite2018}
Extensions to alternative modeling frameworks, such as the binomial-normal hierarchical model (BNHM) \cite{SmithSpiegelhalterThomas1995} and generalized linear mixed models (GLMMs) \cite{stijnen2010}, which directly model binary outcomes rather than relying on the normal approximation for treatment effect estimates, represent another promising direction for future research.

 Next, we would like to point out some possible extensions of our current work. In the present investigation, we focused on the DerSimonian-Laird (DL) type heterogeneity estimator 
 for subgroup-level data, and observed considerable improvement in contrast to the classic (study-level) DL~estimator.
 Many alternative heterogeneity estimators have been proposed (such as the REML and PM estimators);\cite{VeronikiEtAl2016,petropoulou2017} the suggested approach of avoiding zero estimates for the heterogeneity by synthesizing the data at the subgroup level rather than the study level could be easily adopted based on other estimators, and similar benefits may be expected. 
 Also, while our simulations only covered the simple case of dichotomous subgroupings, we expect similar gains when a study may be divided in more than two subgroups.

%%%%%%%%%%%%%%%%%%%%%%%%%%%%%%

\section*{Acknowledgment}
  Support from the \emph{Deutsche Forschungsgemeinschaft (DFG)} is
  gratefully acknowledged (grant numbers \mbox{FR~3070/3-1} and \mbox{FR~3070/3-2}).

\section*{Conflicts of interest}
  The authors have declared no conflict of interest.

%%%%%%%%%%%%%%%%%%%%%%%%%%%%%%

\bibliographystyle{WileyNJD-AMA}
\bibliography{literature,reference}

\clearpage
\appendix

\section*{Web Appendix:
Utilizing subgroup information in random-effects meta-analysis of few studies}

\begin{abstract}
In this Web Appendix, we present the mathematical properties of our
heterogeneity estimator, the rationale for the simulation design, and
the complete simulation results.
\end{abstract}

\maketitle
\section*{Web Appendix A: Properties of the Q-statistic and the proposed heterogeneity estimator using subgroup information}
In the main text, we introduced the three-level hierarchy model for synthesizing the subgroup-level data, and the marginal model is given by

 \begin{equation*}
 \scalebox{0.8}{$\displaystyle
    \left(
        \begin{array}{c}
          y_{i,1} \\ y_{i,2}
        \end{array}\right) \mid \mu,\tau,\Delta,\sigma_\Delta
        \sim 
        \normaldistn \left(
        \left(
        \begin{array}{c}
          \mu \\\mu
        \end{array}\right)+
        \left(
        \begin{array}{c}         
         -(1-p_i)\Delta \\ p_i\Delta
        \end{array}\right)
        ,
        \left(
        \begin{array}{lrll}
          \tau^2+s_{i,1}^2& \tau^2   \\
          \tau^2 & \tau^2+s_{i,2}^2
        \end{array}\right) +
        \left(
        \begin{array}{lrlr}
          (1-p_i)^2\sigma_\Delta^2& -p_i(1-p_i)\sigma_\Delta^2   \\
          -p_i(1-p_i)\sigma_\Delta^2 & p_i^2\sigma_\Delta^2 
        \end{array}\right) \right).
        $}
\end{equation*}
From this, we have
\begin{eqnarray}
\expect(y_{i,1})&=&\mu-(1-p_i)\Delta\ \text{and}\ Var(y_{i,1})=\tau^2+s_{i,1}^2+(1-p_i)^2\sigma_\Delta^2\label{E.yi1}\\
\expect(y_{i,2})&=&\mu+p_i\Delta\ \text{and}\  Var(y_{i,2})=\tau^2+s_{i,2}^2+p_i^2\sigma_\Delta^2\label{E.yi2}\\
Cov(y_{i,1},y_{i,2})&=&\tau^2-p_i(1-p_i)\sigma_\Delta^2.\label{cov.yi1.yi2}
\end{eqnarray}

Considering the variance with respect to the grand population mean $\mu$, we can calculate
\begin{eqnarray}
\label{V.yi1_mu}
\expect[(y_{i,1}-\mu)^2]&=&\expect[y_{i,1}^2]-2\mu\expect[y_{i,1}]+\mu^2\nonumber\\
                        &=&Var(y_{i,1})+\expect[y_{i,1}]^2-2\mu\expect[y_{i,1}]+\mu^2\nonumber\\
                        &=&\tau^2+s_{i,1}^2+(1-p_i)^2\sigma_\Delta^2+[\mu-(1-p_i)\Delta]^2-2\mu[\mu-(1-p_i)\Delta]+\mu^2\nonumber\\
                 &=&\tau^2+(1-p_i)^2(\Delta^2+\sigma_\Delta^2)+s_{i,1}^2
\end{eqnarray}
\begin{eqnarray}
\label{V.yi2_mu}
\expect[(y_{i,2}-\mu)^2]&=&\expect[y_{i,2}^2]-2\mu\expect[y_{i,2}]+\mu^2\nonumber\\
                        &=&Var(y_{i,2})+\expect[y_{i,2}]^2-2\mu\expect[y_{i,2}]+\mu^2\nonumber\\
&=&\tau^2+s_{i,2}^2+p_i^2\sigma_\Delta^2+(\mu+p_i\Delta)^2-2\mu(\mu+p_i\Delta)+\mu^2\nonumber\\
                 &=&\tau^2+p_i^2(\Delta^2+\sigma_\Delta^2)+s_{i,2}^2
\end{eqnarray}

Based on these Equations~(\ref{E.yi1})--(\ref{V.yi2_mu}), we can derive the expectation of $Q_{S}$ as

\begin{eqnarray}
\expect(Q_{S})&=&\expect\left\{\sum_{i=1}^k\sum_{j=1}^2\omega_{i,j}(y_{i,j}-\hat{\mu}_{CE})^2\right\}\nonumber\\
      &=&\expect\left\{\sum_{i=1}^k\sum_{j=1}^2\omega_{i,j}(y_{i,j}-\mu)^2-(\sum_{i=1}^k\sum_{j=1}^2\omega_{i,j})            (\hat{\mu}_{CE}-\mu)^2\right\}\nonumber\\
      &=&\sum_{i=1}^k\omega_{i,1}\expect[(y_{i,1}-\mu)^2]+\omega_{i,2}\expect[(y_{i,2}-\mu)^2]-    
         (\sum_{i=1}^k\sum_{j=1}^2\omega_{i,j})Var(\hat\mu_{CE})\nonumber\\
      &=&\sum_{i=1}^k\omega_{i,1}\left\{\tau^2+(1-p_i)^2(\Delta^2+\sigma_\Delta^2)+s_{i,1}^2\right\}+\omega_{i,2}\left\{\tau^2+p_i^2(\Delta^2+\sigma_\Delta^2)+s_{i,2}^2\right\}\nonumber\\
      &-&(\sum_{i=1}^k\sum_{j=1}^2\omega_{i,j})\dfrac{\sum_{i=1}^k\omega_{i,1}^2Var(y_{i,1})+\omega_{i,2}^2Var(y_{i,2})+2\omega_{i,1}\omega_{i,2}Cov(y_{i,1},y_{i,2})}{(\sum_{i=1}^k\sum_{j=1}^2\omega_{i,j})^2}\nonumber\\
      &=&2k+\tau^2\sum_{i=1}^k\sum_{j=1}^2\omega_{i,j}+(\Delta^2+\sigma_\Delta^2)\sum_{i=1}^k\sum_{j=1}^2\omega_{i,j}p_i(1-p_i)\nonumber\\
      &-&\dfrac{\sum_{i=1}^k\omega_{i,1}^2\left\{\tau^2+s_{i,1}^2+(1-p_i)^2\sigma_\Delta^2\right\}+\omega_{i,2}^2\left\{\tau^2+s_{i,2}^2+p_i^2\sigma_\Delta^2\right\}+2\omega_{i,1}\omega_{i,2}\left\{\tau^2-p_i(1-p_i)\sigma_\Delta^2\right\}}{\sum_{i=1}^k\sum_{j=1}^2\omega_{i,j}}\nonumber\\
      &&(\mbox{Recall that:}\ \omega_i=\sum_{j=1}^2\omega_{i,j},\ \mbox{and}\ \omega_{i,1}=p_i\omega_1,\ \omega_{i,2}=(1-p_i)\omega_i)\nonumber\\
      &=&(2k-1)+\tau^2\left\{\sum_{i=1}^k\sum_{j=1}^2\omega_{i,j}-\dfrac{\sum_{i=1}^k\sum_{j=1}^2\omega_{i,j}^2}{\sum_{i=1}^k\sum_{j=1}^2\omega_{i,j}}\right\}-\tau^2\left\{\dfrac{2\sum_{i=1}^k\omega_{i,1}\omega_{i,2}}{\sum_{i=1}^k\sum_{j=1}^2\omega_{i,j}}\right\}+(\Delta^2+\sigma_\Delta^2)\sum_{i=1}^k\sum_{j=1}^2\omega_{i,j}p_i(1-p_i)\nonumber
\end{eqnarray}
\allowdisplaybreaks

Then we could calculate the expectation of the DerSimonian-Laird type estimator using the subgroup information as
\begin{eqnarray}
\expect(\hat{\tau}_{DLS}^2)&=&\dfrac{\expect(Q_{S})-(2k-1)}  
               {\sum_{i=1}^k\sum_{j=1}^2\omega_{i,j}-\dfrac{\sum_{i=1}^k\sum_{j=1}^2\omega_{i,j}^2}{\sum_{i=1}^k\sum_{j=1}^2\omega_{i,j}}}\nonumber\\
           &=&\tau^2\left\{\dfrac{\sum_{i=1}^k\sum_{j=1}^2\omega_{i,j}-\dfrac{\sum_{i=1}^k\sum_{j=1}^2\omega_{i,j}^2}{\sum_{i=1}^k\sum_{j=1}^2\omega_{i,j}}-\dfrac{2\sum_{i=1}^k\omega_{i,1}\omega_{i,2}}{\sum_{i=1}^k\sum_{j=1}^2\omega_{i,j}}}{\sum_{i=1}^k\sum_{j=1}^2\omega_{i,j}-\dfrac{\sum_{i=1}^k\sum_{j=1}^2\omega_{i,j}^2}{\sum_{i=1}^k\sum_{j=1}^2\omega_{i,j}}} \right\}+
           (\Delta^2+\sigma_\Delta^2)\dfrac{\sum_{i=1}^k\sum_{j=1}^2\omega_{i,j}p_i(1-p_i)}{\sum_{i=1}^k\sum_{j=1}^2\omega_{i,j}-\dfrac{\sum_{i=1}^k\sum_{j=1}^2\omega_{i,j}^2}{\sum_{i=1}^k\sum_{j=1}^2\omega_{i,j}}}\nonumber\\
           &=& \tau^2\left\{1-\dfrac{\dfrac{2\sum_{i=1}^k\omega_{i,1}\omega_{i,2}}{\sum_{i=1}^k\sum_{j=1}^2\omega_{i,j}}}{\dfrac{(\sum_{i=1}^k\sum_{j=1}^2\omega_{i,j})^2-\sum_{i=1}^k\sum_{j=1}^2\omega_{i,j}^2}{\sum_{i=1}^k\sum_{j=1}^2\omega_{i,j}}}\right\}+
           (\Delta^2+\sigma_\Delta^2)\dfrac{\sum_{i=1}^k\sum_{j=1}^2\omega_{i,j}p_i(1-p_i)}{\dfrac{(\sum_{i=1}^k\sum_{j=1}^2\omega_{i,j})^2-\sum_{i=1}^k\sum_{j=1}^2\omega_{i,j}^2}{\sum_{i=1}^k\sum_{j=1}^2\omega_{i,j}}}
           \nonumber\\
           &=&\tau^2\left\{1-\dfrac{2\sum_{i=1}^k\omega_{i,1}\omega_{i,2}}{(\sum_{i=1}^k\sum_{j=1}^2\omega_{i,j})^2-\sum_{i=1}^k\sum_{j=1}^2\omega_{i,j}^2}\right\}+(\Delta^2+\sigma_\Delta^2)\dfrac{\Big(\sum_{i=1}^k\sum_{j=1}^2\omega_{i,j}p_i(1-p_i)\Big)\Big(\sum_{i=1}^k\sum_{j=1}^2\omega_{i,j}\Big)}
           {(\sum_{i=1}^k\sum_{j=1}^2\omega_{i,j})^2-\sum_{i=1}^k\sum_{j=1}^2\omega_{i,j}^2}\nonumber\\
           &=&A\tau^2+B.\nonumber
\end{eqnarray}

Since
\begin{eqnarray}
(\sum_{i=1}^k\sum_{j=1}^2\omega_{i,j})^2-\sum_{i=1}^k\sum_{j=1}^2\omega_{i,j}^2=2\sum_{i=1}^k\omega_{i,1}\omega_{i,2}+2\mathop{\sum}_{i\neq i'}(\omega_{i,1}+\omega_{i,2})(\omega_{i',1}+\omega_{i',2}),\nonumber
\end{eqnarray}
accordingly, we have
\begin{eqnarray}
A &=& 1-\dfrac{\sum_{i=1}^k\omega_{i,1}\omega_{i,2}}{\sum_{i=1}^k\omega_{i,1}\omega_{i,2}+\mathop{\sum}_{i\neq i'}(\omega_{i,1}+\omega_{i,2})(\omega_{i',1}+\omega_{i',2})}\nonumber\\
 &=& 1-\dfrac{1}{1+\dfrac{\sum_{i\neq i'}n_in_{i'}}{\sum_{i=1}^kn_i^2p_i(1-p_i)}}<1.\nonumber
\end{eqnarray}
As long as the number of studies goes to infinity, this term will approximate to 1, when the number of studies is small, it will introduce a negative bias to the $\tau^2$ estimates. On the other hand,
\begin{eqnarray}
B&=&(\Delta^2+\sigma_\Delta^2)\dfrac{\Big(\sum_{i=1}^k\sum_{j=1}^2\omega_{i,j}p_i(1-p_i)\Big)\Big(\sum_{i=1}^k\sum_{j=1}^2\omega_{i,j}\Big)}
           {(\sum_{i=1}^k\sum_{j=1}^2\omega_{i,j})^2-\sum_{i=1}^k\sum_{j=1}^2\omega_{i,j}^2}\nonumber\\
   &=&\dfrac{(\Delta^2+\sigma_\Delta^2)}  
  {2}\dfrac{\Big(\sum_{i=1}^k\sum_{j=1}^2\omega_{i,j}p_i(1-p_i)\Big)\Big(\sum_{i=1}^k\sum_{j=1}^2\omega_{i,j}\Big)}{\sum_{i=1}^k\omega_{i,1}\omega_{i,2}+\mathop{\sum}_{i\neq i'}(\omega_{i,1}+\omega_{i,2})(\omega_{i',1}+\omega_{i',2})}\nonumber\\
  &=&\dfrac{(\Delta^2+\sigma_\Delta^2)}  
  {2}\dfrac{\Big(\sum_{i=1}^kn_ip_i(1-p_i)\Big)\Big(\sum_{i=1}^kn_i\Big)}{\sum_{i=1}^kn_i^2p_i(1-p_i)+\sum_{i\neq i'}n_in_{i'}}>0,\nonumber
\end{eqnarray}
which may introduce a positive bias (which relates to the magnitude of subgroup interaction effects) to the $\tau^2$ estimates.

\section*{Web appendix B: Proof of the simulation settings}
In Section 6.1 of the main text, we described how subgroup level data were generated for each study in the meta-analysis, where each study consists of two subgroups. Here, we briefly prove that the simulated data for each subgroup satisfy the marginal model specified in Section 4.1.

Following the setup in the main text, we suppose study~$i$ has two subgroup options, say age (young or old, Y/O) and gender (male or female, M/F). Age and sex together define four mutually exclusive subgroups that constitute the entire study population. Based on four subgroup-specific effects defined in Equation (22) of the main text, we show the marginal subgroup effects are obtained as weighted averages. For example, 

\begin{eqnarray}
\theta_{\young,i}&=&\dfrac{p_{\young\male,i}\,\theta_{\young\male,i}+p_{\young\female,i}\,\theta_{\young\female,i}}{p_{\young,i}}\nonumber\\
            &=&\dfrac{p_{\young\male,i}\,\Big[\theta_i-(1-p_{\young,i})\,\delta_{\mathrm{age},i}-(1-p_{\male,i})\,\delta_{\mathrm{gender},i}\Big]+p_{\young\female,i}\,\Big[\theta_i-(1-p_{\young,i})\,\delta_{\mathrm{age},i}+p_{\male,i}\delta_{\mathrm{gender},i}\Big]}{p_{\young,i}}\nonumber\\
            &=&\dfrac{(p_{\young\male,i}+p_{\young\female,i})\Big[(\theta_i-(1-p_{\young,i})\,\delta_{\mathrm{age},i}\Big]+\Big[p_{\young\male,i}\,(-1+p_{\male,i})+p_{\young\female,i}\,p_{\male,i}\Big]\delta_{\mathrm{gender},i}}{p_{\young,i}}\nonumber\\\label{theta_Y}
            &=&\theta_i-(1-p_{\young,i})\,\delta_{\mathrm{age},i}\\
            [5pt]
\theta_{\old,i}&=&\dfrac{p_{\old\male,i}\theta_{\old\male,i}+p_{\old\female,i}\theta_{\old\female,i}}{p_{\old,i}}\nonumber\\
            &=&\dfrac{p_{\old\male,i}\,\Big[\theta_i+p_{\young,i}\delta_{\mathrm{age},i}-(1-p_{\male,i})\,\delta_{\mathrm{gender},i}\Big]+p_{\old\female,i}\,(\theta_i+p_{\young,i}\delta_{\mathrm{age},i}+p_{\male,i}\delta_{\mathrm{gender},i})}{p_{\old,i}}\nonumber\\
            &=&\dfrac{(p_{\old\male,i}+p_{\old\female,i})\,(\theta_i+p_{\young,i}\delta_{\mathrm{age},i})+\Big[p_{\old\male,i}(-1+p_{\male,i})+p_{\old\female,i}\,p_{\male,i}\Big]\delta_{\mathrm{gender},i}}{p_{\old,i}}\nonumber\\
            &=&\theta_i+p_{\young,i}\delta_{\mathrm{age},i}  \label{theta_O}
\end{eqnarray}

Similarly, we can show that 
\begin{eqnarray}
\theta_{\male,i}&=&\dfrac{p_{\young\male,i}\theta_{\young\male,i}+p_{\old\male,i}\theta_{\old\male,i}}{p_{\male,i}}\nonumber\\
            &=&\dfrac{p_{\young\male,i}\Big[\theta_i-(1-p_{\young,i})\,\delta_{\mathrm{age},i}-(1-p_{\male,i})\,\delta_{\mathrm{gender},i}\Big]+p_{\old\male,i}\Big[\theta_i+p_{\young,i}\delta_{\mathrm{age},i}-(1-p_{\male,i})\,\delta_{\mathrm{gender},i}\Big]}{p_{\male,i}}\nonumber\\
            &=&\dfrac{(p_{\young\male,i}+p_{\old\male,i})\Big[(\theta_i-(1-p_{\male,i})\,\delta_{\mathrm{gender},i}\Big]+\Big[p_{\young\male,i}(-1+p_{\young,i})+p_{\old\male,i}p_{\young,i}\Big]\delta_{\mathrm{age},i}}{p_{\male,i}}\nonumber\\
            &=&\theta_i-(1-p_{\male,i})\,\delta_{\mathrm{gender},i}\\
            [5pt]
\theta_{\female,i}&=&\dfrac{p_{\young\female,i}\theta_{\young\female,i}+p_{\old\female,i}\theta_{\old\female,i}}{p_{\female,i}}\nonumber\\
            &=&\dfrac{p_{\young\female,i}\Big[\theta_i-(1-p_{\young,i})\,\delta_{\mathrm{age},i}+p_{\male,i}\,\delta_{\mathrm{gender},i}\Big]+p_{\old\female,i}\,(\theta_i+p_{\young,i}\delta_{\mathrm{age},i}+p_{\male,i}\,\delta_{\mathrm{gender},i})}{p_{\female,i}}\nonumber\\
            &=&\dfrac{(p_{\young\female,i}+p_{\old\female,i})\,(\theta_i+p_{\male,i}\,\delta_{\mathrm{gender},i})+\Big[p_{\young\female,i}\,(-1+p_{\young,i})+p_{\old\female,i}\,p_{\young,i}\Big]\delta_{\mathrm{age},i}}{p_{\female,i}}\nonumber\\
            &=&\theta_i+p_{\male,i}\delta_{\mathrm{gender},i}  \label{theta_F}
\end{eqnarray}

As shown in Equations (\ref{theta_Y})–(\ref{theta_F}), the marginalized subgroup effects follow the same form as Model (8) in Section 4.2. Consequently, the simulated data conform to the assumed marginal model and are therefore suitable for assessing the performance of the proposed methods.

\section*{Web appendix C: Table and figures for the complete simulation results}
In this appendix, we present the complete simulation results corresponding to Section 5 of the main text. The layout of the figures under the various simulation scenarios is summarized in Table (\ref{sim:layout}).
Since the proposed methods involve a subgroup selection procedure, their post-selection performance is of interest. Therefore, we focus on settings in which subgroup 1 is expected to be selected for the potential greater benefits than subgroup 2. Nevertheless, due to sampling variability, subgroup 1 is not always selected in the simulation studies, as will be shown in the results.
To simplify the presentation, we summarize the results according to each ($\Delta_1$,$\sigma_{\Delta_1}$) combination that corresponds to the largest potential benefit our methods can obtain from the optimal subgroup. To understand how prevalence of one subgroup is going to change the performance of our methods, we present the figures of results according to different values of prevalence $p_1$, since subgroup 1 is potentially the optimal subgroup. For each value of $p_1$, we summarize the performance of the proposed methods and several standard approaches for estimating the between-study heterogeneity ($\tau^2$) using two figures as in the main text:

1. Bias in estimating the between-study heterogeneity parameter ($\tau$) for both study-level estimators and the proposed subgroup-level estimators;
2. The fraction of zero estimates produced by the different estimators.
3. Mean squared error of different $\tau$ estimators.

For the overall treatment effect ($\mu$), we present four figures:

1. Bias of different treatment effect estimators;
2. Monte Carlo errors of these different estimators;
3. Coverage probabilities of the 95\% confidence intervals based on study-level and subgroup-level analyses;
4. Mean lengths of the corresponding confidence intervals.

The third figure for $\tau$ and the first two figures for $\mu$ are not included in the main text, as they provide less insight into the differences in performance among the competing methods. Finally, we present additional properties of the subgroup selection procedure.

\begin{table}[ht]
\centering
\caption{Layout of the simulation results}
\label{sim:layout}

\begin{tabular}{ccccc}
\toprule
Prevalence ($p_i$) & $K$ & Parameter & Summary & No. of figures \\
\midrule

\multirow[t]{7}{*}{$1/4$}
& \multirow[t]{3}{*}{}
& \multirow[t]{3}{*}{$\tau^2$}
& Bias & S1 \\
&&& FOZ & S2 \\
&&& MSE & S3 \\
& & \multirow[t]{4}{*}{$\mu$}
& Bias & S4 \\
&&& MCSE & S5 \\
&&& CP & S6 \\
&&& LOCI & S7 \\

\midrule

\multirow[t]{7}{*}{$1/3$}
& \multirow[t]{3}{*}{}
& \multirow[t]{3}{*}{$\tau^2$}
& Bias & S8 \\
&&& FOZ & S9 \\
&&& MSE & S10 \\
& & \multirow[t]{4}{*}{$\mu$}
& Bias & S11 \\
&&& MCSE & S12 \\
&&& CP & S13 \\
&&& LOCI & S14 \\

\midrule

\multirow[t]{7}{*}{$1/2$}
& \multirow[t]{3}{*}{}
& \multirow[t]{3}{*}{$\tau^2$}
& Bias & S15 \\
&&& FOZ & S16 \\
&&& MSE & S17 \\
& & \multirow[t]{4}{*}{$\mu$}
& Bias & S18 \\
&&& MCSE & S19 \\
&&& CP & S20 \\
&&& LOCI & S21 \\

\midrule

& \multirow[t]{2}{*}{$2$}
& & CSR & S22 \\
& & $\mu$ & CP & S23 \\
& \multirow[t]{2}{*}{$3$}
& & CSR & S24 \\
& & $\mu$ & CP & S25 \\
& \multirow[t]{2}{*}{$5$}
& & CSR & S26 \\
& & $\mu$ & CP & S27 \\

\midrule

& & $A$ & DOA & S28 \\

\bottomrule
\end{tabular}

\vspace{2mm}

\begin{minipage}{0.90\textwidth}
\footnotesize
FOZ, fraction of zero estimates;
MSE, mean squared error;
MCSE, monte carlo standard error;
LOCI, length of confidence intervals;
CSR, correct selection rate;
DOA, distribution of the adjustment term $A$.
\end{minipage}

\end{table}

\renewcommand{\thefigure}{S\arabic{figure}}

\begin{figure}[t]
\centering\includegraphics[width=0.95\textwidth]{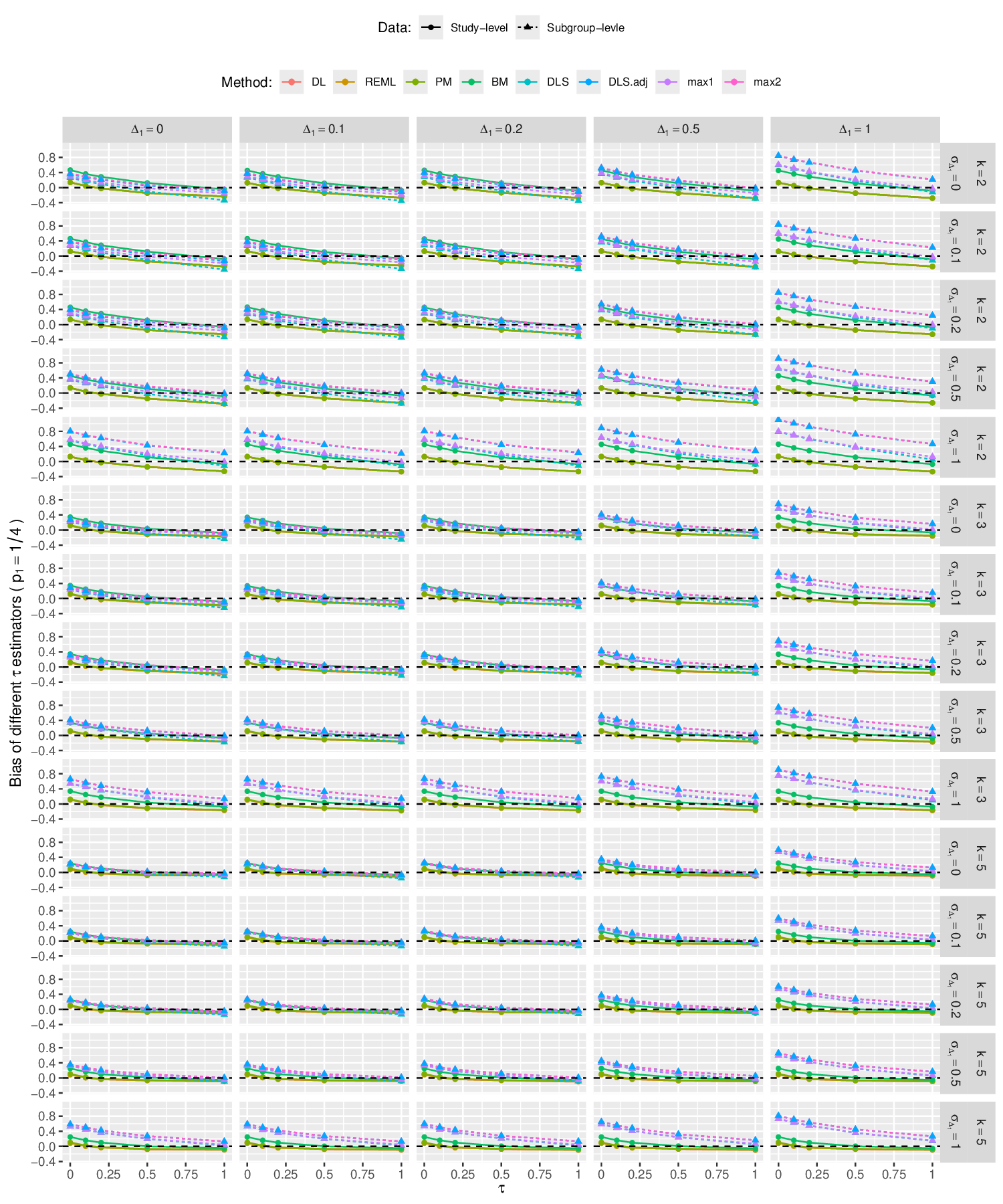}
\caption{\it{
Bias in estimating the between-study heterogeneity $\tau$ for several study-level estimators as well as the newly proposed
estimators based on subgroup-level data ($p_1$=1/4)}}
\label{fig s1}
\end{figure}

\begin{figure}[t]
\centering\includegraphics[width=0.95\textwidth]{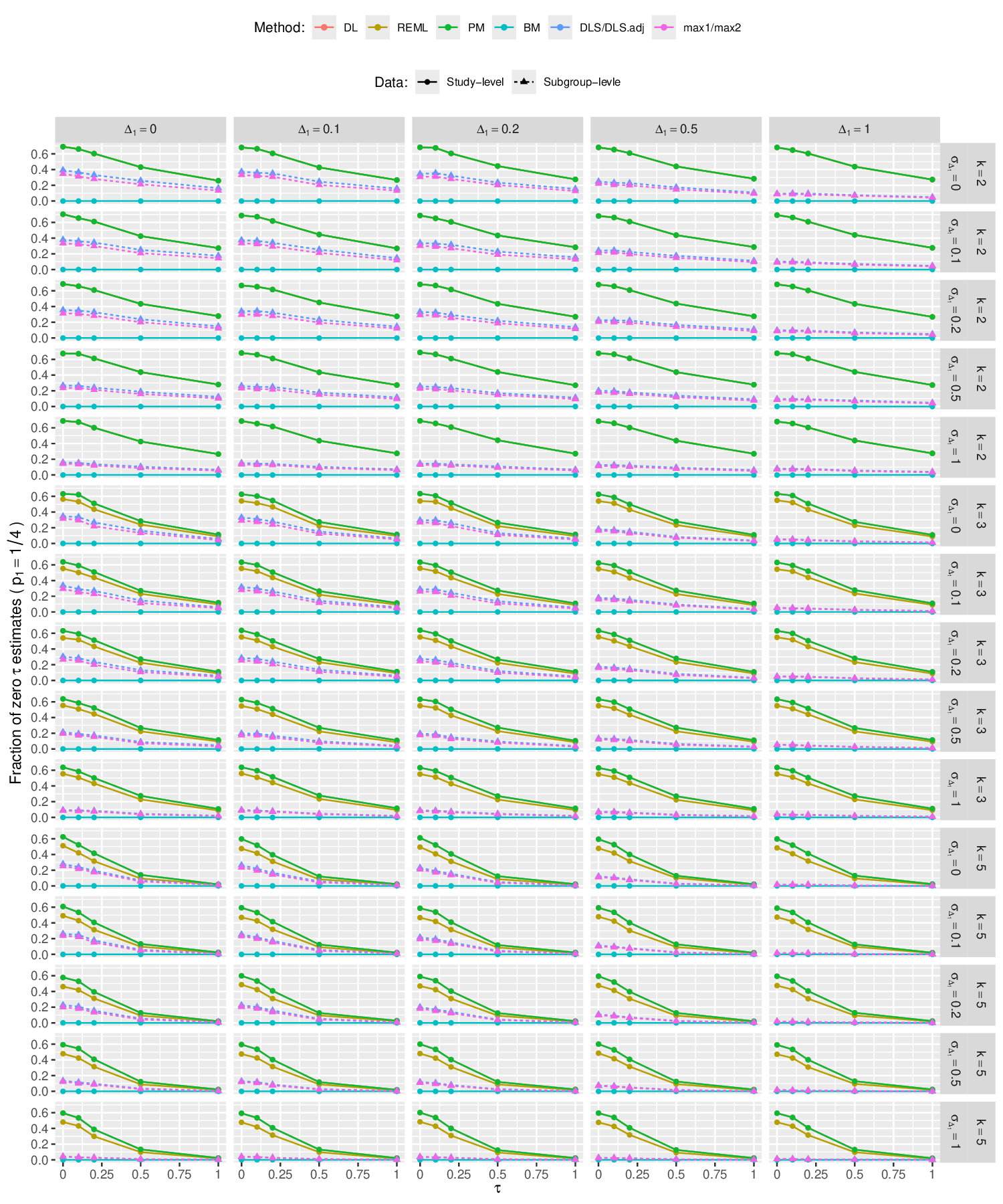}
\caption{\it{
Fraction of zero estimates of the between-study heterogeneity $\tau$ for the classic estimators using study-level data and the newly proposed estimators using subgroup-level data ($p_1$=1/4)}}
\label{fig s2}
\end{figure}

\begin{figure}[t]
\centering\includegraphics[width=0.95\textwidth]{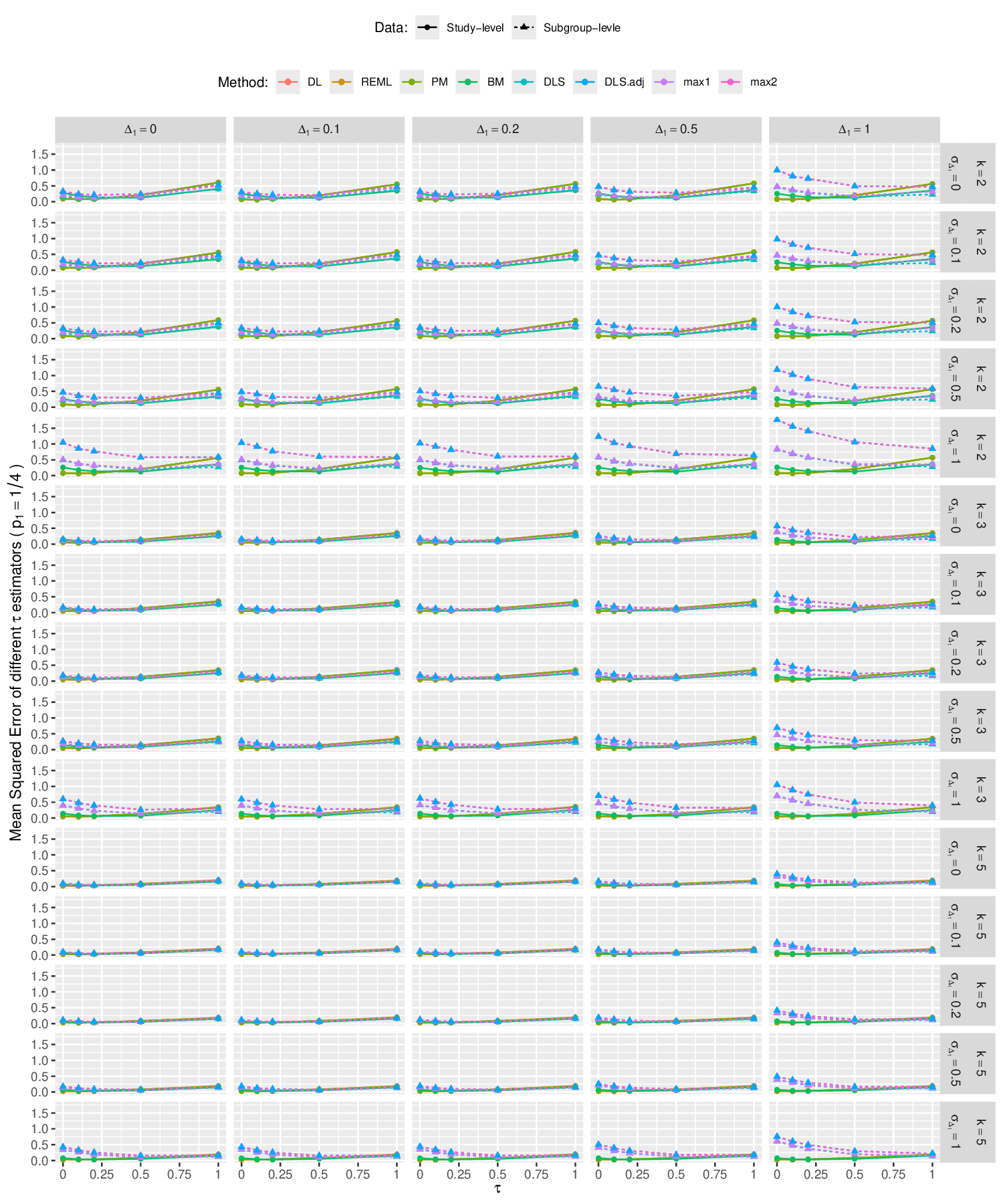}
\caption{\it{
Mean squared errors of the between-study heterogeneity $\tau$ for the classic estimators using study-level data and the newly proposed estimators using subgroup-level data ($p_1$=1/4)}}
\label{fig s3}
\end{figure}

\begin{figure}[t]
\centering\includegraphics[width=0.95\textwidth]{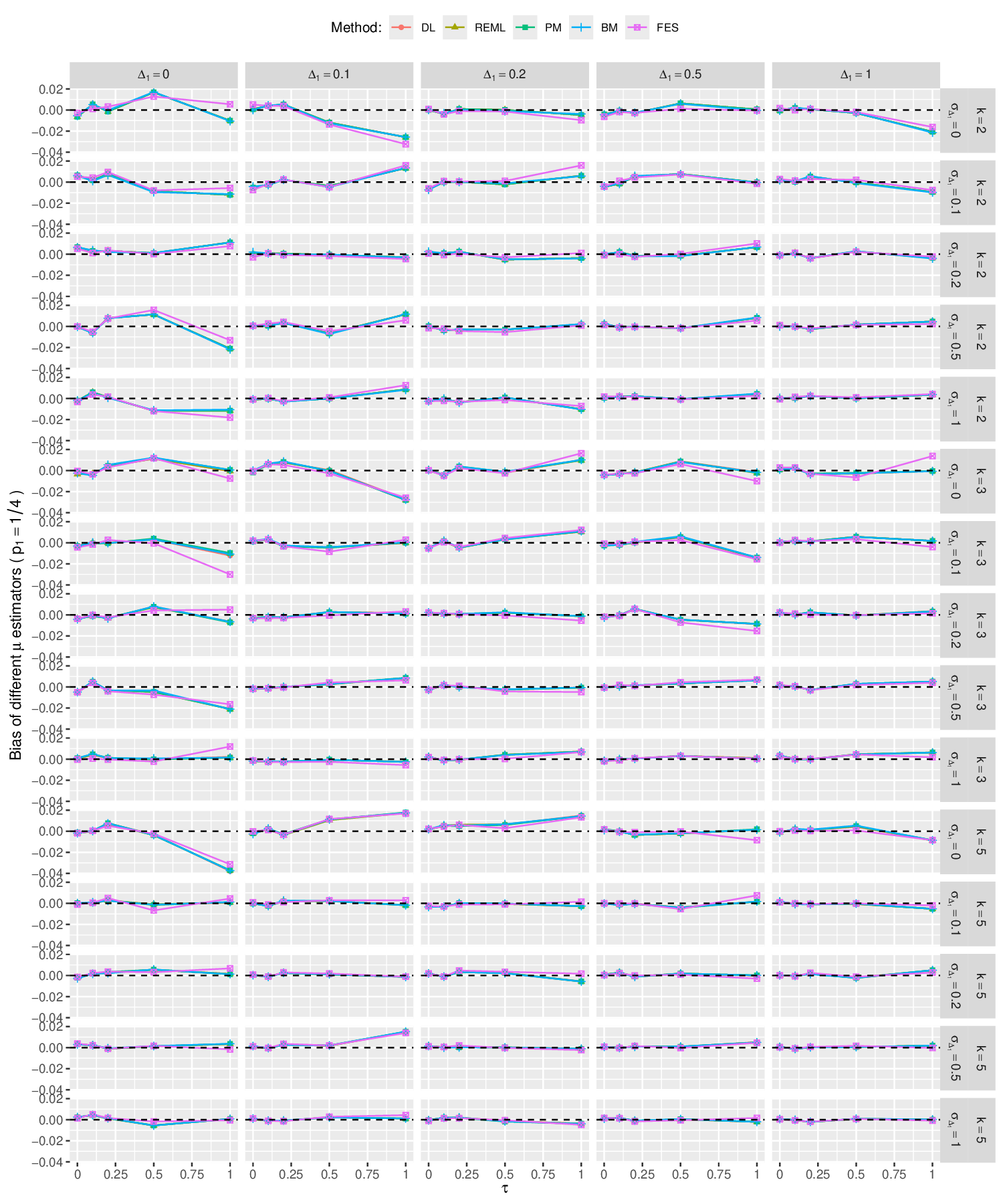}
\caption{\it{
Bias in estimating the overall treatment effect $\mu$ for several study-level estimators and the newly proposed
estimators based on subgroup-level data ($p_1$=1/4)}}
\label{fig s4}
\end{figure}

\begin{figure}[t]
\centering\includegraphics[width=0.95\textwidth]{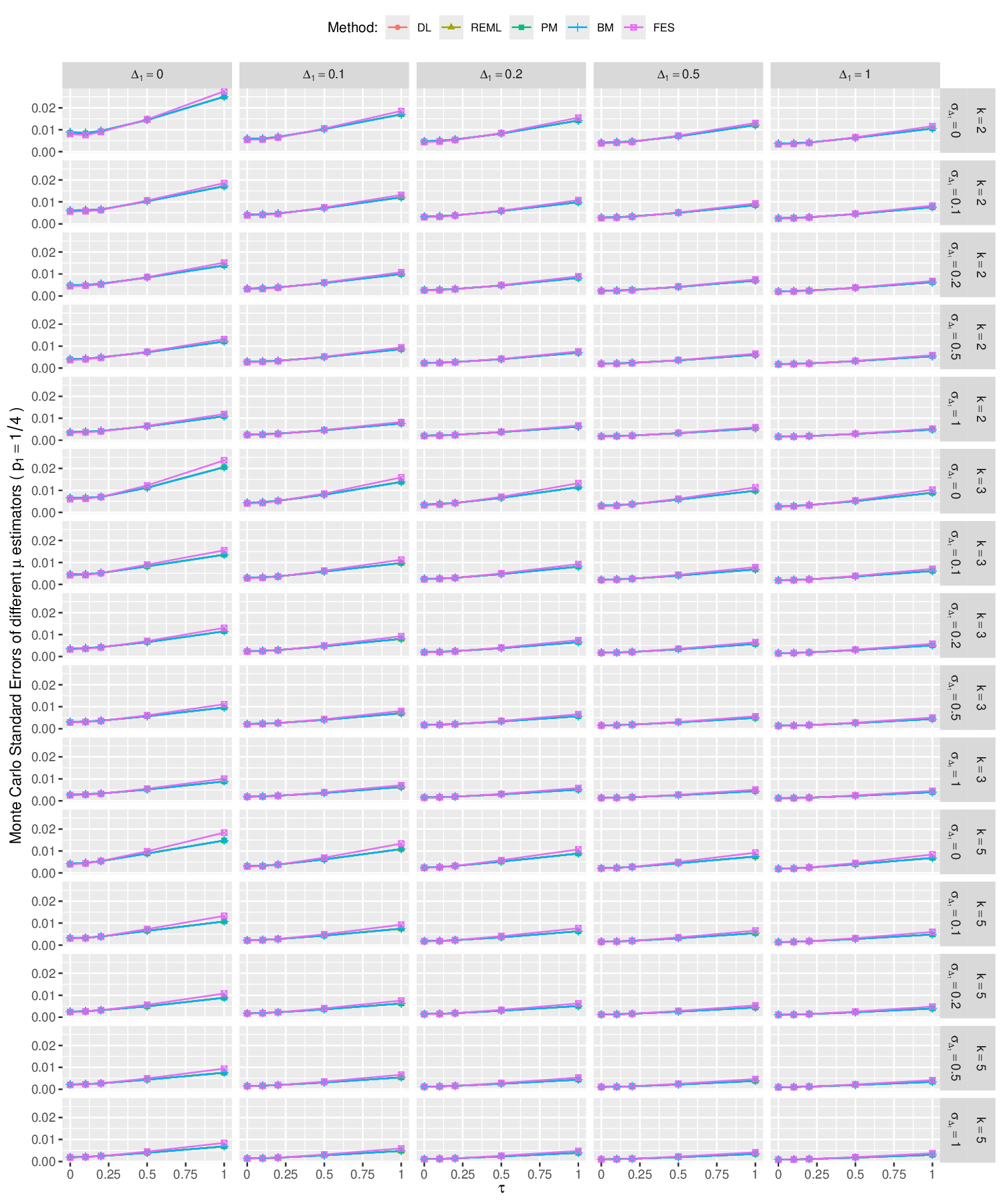}
\caption{\it{
Monte Carlo Standard Errors for different $\mu$ estimators using study-level data and the newly proposed estimators based on subgroup-level data ($p_1$=1/4)}}
\label{fig s5}
\end{figure}

\begin{figure}[t]
\centering\includegraphics[width=0.95\textwidth,height=0.95\textheight]{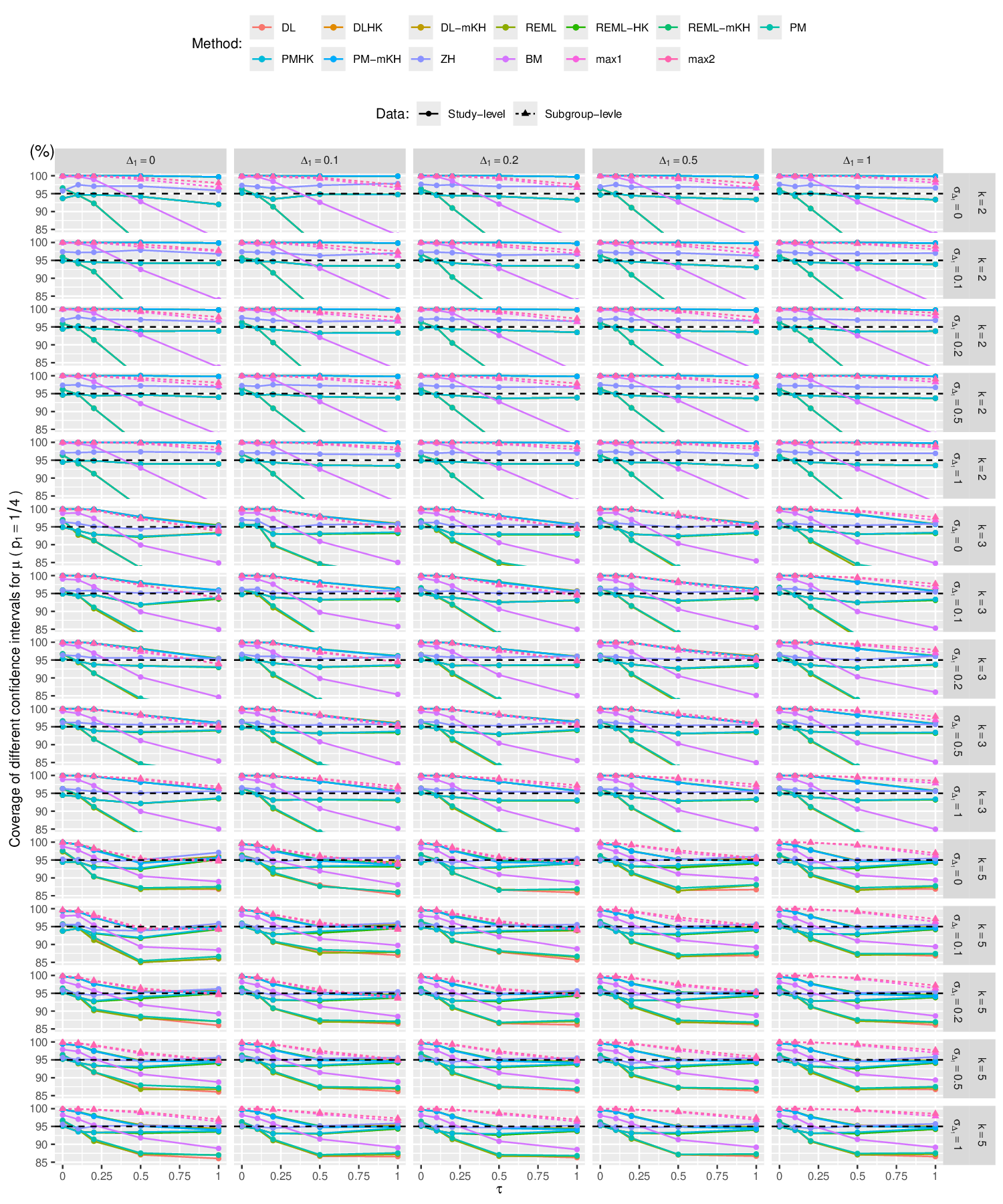}
\caption{\it{
Coverage percentages for the overall effect $\mu$ with 95\% confidence intervals based on study-level data or based on subgroup-level data ($p_1$=1/4)}}
\label{fig s6}
\end{figure}

\begin{figure}[t]
\centering\includegraphics[width=0.95\textwidth]{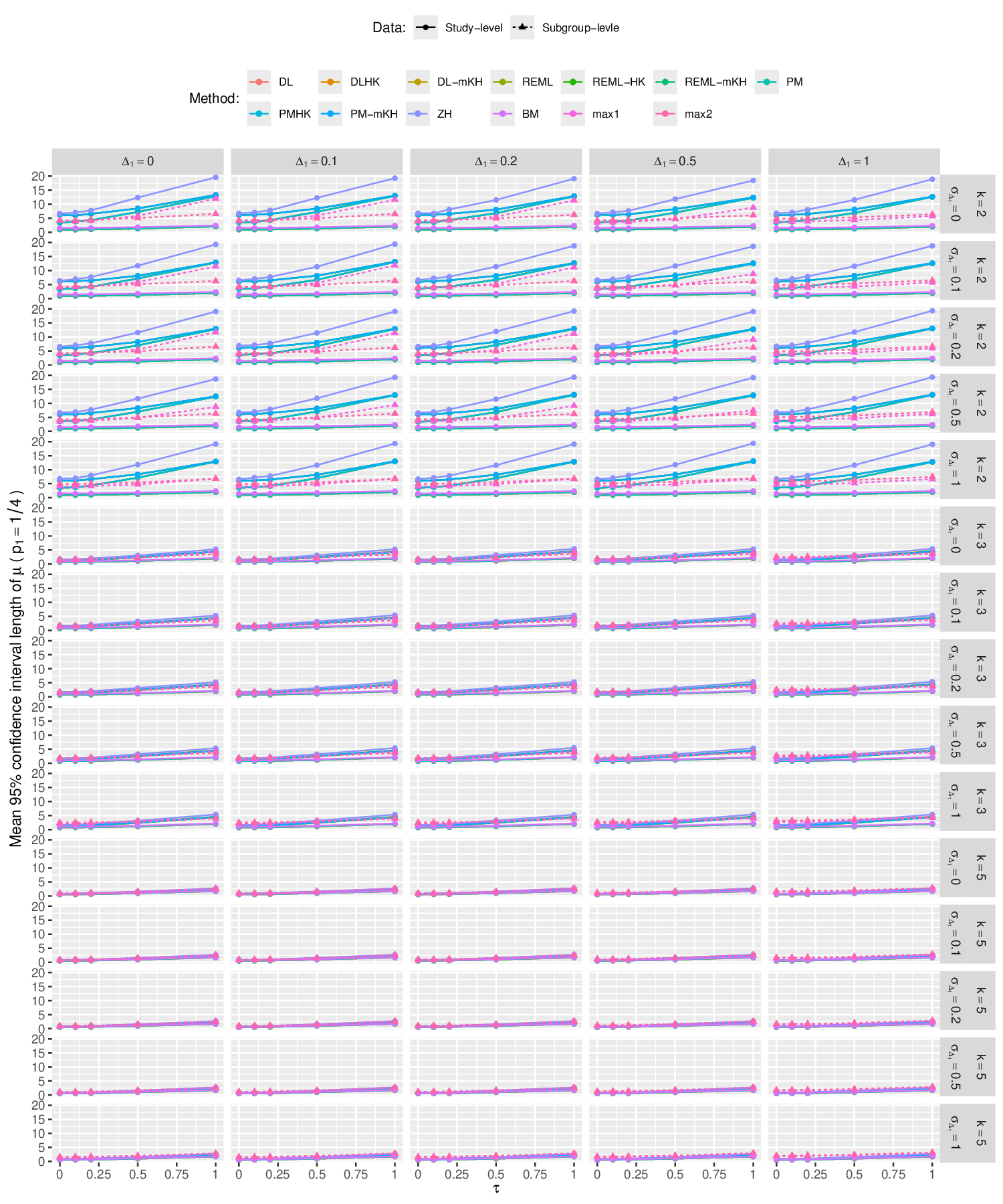}
\caption{\it{
Mean lengths of confidence intervals for the overall mean effect $\mu$  using different methods ($p_1$=1/4)}}
\label{fig s7}
\end{figure}

\begin{figure}[t]
\centering\includegraphics[width=0.95\textwidth]{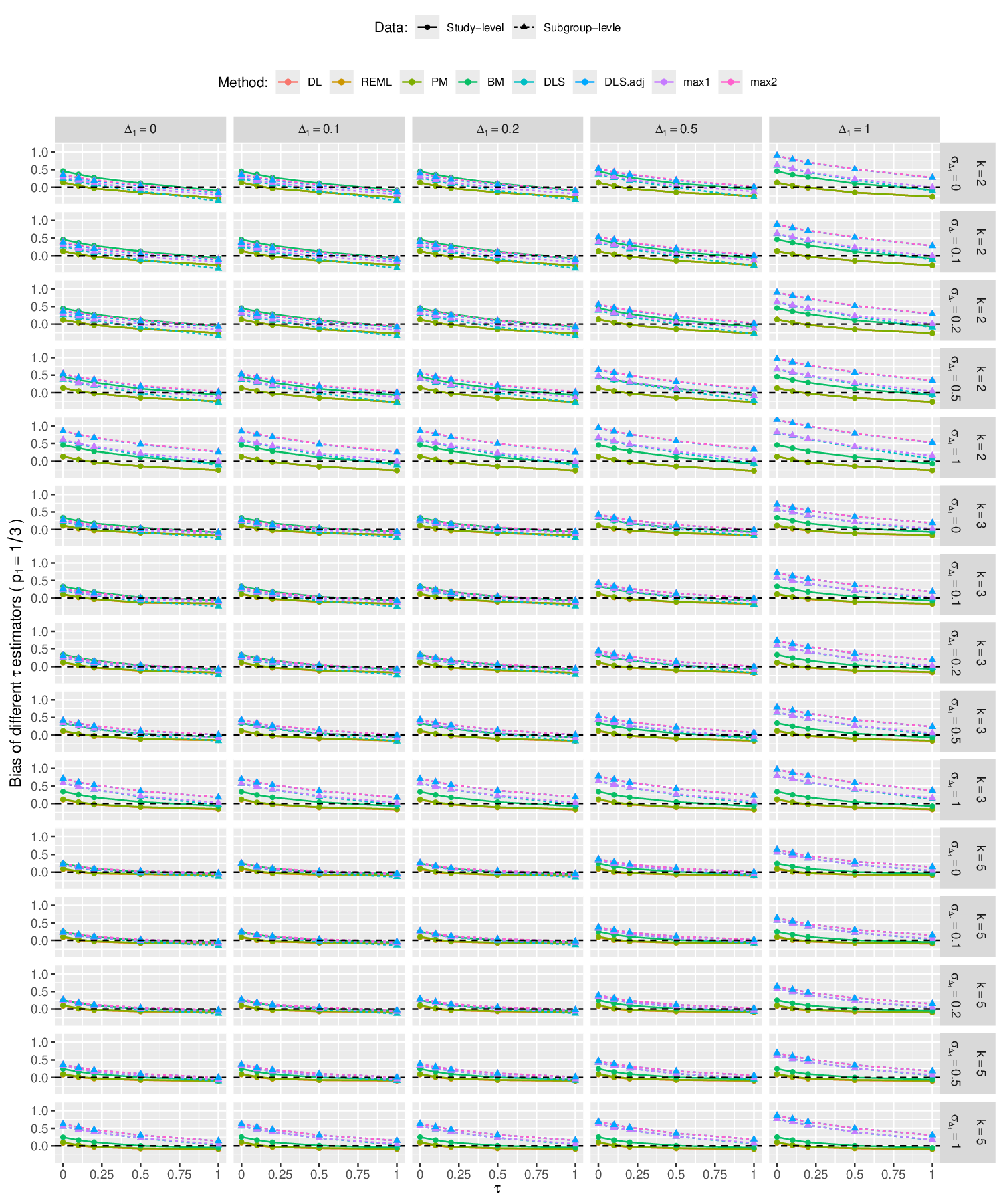}
\caption{\it{Bias in estimating the between-study heterogeneity $\tau$ for several study-level estimators as well as the newly proposed
estimators based on subgroup-level data ($p_1$=1/3)}}
\label{fig s8}
\end{figure}

\begin{figure}[t]
\centering\includegraphics[width=0.95\textwidth]{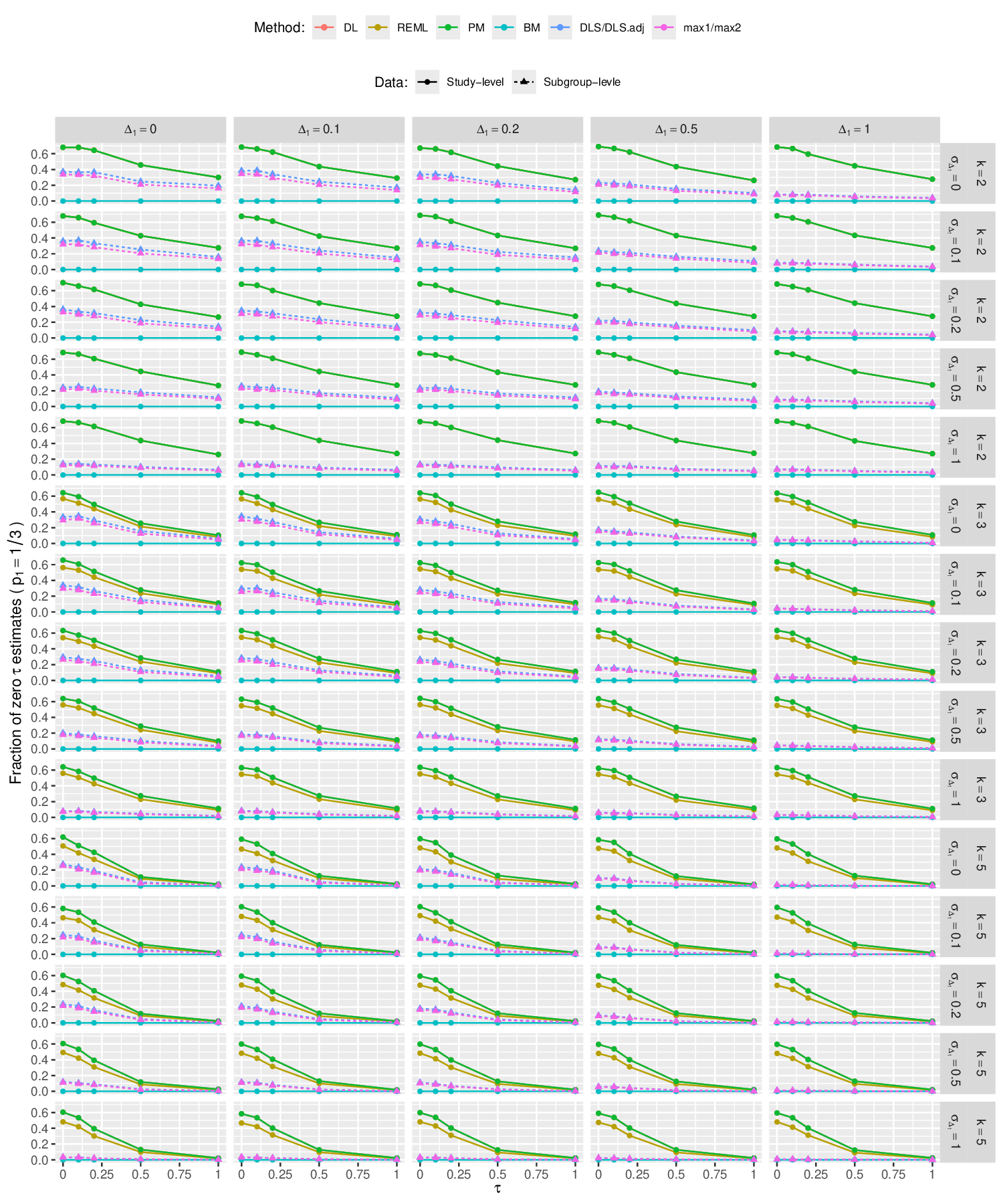}
\caption{\it{
Fraction of zero estimates of the between-study heterogeneity $\tau$ for the classic estimator using study-level data and the newly proposed estimators using subgroup-level data ($p_1$=1/3)}}
\label{fig s9}
\end{figure}

\begin{figure}[t]
\centering\includegraphics[width=0.95\textwidth]{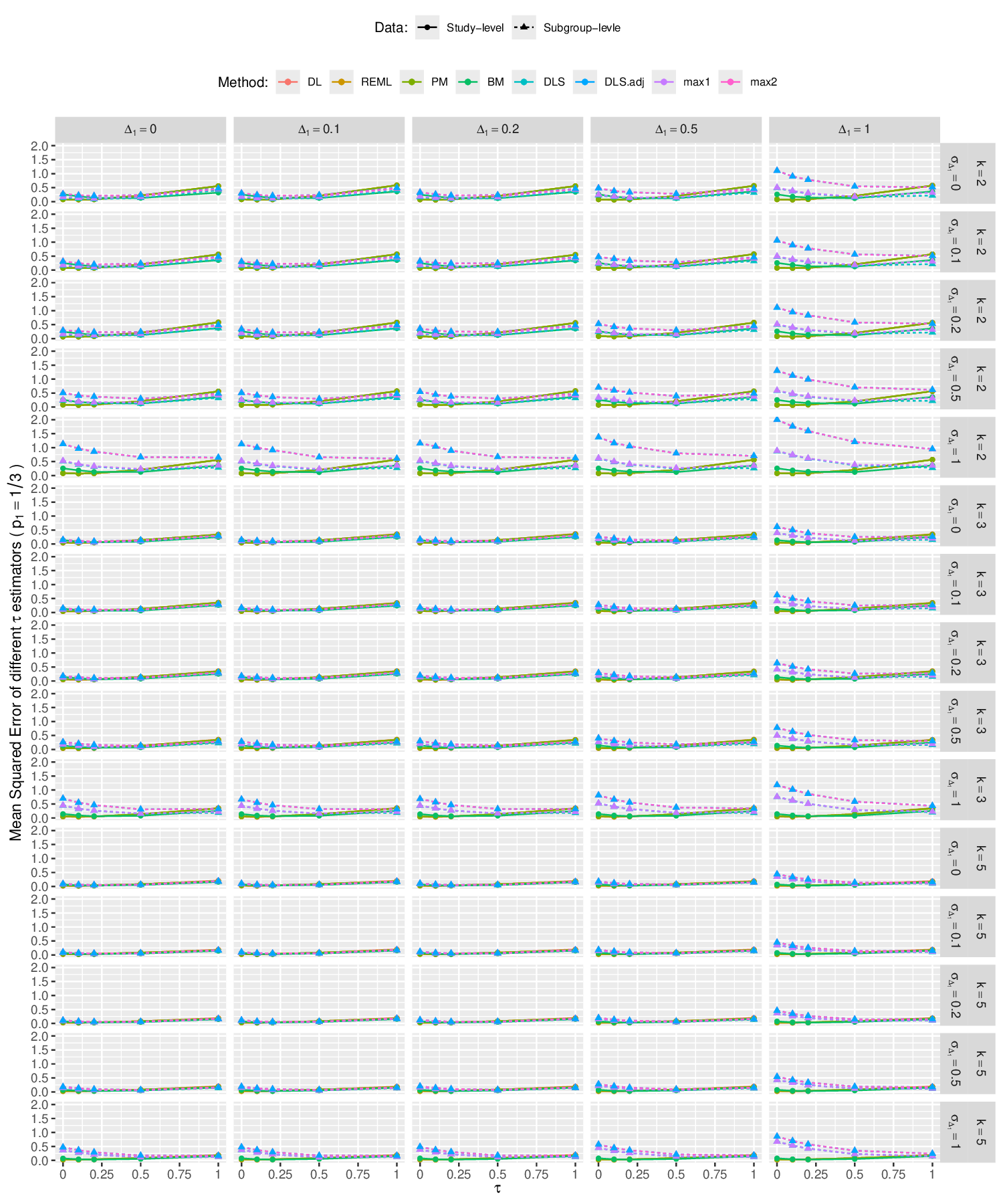}
\caption{\it{
Mean squared errors of the between-study heterogeneity $\tau$ for the classic estimators using study-level data and the newly proposed estimators using subgroup-level data ($p_1$=1/3)}}
\label{fig s10}
\end{figure}

\begin{figure}[t]
\centering\includegraphics[width=0.95\textwidth]{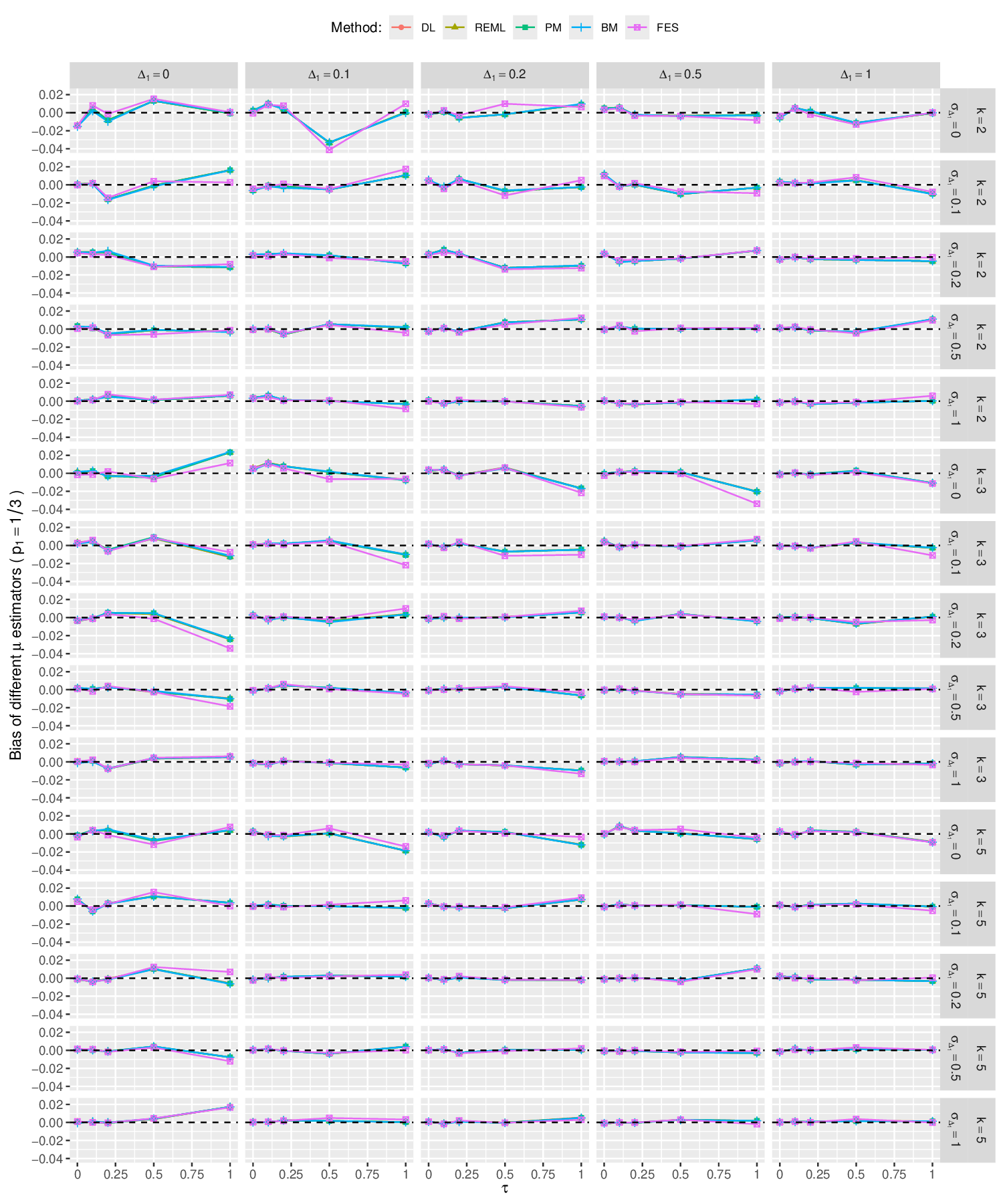}
\caption{\it{
Bias in estimating the overall treatment effect $\mu$ for several study-level estimators and the newly proposed
estimators based on subgroup-level data ($p_1$=1/3}}
\label{fig s11}
\end{figure}

\begin{figure}[t]
\centering\includegraphics[width=0.95\textwidth]{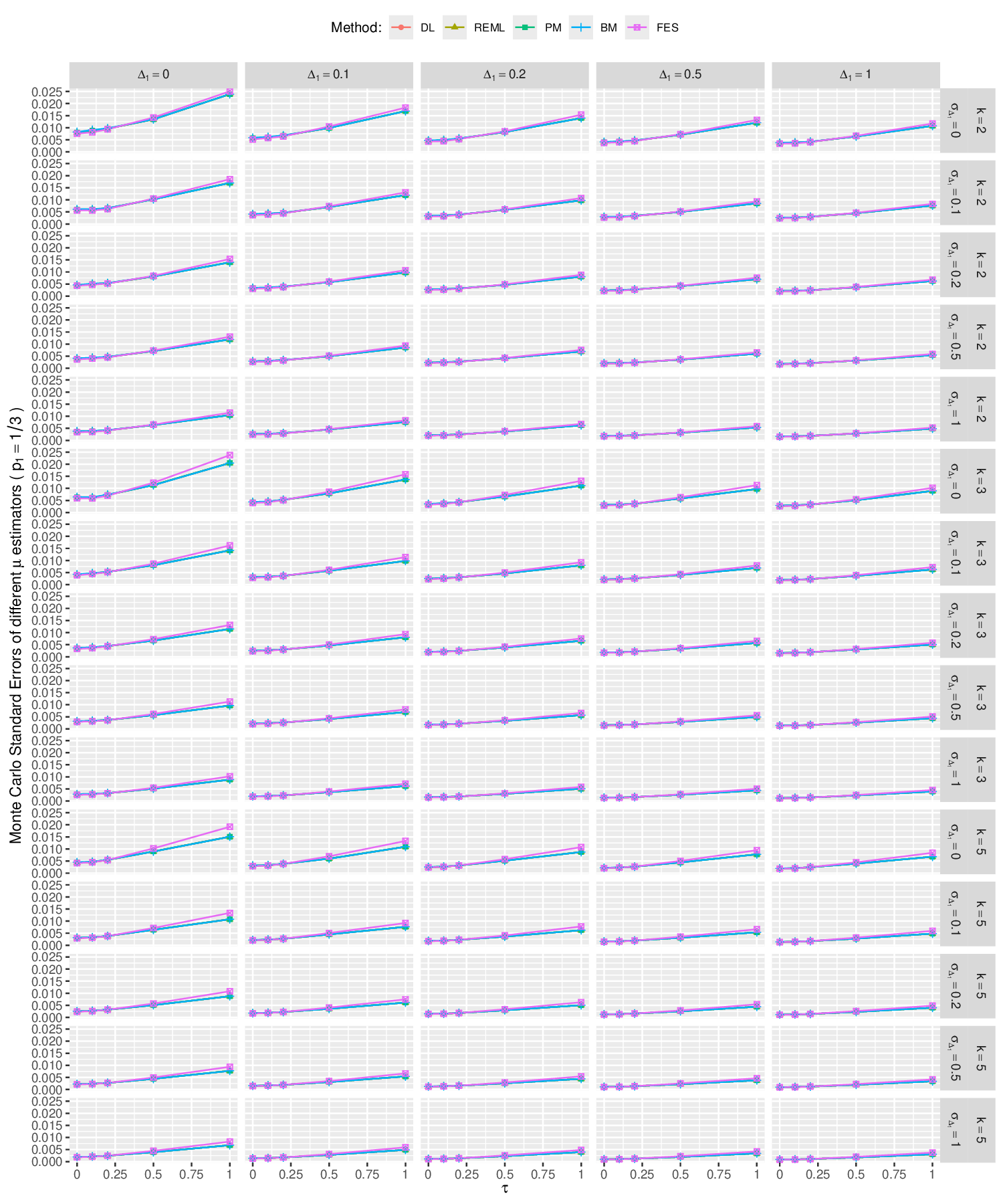}
\caption{\it{
Monte Carlo Standard Errors for different $\mu$ estimators using study-level data and the newly proposed estimators based on subgroup-level data ($p_1$=1/3)}}
\label{fig s12}
\end{figure}

\begin{figure}[t]
\centering\includegraphics[width=0.95\textwidth,height=0.95\textheight]{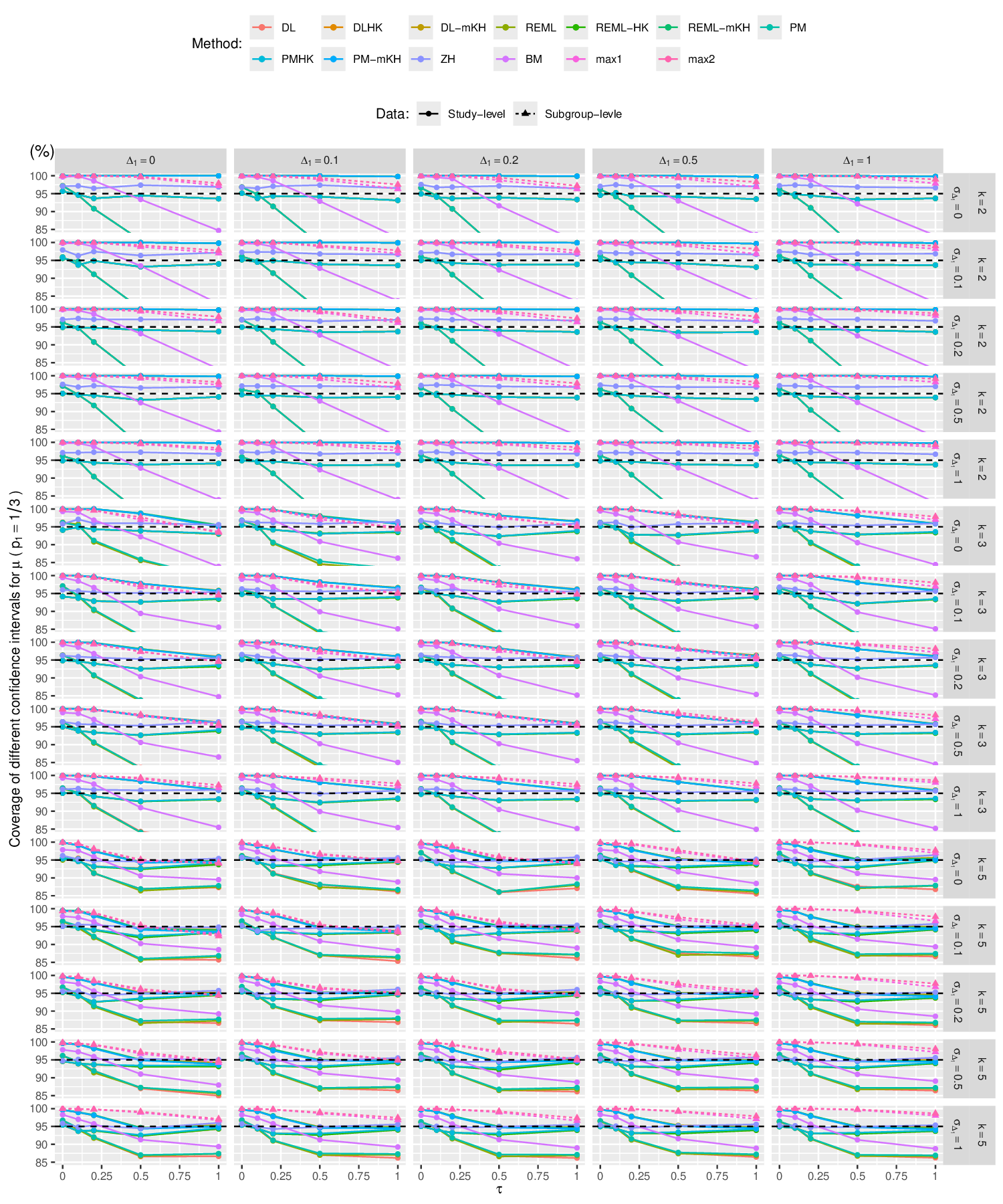}
\caption{\it{
Coverage percentages for the overall effect $\mu$ with 95\% confidence intervals based on study-level data or based on subgroup-level data ($p_i$=1/3)}}
\label{fig s13}
\end{figure}

\begin{figure}[t]
\centering\includegraphics[width=0.95\textwidth]{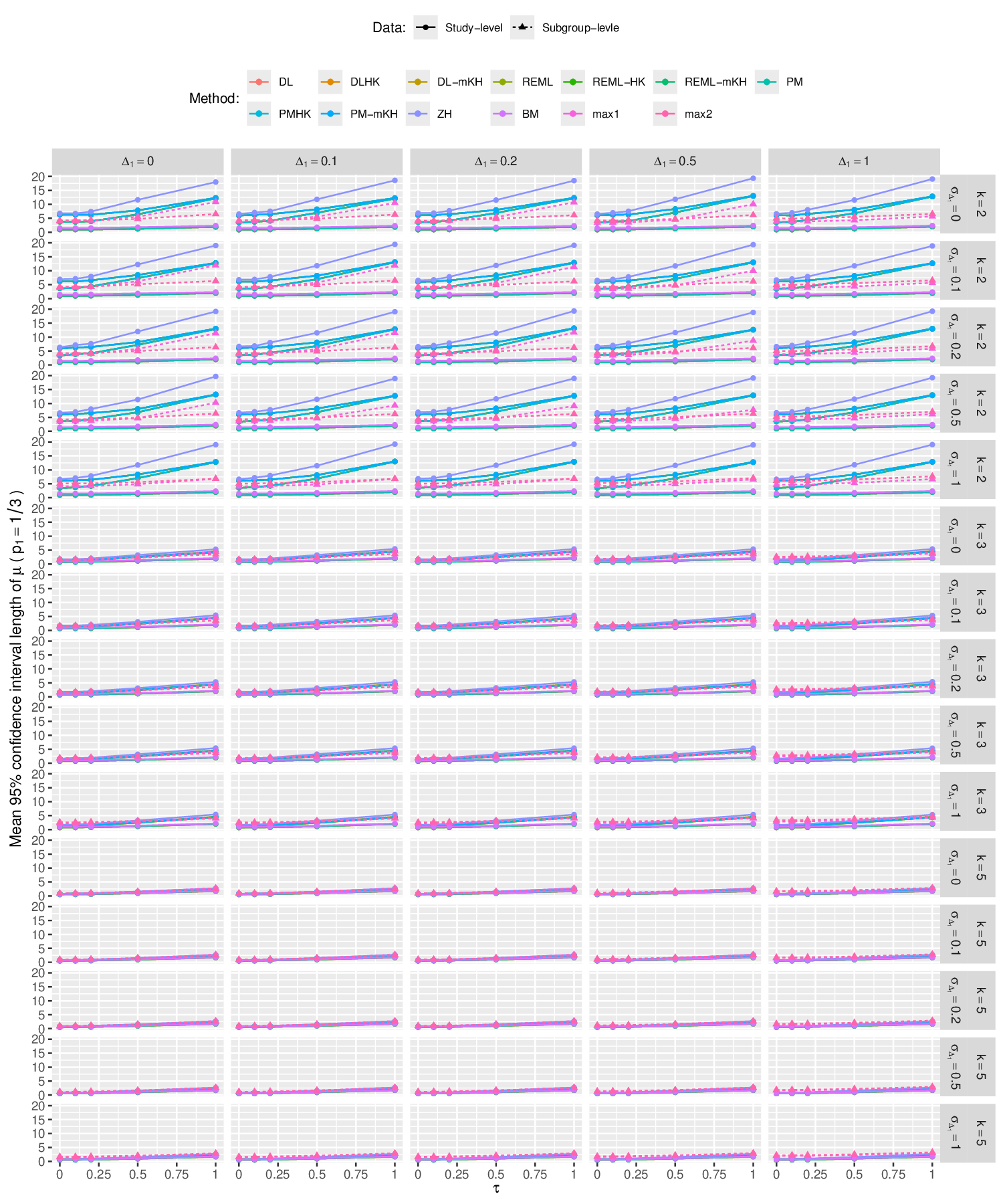}
\caption{\it{
Mean lengths of confidence intervals for the overall mean effect $\mu$ using different methods ($p_1$=1/3)}}
\label{fig s14}
\end{figure}

\begin{figure}[t]
\centering\includegraphics[width=0.95\textwidth]{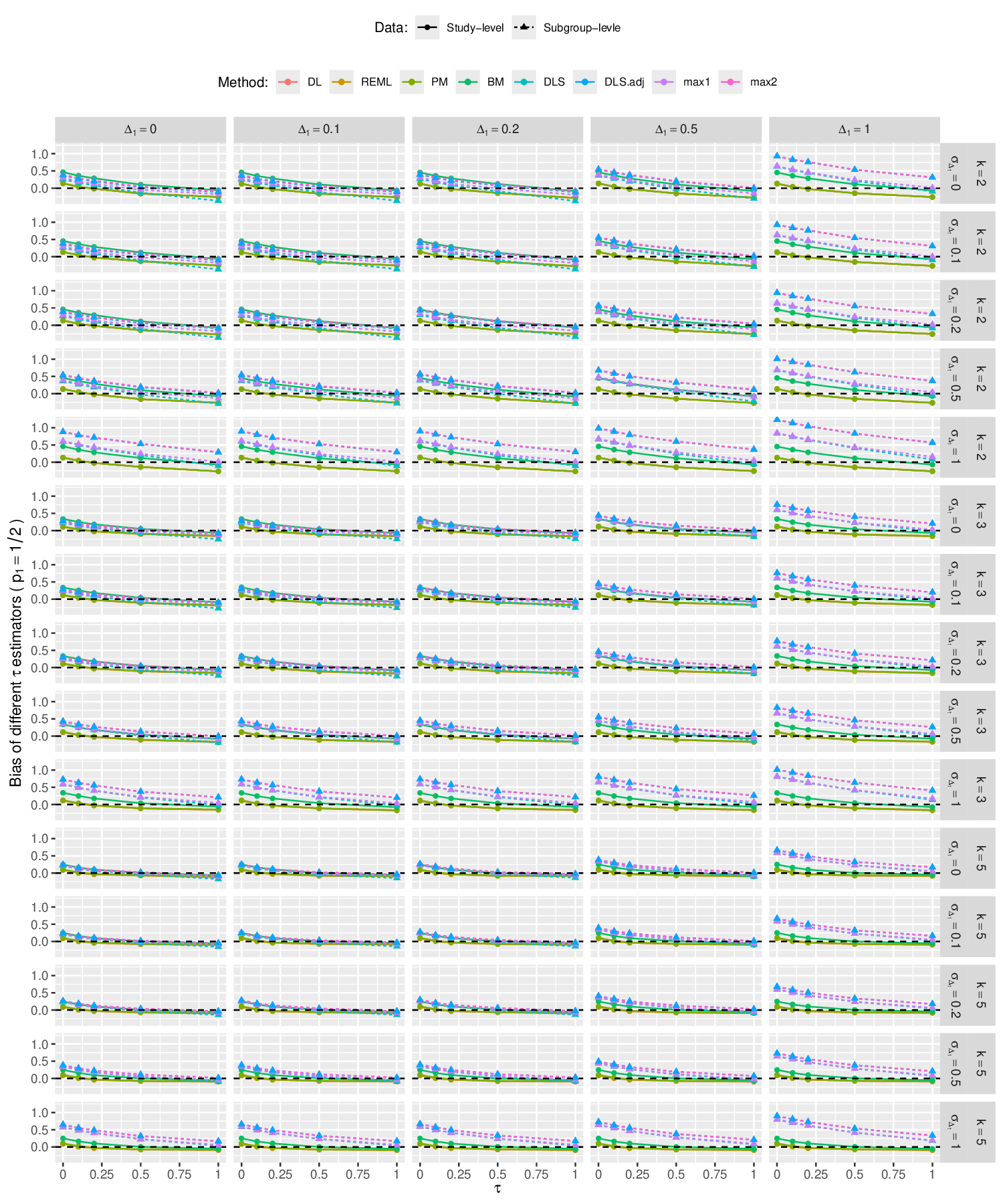}
\caption{\it{Bias in estimating the between-study heterogeneity $\tau$ for several study-level estimators as well as the newly proposed
estimators based on subgroup-level data ($p_1$=1/2)}}
\label{fig s15}
\end{figure}

\begin{figure}[t]
\centering\includegraphics[width=0.95\textwidth]{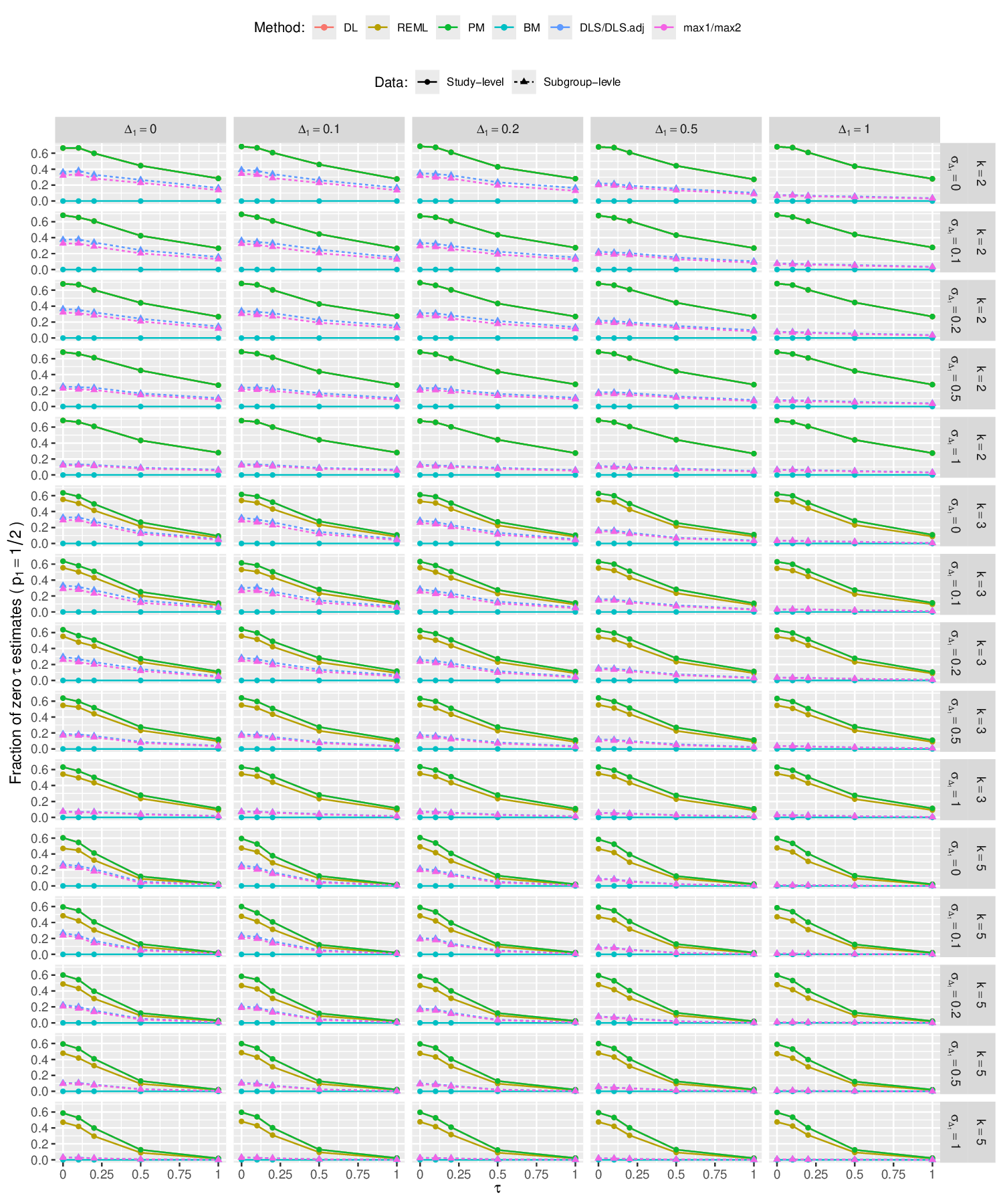}
\caption{\it{
Fraction of zero estimates of the between-study heterogeneity $\tau$ for the classic estimator using study-level data and the newly proposed estimators using subgroup-level data ($p_1$=1/2)}}
\label{fig s16}
\end{figure}

\begin{figure}[t]
\centering\includegraphics[width=0.95\textwidth]{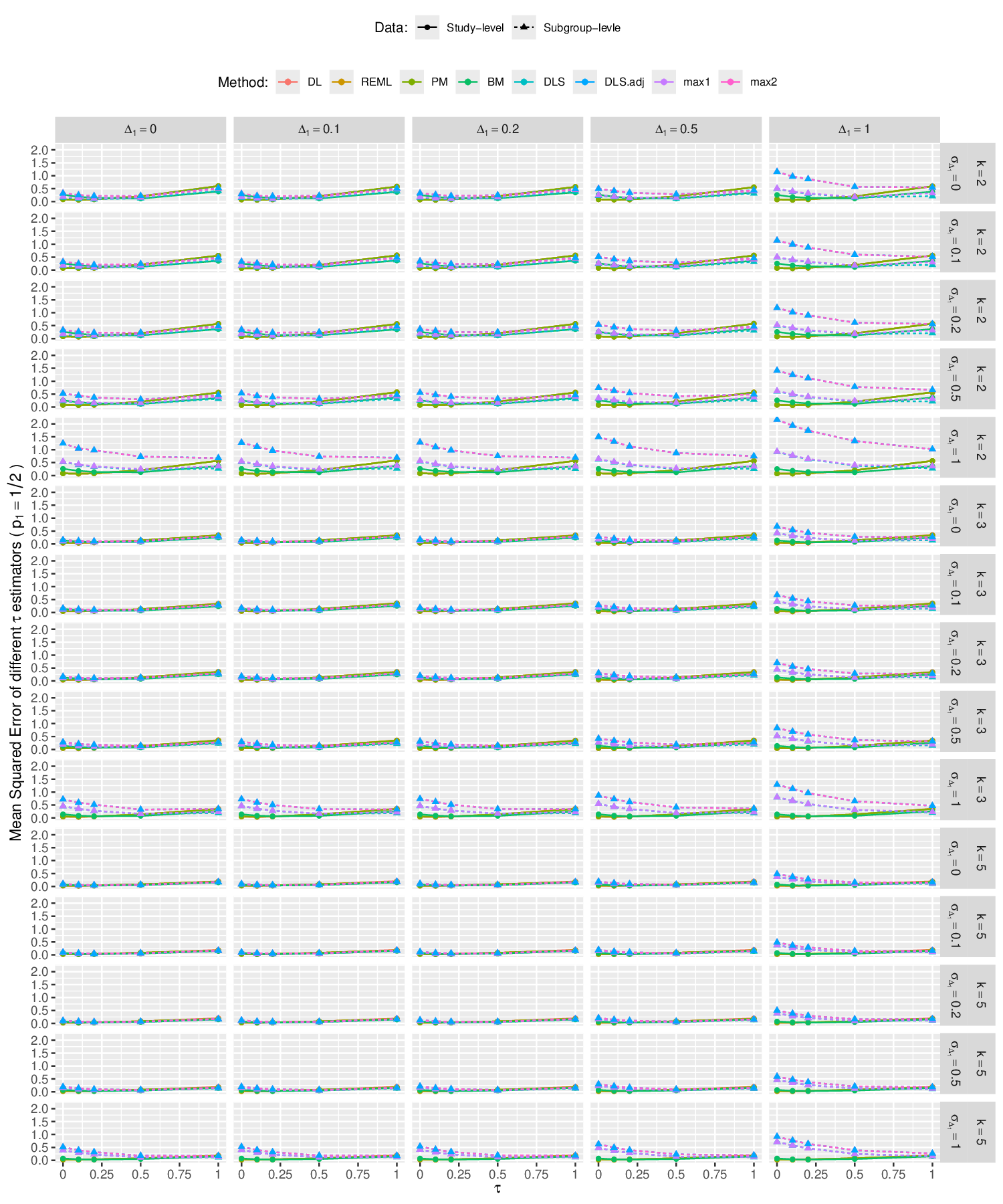}
\caption{\it{
Mean squared errors of the between-study heterogeneity $\tau$ for the classic estimators using study-level data and the newly proposed estimators using subgroup-level data ($p_1$=1/2)}}
\label{fig s17}
\end{figure}

\begin{figure}[t]
\centering\includegraphics[width=0.95\textwidth]{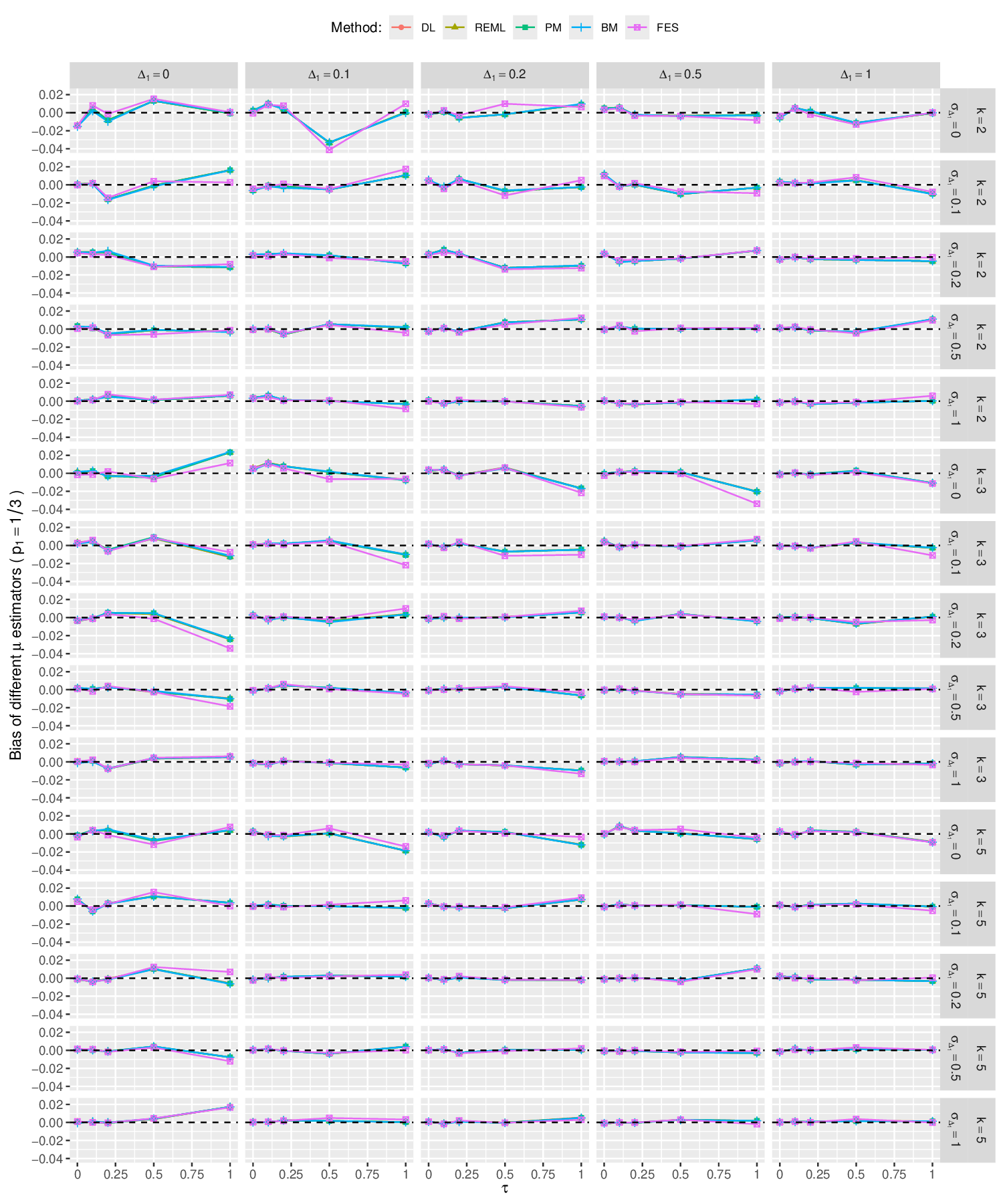}
\caption{\it{
Bias in estimating the overall treatment effect $\mu$ for several study-level estimators and the newly proposed
estimators based on subgroup-level data ($p_1$=1/2}}
\label{fig s18}
\end{figure}

\begin{figure}[t]
\centering\includegraphics[width=0.95\textwidth]{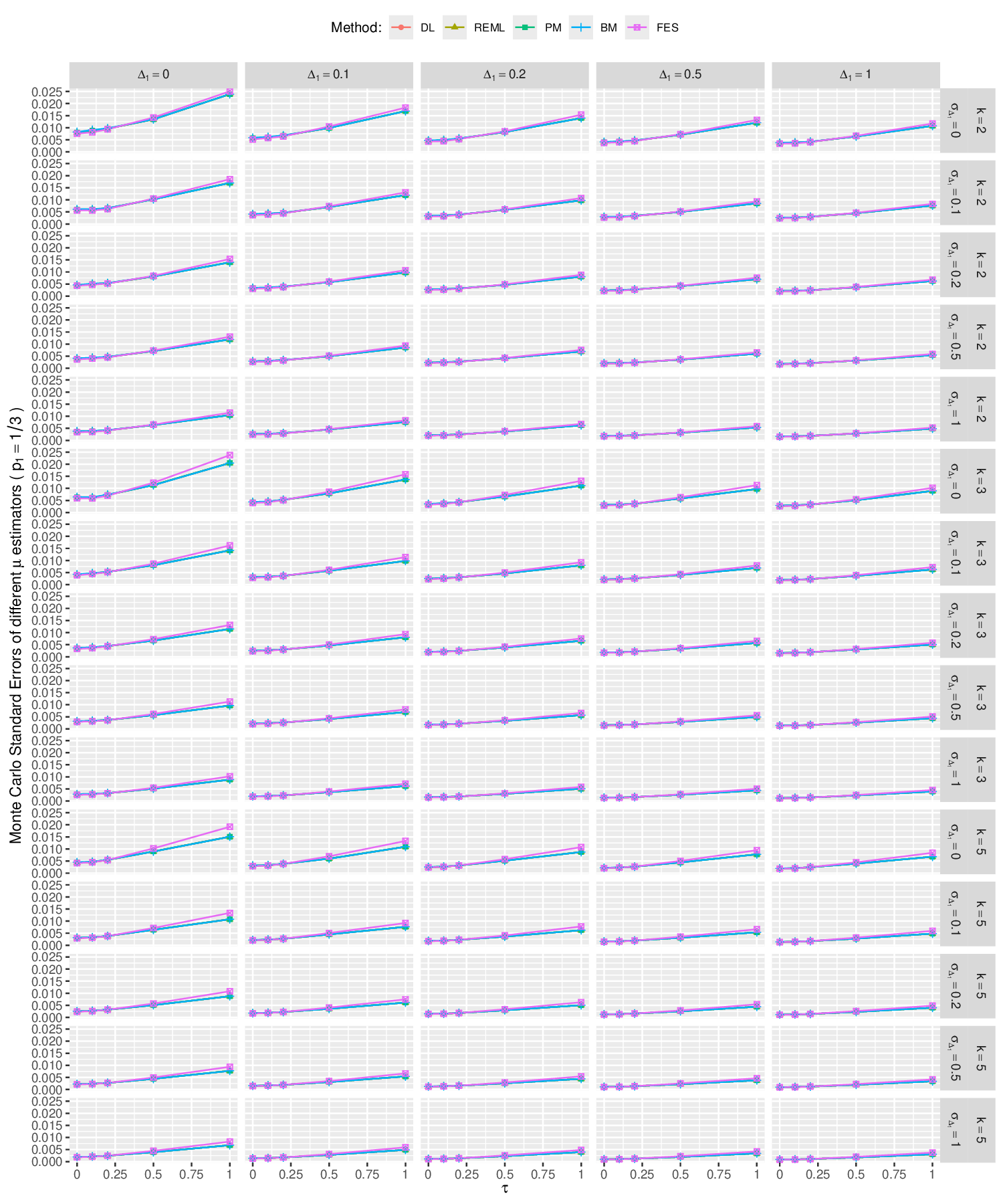}
\caption{\it{
Monte Carlo Standard Errors for different $\mu$ estimators using study-level data and the newly proposed estimators based on subgroup-level data ($p_1$=1/2)}}
\label{fig s19}
\end{figure}

\begin{figure}[t]
\centering\includegraphics[width=0.95\textwidth]{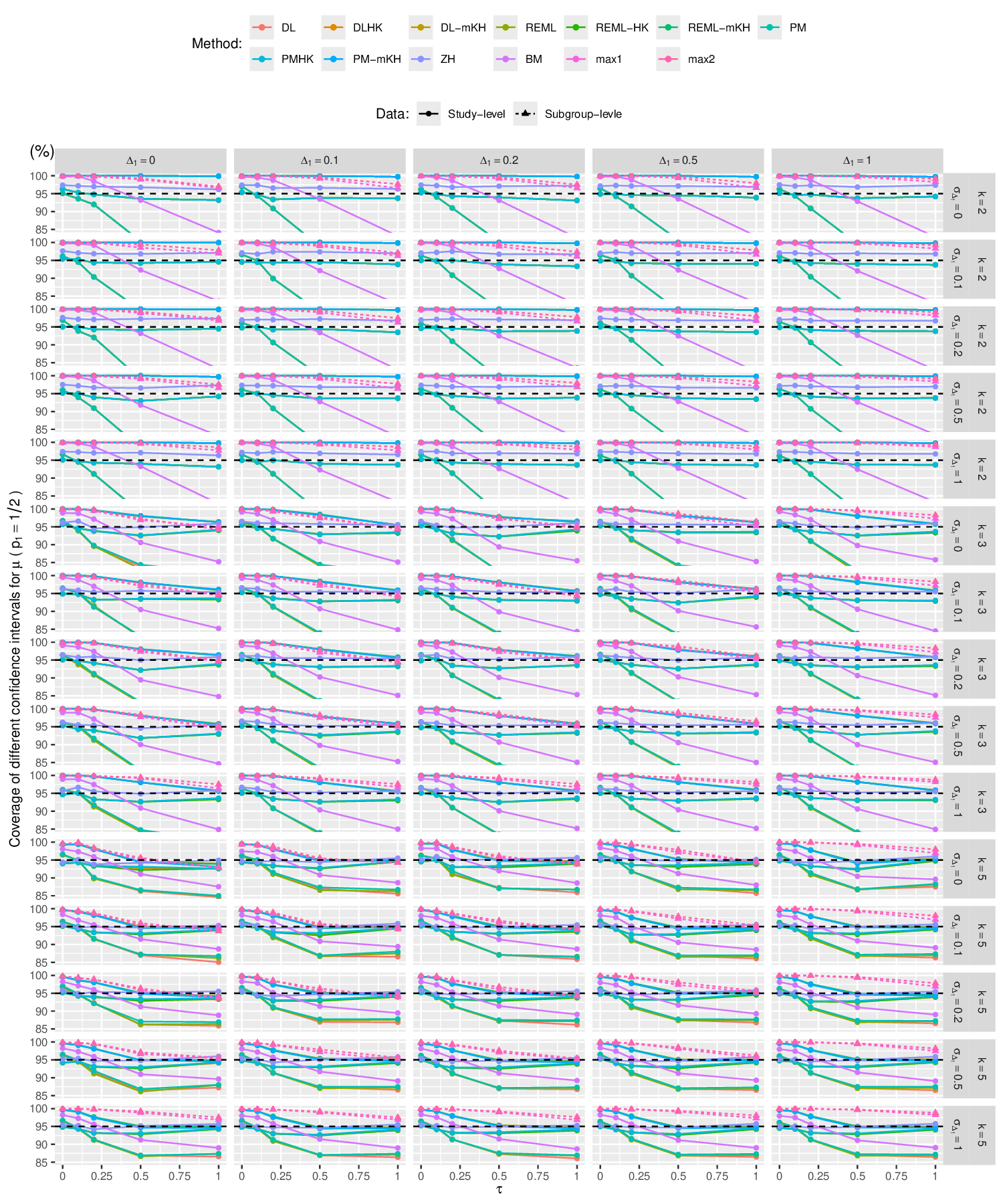}
\caption{\it{
Coverage percentages for the overall effect $\mu$ with 95\% confidence intervals based on study-level data or based on subgroup-level data ($p_1$=1/2)}}
\label{fig s20}
\end{figure}

\begin{figure}[t]
\centering\includegraphics[width=0.95\textwidth]{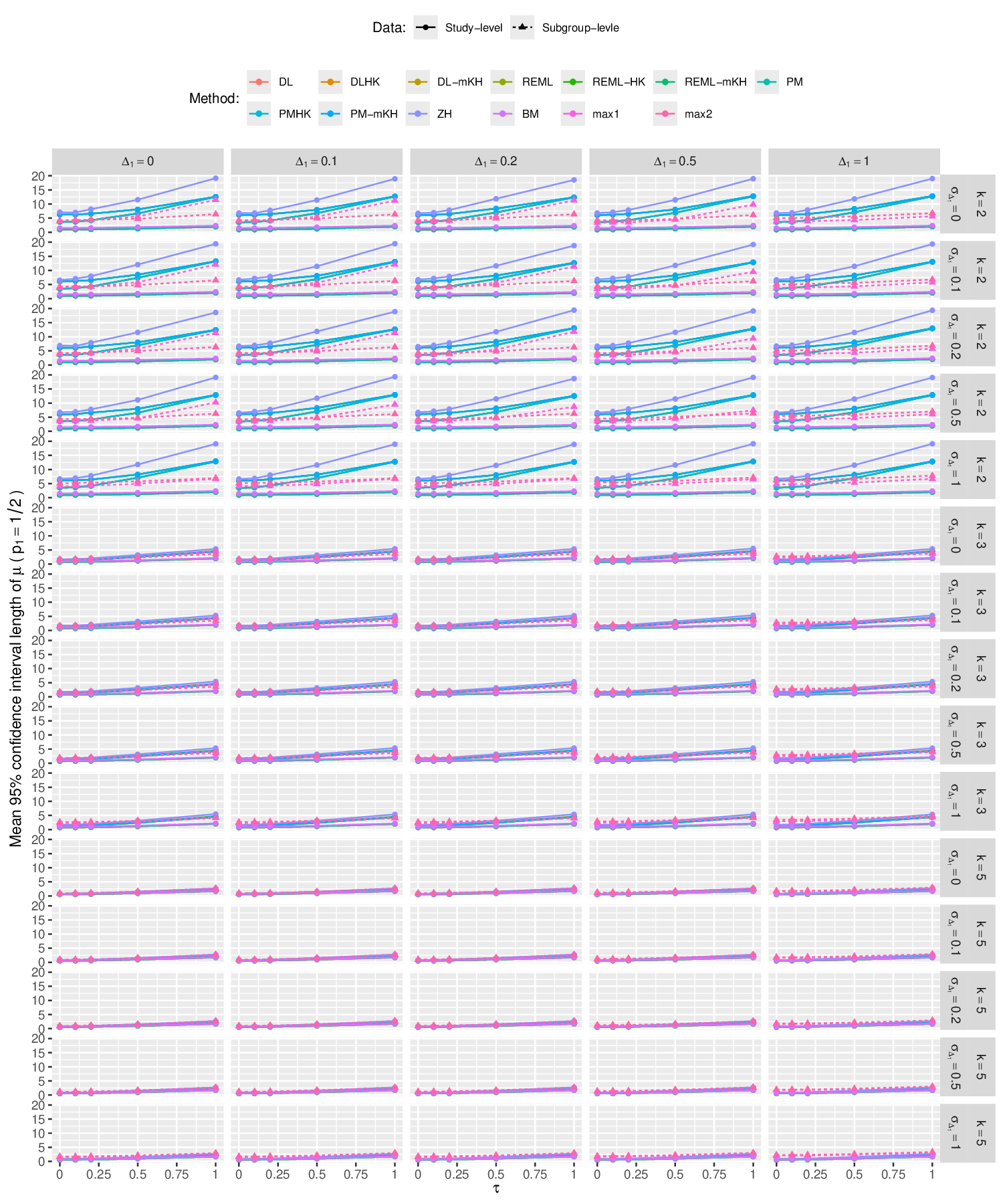}
\caption{\it{
Mean lengths of confidence intervals for the overall mean effect $\mu$ using different methods ($p_1$=1/2)}}
\label{fig s21}
\end{figure}

\begin{figure}[t]
\centering\includegraphics[width=0.95\textwidth]{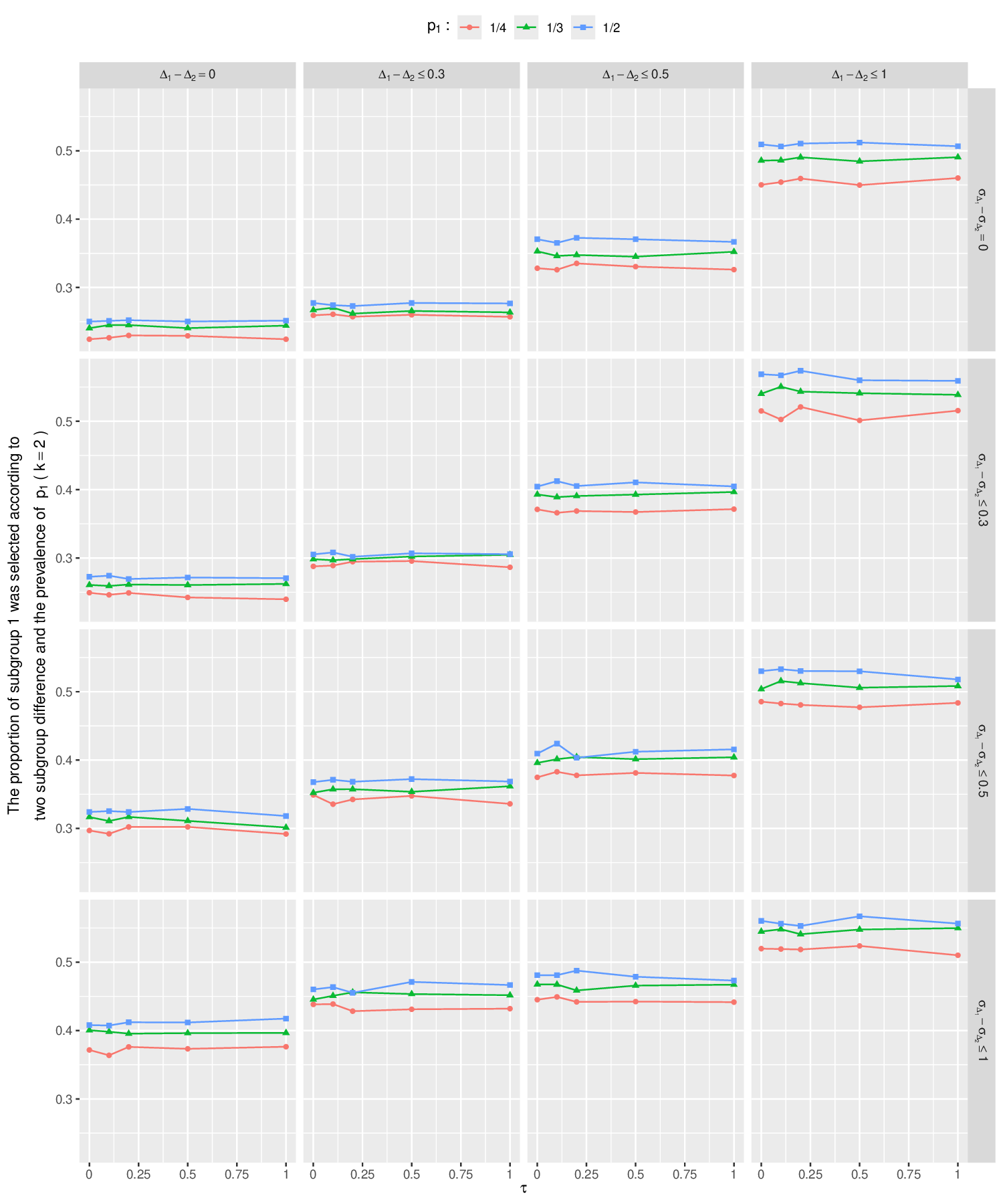}
\caption{\it{ The empirical selection rate of subgroup 1 depending on the value of subgroup effect differences and its prevalence (k=2)}}
\label{fig s22}
\end{figure}

\begin{figure}[t]
\centering\includegraphics[width=0.95\textwidth]{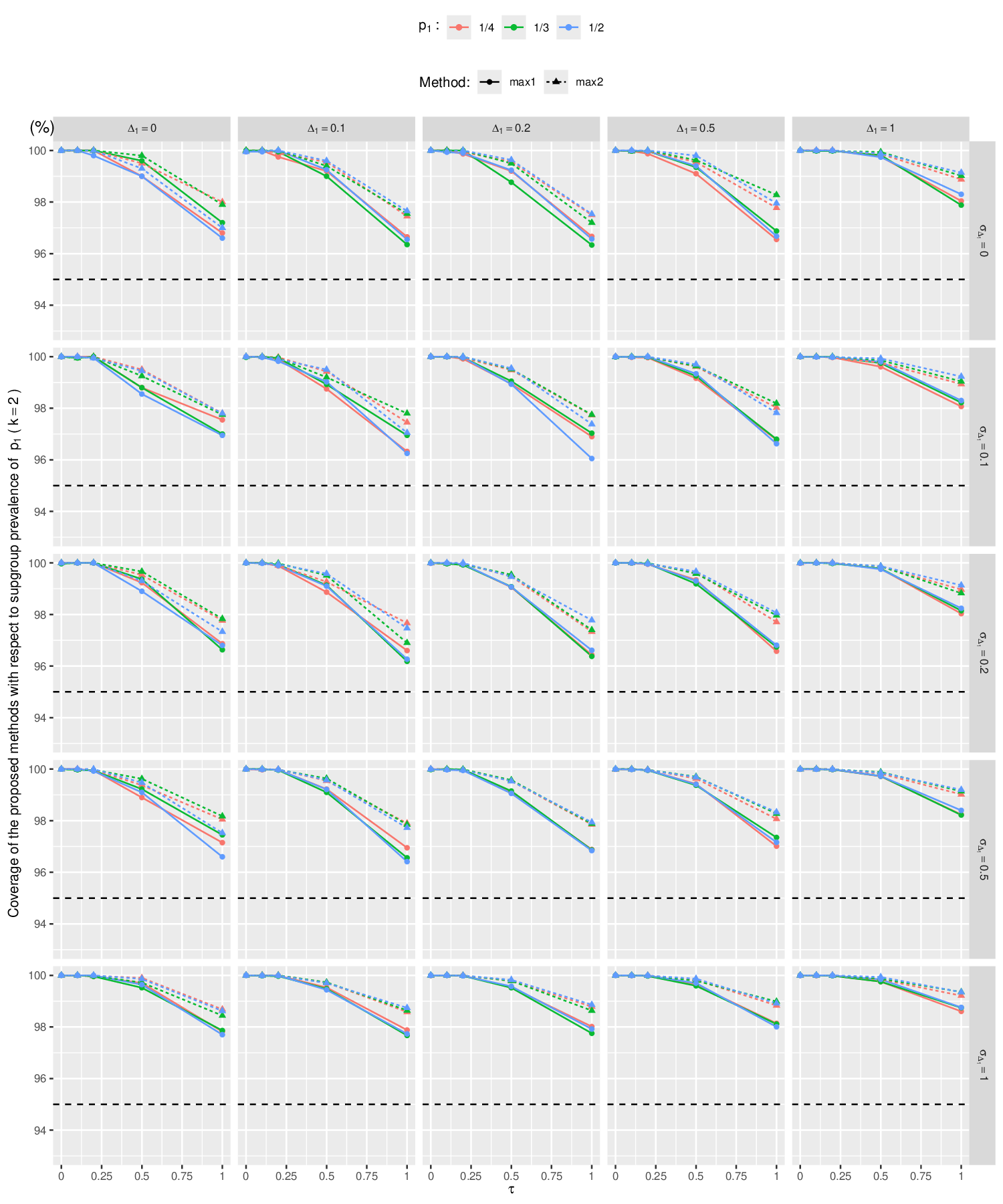}
\caption{\it{ Coverage percentages for the overall mean effect~$\mu$ with 95\%~confidence intervals based on subgroup level data depending on the subgroup prevalence~$p_i$ ($k=2$~studies).}}
\label{fig s23}
\end{figure}

\begin{figure}[t]
\centering\includegraphics[width=0.95\textwidth]{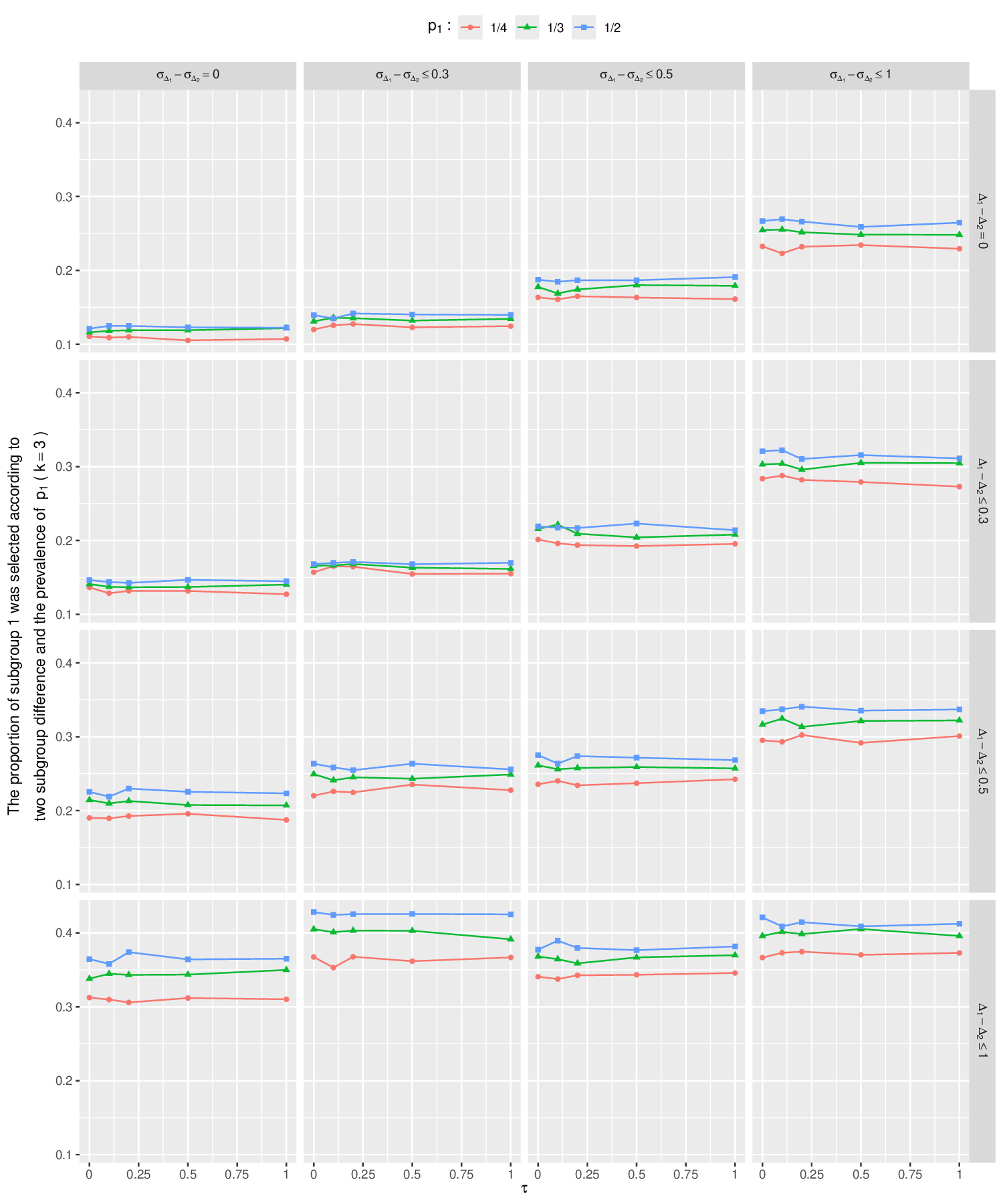}
\caption{\it{ The empirical selection rate of subgroup 1 depending on the value of subgroup effect differences and its prevalence (k=3)}}
\label{fig s24}
\end{figure}

\begin{figure}[t]
\centering\includegraphics[width=0.95\textwidth]{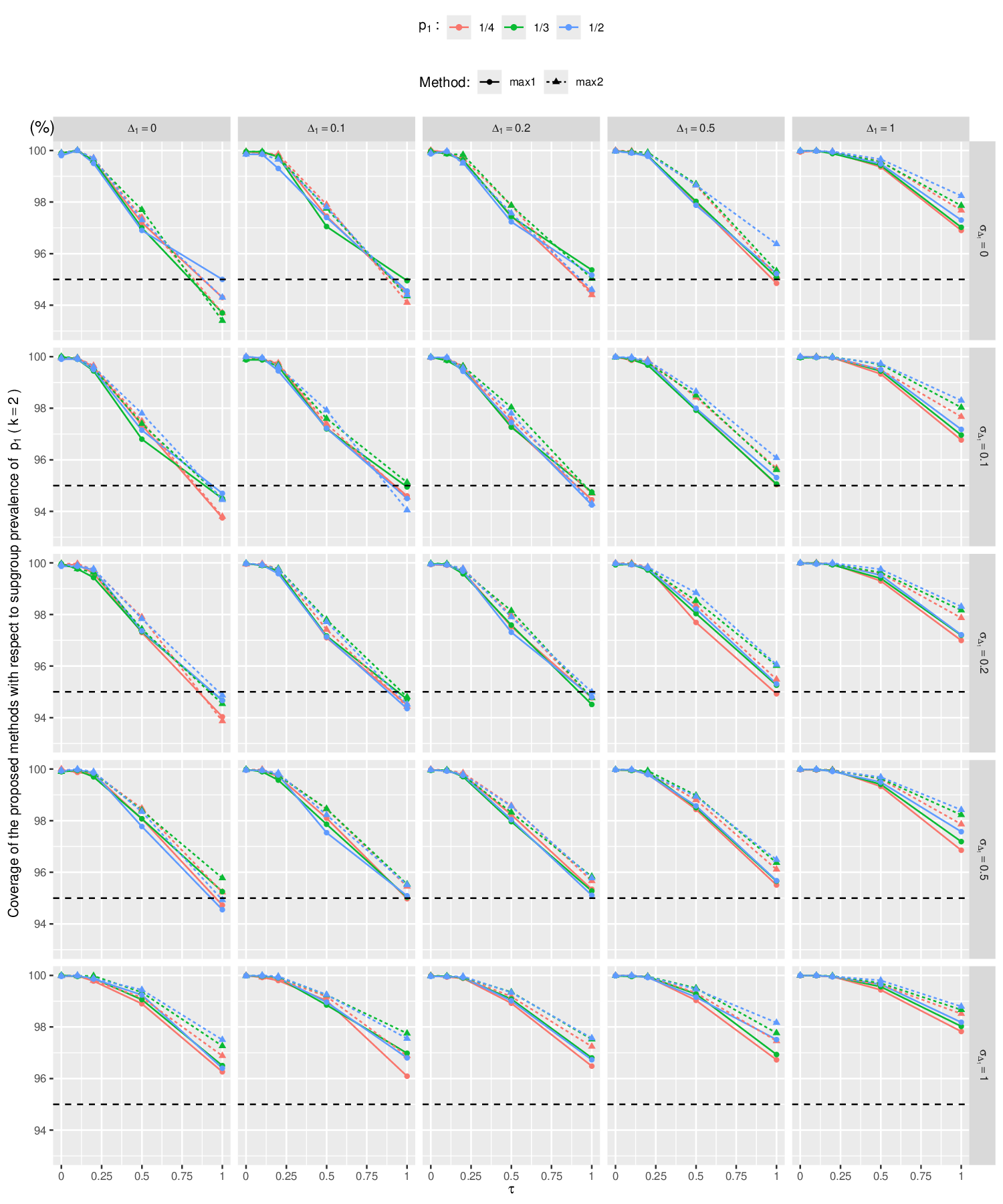}
\caption{\it{ Coverage percentages for the overall mean effect~$\mu$ with 95\%~confidence intervals based on subgroup level data depending on the subgroup prevalence~$p_i$ ($k=3$~studies).}}
\label{fig s25}
\end{figure}

\begin{figure}[t]
\centering\includegraphics[width=0.95\textwidth]{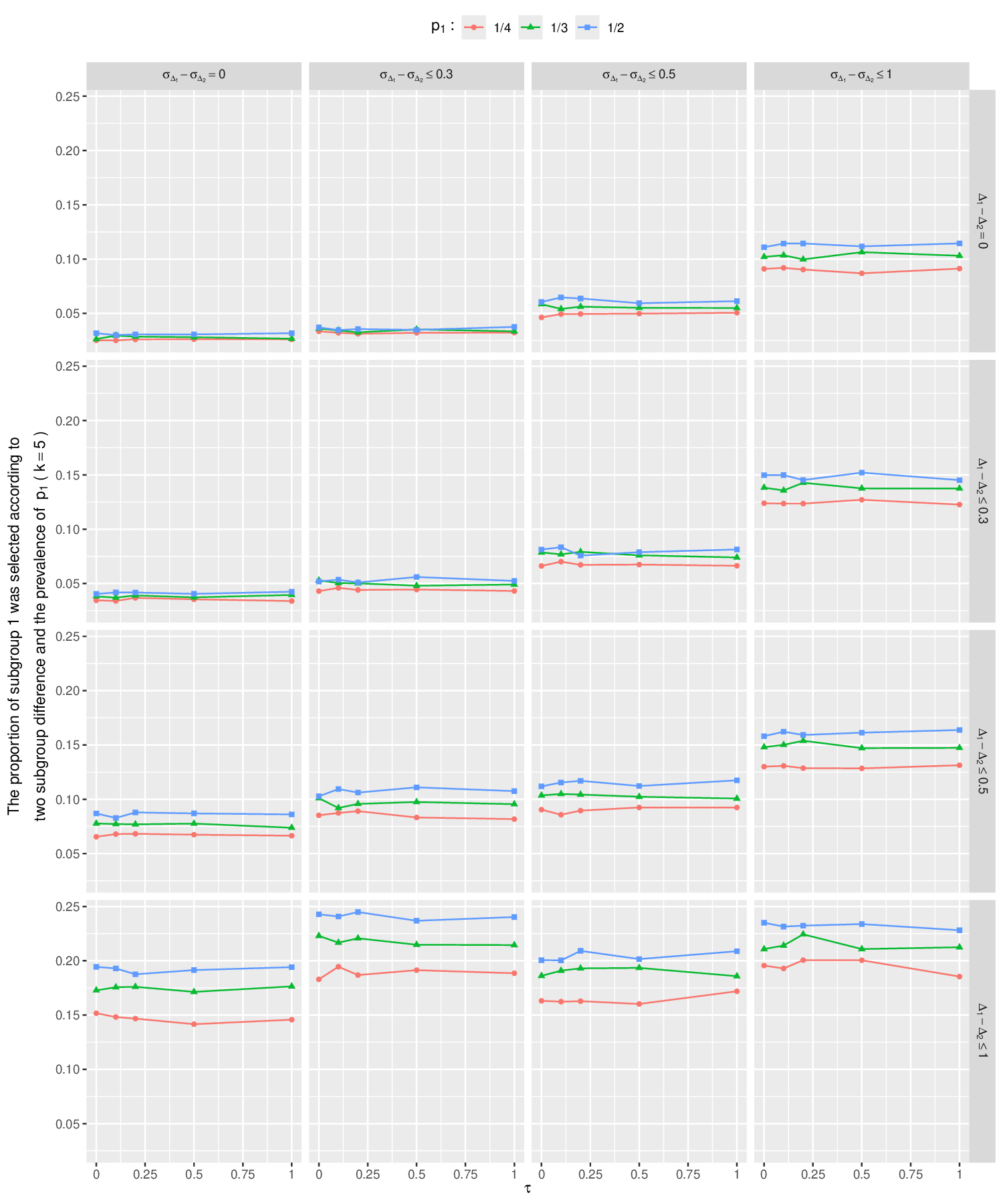}
\caption{\it{ The empirical selection rate of subgroup 1 depending on the value of subgroup effect differences and its prevalence (k=5)}}
\label{fig s26}
\end{figure}

\begin{figure}[t]
\centering\includegraphics[width=0.95\textwidth]{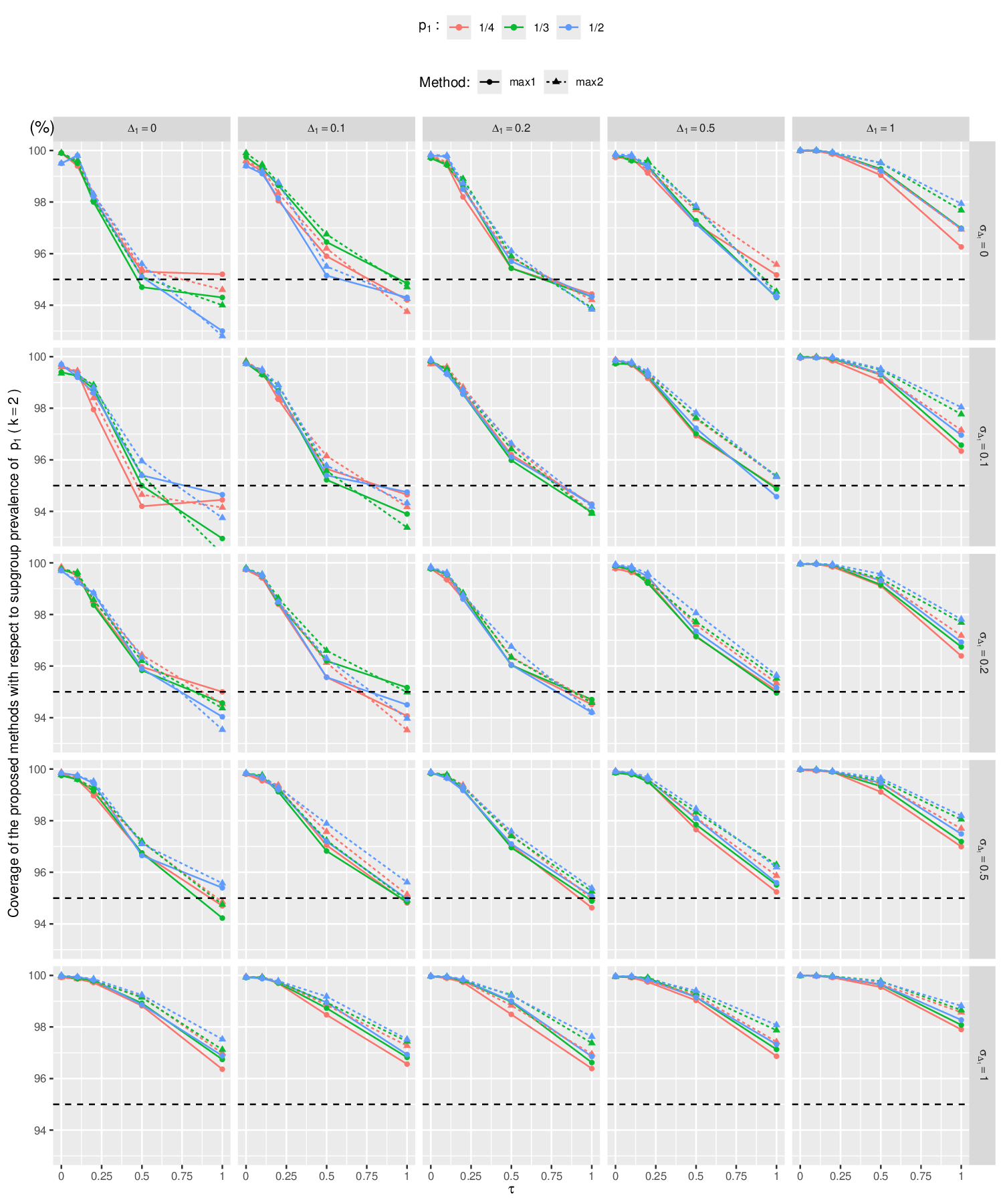}
\caption{\it{ Coverage percentages for the overall mean effect~$\mu$ with 95\%~confidence intervals based on subgroup level data depending on the subgroup prevalence~$p_i$ ($k=5$~studies).}}
\label{fig s27}
\end{figure}

\begin{figure}[t]
\centering\includegraphics[width=0.95\textwidth]{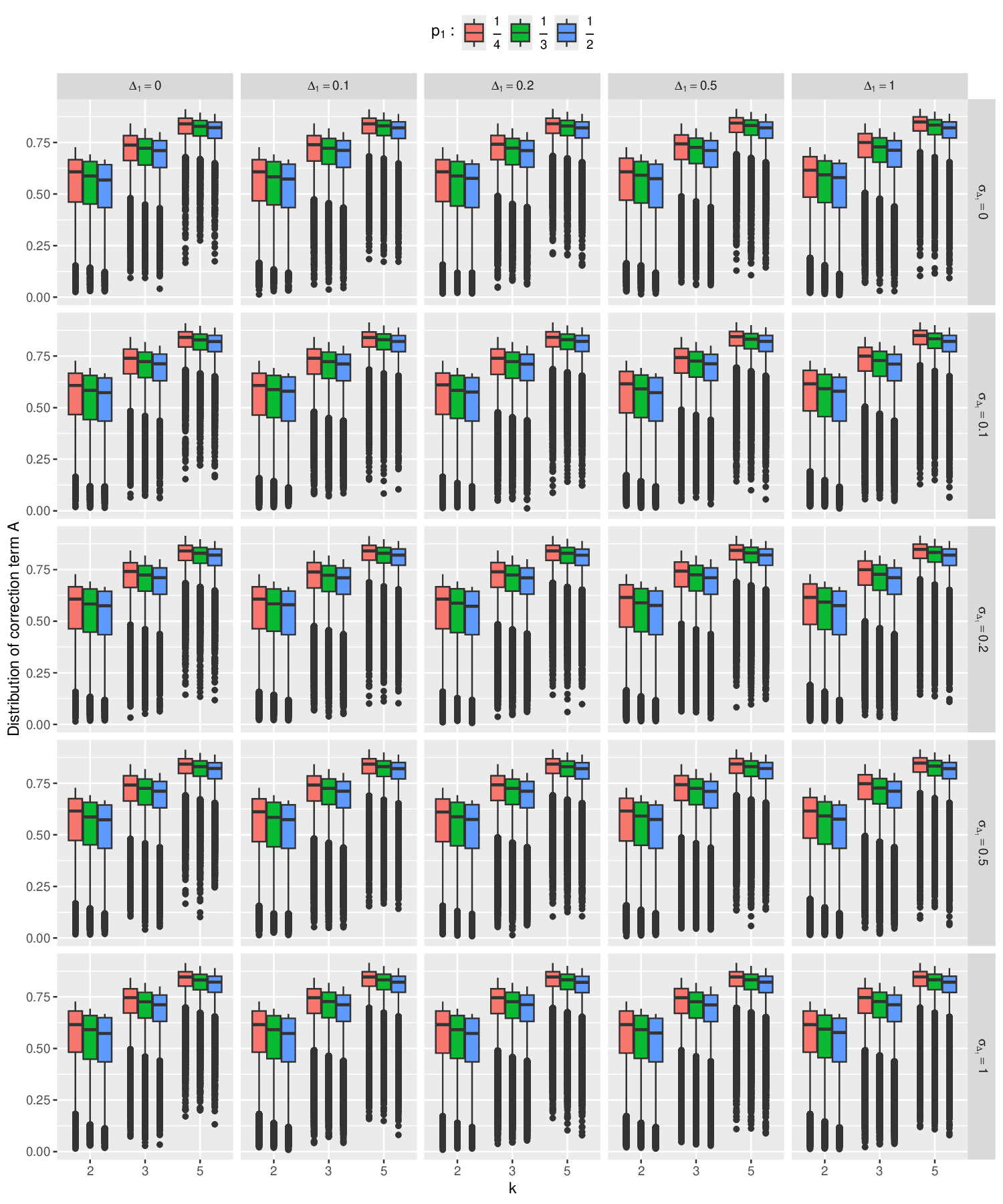}
\caption{\it{ The empirical distribution of the adjustment term A}}
\label{fig s28}
\end{figure}

\FloatBarrier

\section*{Web Appendix D: Data structure required for applying the proposed method}
In this appendix, we present the data structure of the examples used in the main text to illustrate the data required for applying the proposed methods.

\begin{table}[h]
\centering
\caption{Study-level and subgroup-level data used for the two RESPIRE trials.}
\label{tab:respire_data}
\begin{tabular}{lccccl}
\toprule
Data type & Study & Treatment & $\log HR$ & $s_i$ & Subgroup \\
\midrule
\multirow{4}{*}{Study-level}
& RESPIRE 1 & 14 days & -0.6349 & 0.1781 &  \\
& RESPIRE 2 &  & -0.1393 & 0.1698 &  \\
& RESPIRE 1 & 28 days & -0.3147 & 0.1697 &  \\
& RESPIRE 2 &  & -0.3425 & 0.1794 &  \\
\midrule
\multirow{20}{*}{Subgroup-level}
& RESPIRE 1 & 14 days & -0.4943 & 0.2873 & Age ($<65$) \\
& RESPIRE 2 &  & -0.3711 & 0.2200 &  \\
& RESPIRE 1 &  & -0.7133 & 0.2372 & Age ($\geq 65$) \\
& RESPIRE 2 &  &  0.1740 & 0.2704 &  \\
& RESPIRE 1 & 28 days & -0.2744 & 0.2755 & Age ($<65$) \\
& RESPIRE 2 &  & -0.5978 & 0.2371 &  \\
& RESPIRE 1 &  & -0.3011 & 0.2169 & Age ($\geq 65$) \\
& RESPIRE 2 & & -0.0513 & 0.2860 &  \\
\cmidrule{2-6}
& RESPIRE 1 & 14 days & -0.7133 & 0.2216 & Sex (Female) \\
& RESPIRE 2 &  & -0.4463 & 0.2327 &  \\
& RESPIRE 1 &  & -0.4620 & 0.3181 & Sex (Male) \\
& RESPIRE 2 &  &  0.3988 & 0.2899 &  \\
& RESPIRE 1 & 28 days & -0.2485 & 0.1990 & Sex (Female) \\
& RESPIRE 2 &  & -0.5447 & 0.2434 &  \\
& RESPIRE 1 &  & -0.5621 & 0.3467 & Sex (Male)  \\
& RESPIRE 2 &  &  0.0198 & 0.3018 &  \\
\cmidrule{2-6}
& RESPIRE 2 & 14 days & -0.1985 & 0.1957 & Race (White) \\
& RESPIRE 2 &  &  0.0677 & 0.3486 &  \\
& RESPIRE 2 & 28 days & -0.3011 & 0.1974 & Race (Nonwhite)\\
& RESPIRE 2 &  & -0.8210 & 0.4450 &  \\
\bottomrule
\end{tabular}
\end{table}

\begin{table}[h]
\centering
\caption{Study-level and subgroup-level data used for the SGLT2 studies.}
\label{tab:sglt2_data}
\begin{tabular}{lllr r}
\toprule
Data type & Study & Subgroup & $\log HR$ & $s_i$ \\
\midrule
\multirow{6}{*}{Study-level}
& CANVAS program &  & -0.1165 & 0.1422 \\
& CREDENCE &  & -0.2614 & 0.1209 \\
& DAPA-CKD &  & -0.1278 & 0.1094 \\
& DECLARE-TIMI 58 &  & -0.4005 & 0.1795 \\
& EMPA-REG OUTCOME &  & -0.1863 & 0.1122 \\
& VERTIS CV &  & -0.1054 & 0.0988 \\
\midrule
\multirow{32}{*}{Subgroup-level}
& CANVAS program & HbA1c ($\leq 8\%$)& -0.0943 & 0.2186 \\
&  & HbA1c (>8\%)& -0.1508 & 0.1872 \\
&  & Heart failure (No)& -0.1054 & 0.1648 \\
& & Heart failure (Yes)& -0.1393 & 0.2836 \\
&  & Diuretic use (No)& -0.0726 & 0.1982 \\
&  & Diuretic use (Yes)& -0.1744 & 0.2048 \\
\cmidrule{2-5}
& CREDENCE & eGFR ($\geq60$)& 0.0583 & 0.2414 \\
&  & eGFR (<60)& -0.3711 & 0.1379 \\
&  & Heart failure (No)& -0.4155 & 0.1384 \\
&  & Heart failure (Yes)& 0.3148 & 0.2620 \\
&  & Diuretic use (No)& -0.1744 & 0.1704 \\
&  & Diuretic use (Yes)& -0.3567 & 0.1743 \\
\cmidrule{2-5}
& DAPA-CKD & eGFR ($\geq60$)& 0.0677 & 0.4106 \\
&  & eGFR (<60)& -0.1393 & 0.1143 \\
&  & Heart failure (No)& -0.1054 & 0.1208 \\
&  & Heart failure (Yes)& -0.2614 & 0.2726 \\
&  & Diuretic use (No)& -0.3285 & 0.1620 \\
&  & Diuretic use (Yes)& 0.0392 & 0.1455 \\
\cmidrule{2-5}
& DECLARE-TIMI 58 & Heart failure (No)& -0.4308 & 0.2017 \\
&   & Heart failure (Yes)& -0.2485 & 0.3814 \\
&   & Diuretic use (No)& -0.3011 & 0.2425 \\
&   & Diuretic use (Yes)& -0.5276 & 0.2678 \\
\cmidrule{2-5}
& EMPA-REG OUTCOME & Heart failure (No)& -0.1278 & 0.1213 \\
&  & Heart failure (Yes)& -0.5108 & 0.2948 \\
&  & Diuretic use (No)& -0.1508 & 0.1539 \\
&  & Diuretic use (Yes)& -0.2231 & 0.1656 \\
\cmidrule{2-5}
& VERTIS CV & eGFR ($\geq60$)& -0.0834 & 0.1181 \\
&  & eGFR ($<60$)& -0.1393 & 0.1748 \\
&  & Heart failure (No)& -0.1625 & 0.1241 \\
&  & Heart failure (Yes)& 0.0100 & 0.1608 \\
&  & Diuretic use (No)& -0.1054 & 0.1310 \\
&  & Diuretic use (Yes)& -0.1054 & 0.1508 \\
\bottomrule
\end{tabular}
\end{table}
\end{document}